\documentclass[twoside]{article}
\usepackage{amsmath,amsfonts,amssymb}
\usepackage{qic,epsfig}
\usepackage{cite}
\usepackage{color,colortbl}
\usepackage{graphicx,booktabs}
\usepackage{array,latexsym}
\usepackage{textcomp,booktabs}
\usepackage[usenames,dvipsnames]{xcolor}
\usepackage{graphicx}
\usepackage{tikz}

\renewcommand{\theequation}{\arabic{section}.\arabic{equation}}

\newtheorem{thm}{Theorem}
\newtheorem{cor}{Corollary}
\newtheorem{lem}{Lemma}
\newtheorem{defn}{Definition}

\renewenvironment{proof}{\par\noindent{\bf Proof.}}{$\quad\Box$\par}
\newcommand{\ket}[1]{| #1 \rangle}
\newcommand{\bra}[1]{\langle #1 |}

\begin{document}
\setlength{\textheight}{8.0truein}    

\runninghead{Quotient Algebra Partition and Cartan Decomposition
for $su(N)$ IV}
            {Zheng-Yao Su and Ming-Chung Tsai}

\normalsize\textlineskip \thispagestyle{empty}
\setcounter{page}{1}

\vspace*{0.88truein}

\alphfootnote

\fpage{1}

\centerline{\bf Quotient Algebra Partition and Cartan
Decomposition for $su(N)$ IV} \vspace*{0.035truein}
\centerline{\footnotesize Zheng-Yao Su\footnote{Email:
zsu@nchc.narl.org.tw
}\hspace{.15cm} and Ming-Chung Tsai}\centerline{\footnotesize\it
National Center for High-Performance
 Computing,}
 \centerline{\footnotesize\it National Applied Research Laboratories,
 Taiwan, R.O.C.}

\vspace*{0.21truein}

 \abstracts{Else from the quotient algebra partition considered in the preceding episodes~\cite{Su,SuTsai1,SuTsai2},
            two kinds of partitions
            on unitary Lie algebras are created by 
             {\em nonabelian} bi-subalgebras. 
            It is of interest that there exists a {\em partition duality} between
            the two kinds of partitions.
            With an application of an appropriate coset rule,
            the two partitions return to a quotient algebra partition when the generating bi-subalgebra is {\em abelian}.
            Procedures are proposed to {\em merge} or {\em detach} a co-quotient algebra,
            which help deliver type-{\bf AIII} Cartan decompositions
            of more varieties.
            In addition, every Cartan decomposition is obtainable
            from the quotient algebra partition of the highest rank.
            Of significance is the universality of the quotient algebra partition to classical and exceptional Lie algebras.}{}{} 

 \section{Introduction}\label{secintro}
  The main purpose of this episode is to give an examination in depth
  for admissible partitions generated by nonabelian subalgebras.
  As pointed out earlier~\cite{SuTsai1},  a nonabelian subalgebra
  is eligible as well to create a bi-subalgebra 
  and a commutator partitions over a unitary Lie algebra.
  An elegant relation of duality between these two kinds of partitions is
  revealed and expounded in detail.
  Importantly, the partitions of these two kinds dual to each other
  merge into one quotient algebra partition when the associated bi-subalgebra
  recovers to be abelian.
  It is asserted that all Cartan decompositions of three types are acquirable recursively within the quotient algebra partition of the highest rank.
  Moreover, the structure of quotient algebra partition
  is universal to classical and exceptional Lie algebras.

 \section{Bi-Subalgebras of $su(2^p)$\label{secbisubalginsu}}
  Writing its generators in the $s$-representation unveils
  a simple group structure embedded in the Lie algebra $su(2^p)$,
  {\em cf.} Appendix~B in~\cite{Su}.
 \vspace{6pt}
 \begin{lemma}\label{lemisosu2^p}
  The set of spinor generators of $su(2^p)$ forms an abelian group isomorphic to $Z^{2p}_2$
  under the bi-addition $\diamond$:
  $\forall\hspace{2pt}{\cal S}^{\zeta}_{\alpha},{\cal S}^{\eta}_{\beta}\in{su(2^p)}$,
  ${\cal S}^{\zeta}_{\alpha}\diamond{\cal S}^{\eta}_{\beta}\equiv{\cal S}^{\zeta+\eta}_{\alpha+\beta}\in{su(2^p)}$.
 \end{lemma}
 \vspace{3pt}
 \begin{proof}
  This lemma is easily asserted by one-to-one mapping each spinor generator
  ${\cal S}^{\zeta}_{\alpha}$ 
  to the concatenated string $\zeta\circ\alpha\in Z^{2p}_2$, and {\em vice versa}. 
 \end{proof}
 \vspace{6pt}
  With this structure, the concept of a bi-subalgebra introduced in~\cite{SuTsai1}
  is naturally generalised from a Cartan subalgebra to the whole algebra $su(2^p)$.
 \vspace{6pt}
 \begin{defn}\label{defbisubalgsu}
  A set of spinor generators of $su(2^p)$ is called a bi-subalgebra
  of $su(2^p)$, denoted as $\mathcal{B}_{su}$,  if
  ${\cal S}^{\zeta+\eta}_{\alpha+\beta}\in\mathcal{B}_{su}$,
  $\forall\hspace{2pt}{\cal S}^{\zeta}_{\alpha},{\cal S}^{\eta}_{\beta}\in\mathcal{B}_{su}$.
 \end{defn}
 \vspace{6pt}
  Intentionally the notation $\mathcal{B}_{su}$ distinguishes bi-subalgebras of $su(2^p)$
  from those of a Cartan subalgebra~\cite{SuTsai1,SuTsai2}. 
  Similarly, these bi-subalgebras can be {\em ordered} by {\em the degree of maximality}.
 \vspace{6pt}
 \begin{defn}\label{defmaxbisubalgsu}
  A bi-subalgebra $\mathcal{B}_{su}$ is maximal in $su(2^p)$
  if ${\cal S}^{\zeta+\eta}_{\alpha+\beta}\in\mathcal{B}_{su}$,
  $\forall\hspace{2pt}{\cal S}^{\zeta}_{\alpha},{\cal S}^{\eta}_{\beta}\in\mathcal{B}^c_{su}={su(2^p)}-\mathcal{B}_{su}$.
 \end{defn}
 \vspace{6pt}
  The ordering is recursive.
 \vspace{6pt}
 \begin{defn}\label{defrthbisubinsu}
  Denoted as $\mathcal{B}^{[r]}_{su}$, an $r$-th maximal bi-subalgebra of $su(2^p)$
  is a proper maximal  bi-subalgebra of an $(r-1)$-th maximal bi-subalgebra $\mathcal{B}^{[r-1]}_{su}\subset{su(2^p)}$,
  $1<r\leq 2p$.
 \end{defn}
 \vspace{6pt}
  The order $r$ of a such bi-subalgebras ranges from $0$ to $2p$,
  initially with the zeroth maximal $\mathcal{B}^{[0]}_{su}=su(2^p)$ and
  ending at the identity
  $\mathcal{B}^{[2p]}_{su}=\{{\cal S}^{\bf 0}_{\hspace{.01cm}{\bf 0}}\}$.
  Notice that every bi-subalgebra of order less than $p$ is nonabelian,
  and that of order $r\geq p$ is not necessarily abelian.
  The total number of spinors in a bi-subalgebra is acquired
  with the isomorphism of $su(2^p)$ and $Z^{2p}_2$.
 \vspace{6pt}
 \begin{lemma}\label{lemnumbi-subinsu}
  A bi-subalgebra $\mathcal{B}_{su}$ of $su(2^p)$ is an $r$-th
  maximal bi-subalgebra in $su(2^p)$ if and only if it consists of a number $2^{2p-r}$
  of spinor generators.
 \end{lemma}
 \vspace{6pt}

  Most of assertions regarding bi-subalgebras of a Cartan subalgebra
  $\mathfrak{C}\subset{su(2^p)}$
  remain valid to those of an $r$-th maximal bi-subalgebra
  $\mathcal{B}^{[r]}_{su}\subset{su(2^p)}$ as in the following.
  Their proofs are omitted because
  each of which can be carried out through the same procedure
  as that in the corresponding assertion in~\cite{SuTsai1,SuTsai2} except replacing
  $\mathfrak{C}$ with $\mathcal{B}^{[r]}_{su}$.
  Let it begin with the feature that
  a $3$rd member is obtainable from every two bi-subalgebras. 
 \vspace{6pt}
 \begin{lemma}\label{lem3rdmaxbisub}
  Derived from every two maximal bi-subalgebras $\mathcal{B}_{su,1}$ and $\mathcal{B}_{su,2}$ of
  an $r$-th maximal bi-subalgebra $\mathcal{B}^{[r]}_{su}$ in $su(2^p)$, the subspace
  $\mathcal{B}=(\mathcal{B}_{su,1}\cap\mathcal{B}_{su,2})\cup(\mathcal{B}^c_{su,1}\cap\mathcal{B}^c_{su,2})$
  is also a maximal bi-subalgebra of $\mathcal{B}^{[r]}_{su}$,
  here $\mathcal{B}^c_{su,1}=\mathcal{B}^{[r]}_{su}-\mathcal{B}_{su,1}$ and
  $\mathcal{B}^c_{su,2}=\mathcal{B}^{[r]}_{su}-\mathcal{B}_{su,2}$.
 \end{lemma}
 \vspace{6pt}
  An abelian group associated to $\mathcal{B}^{[r]}_{su}$ is needed for a purpose of partitioning.
 \vspace{6pt}
 \begin{lemma}\label{lemabeGinsu}
  Given an $r$-th maximal bi-subalgebra $\mathcal{B}^{[r]}_{su}$ of $su(2^p)$,
  the set $\mathcal{G}(\mathcal{B}^{[r]}_{su})=
  \{\mathcal{B}_{su,i}:
  \mathcal{B}_{su,i}\text{ is a maximal bi-subalgebra of }\mathcal{B}^{[r]}_{su},\hspace{2pt}0\leq i<2^{2p-r}\}$
  forms an abelian group isomorphic to $Z^{2p-r}_2$ under the
  $\sqcap$-operation, that is,
  $\forall\hspace{2pt}\mathcal{B}_{su,i},\mathcal{B}_{su,j}\in\mathcal{G}(\mathcal{B}^{[r]}_{su})$,
  $\mathcal{B}_{su,i}\sqcap\mathcal{B}_{su,j}=(\mathcal{B}_{su,i}\cap\mathcal{B}_{su,j})\cup(\mathcal{B}^c_{su,i}\cap\mathcal{B}^c_{su,j})\in\mathcal{G}(\mathcal{B}^{[r]}_{su})$,
  where $\mathcal{B}^c_{su,i}=\mathcal{B}^{[r]}_{su}-\mathcal{B}_{su,i}$, $\mathcal{B}^c_{su,j}=\mathcal{B}^{[r]}_{su}-\mathcal{B}_{su,j}$
  and $\mathcal{B}_{su,0}=\mathcal{B}^{[r]}_{su}$ is the group identity, $0\leq i,j<2^{2p-r}$.
 \end{lemma}
 \vspace{6pt}
  Every spinor commutes with a unique maximal
  bi-subalgebra $\mathcal{B}_{su}$ of $\mathcal{B}^{[r]}_{su}$
  but with none part of its complement.
 \vspace{6pt}
 \begin{lemma}\label{lemspinorcommmax}
  For an $r$-th maximal bi-subalgebra $\mathcal{B}^{[r]}_{su}$ of
  $su(2^p)$ and a spinor generator ${\cal S}^{\zeta}_{\alpha}\in{su(2^p)}$, $p>1$,
  there exists a unique maximal bi-subalgebra $\mathcal{B}_{su}\in\mathcal{G}(\mathcal{B}^{[r]}_{su})$
  of $\mathcal{B}^{[r]}_{su}$ such that $[{\cal S}^{\zeta}_{\alpha},\mathcal{B}_{su}]=0$
  and $[{\cal S}^{\zeta}_{\alpha},{\cal S}^{\eta}_{\beta}]\neq 0$
  for all ${\cal S}^{\eta}_{\beta}\in\mathcal{B}^c_{su}
  =\mathcal{B}^{[r]}_{su}-\mathcal{B}_{su}$. 
 \end{lemma}
 \vspace{3pt}
 \vspace{6pt}
 \begin{lemma}\label{lemspinorinBcomm}
  If a spinor ${\cal S}^{\zeta}_{\alpha}\in{su(2^p)}$ commuting
  with a maximal bi-subalgebra $\mathcal{B}_{su}\in\mathcal{G}(\mathcal{B}^{[r]}_{su})$
  of an $r$-th maximal bi-subalgebra $\mathcal{B}^{[r]}_{su}\subset{su(2^p)}$,
  i.e., $[{\cal S}^{\zeta}_{\alpha},\mathcal{B}_{su}]=0$,
  any generator ${\cal S}^{\eta}_{\beta}$
  commuting with ${\cal S}^{\zeta}_{\alpha}$
  must be in $\mathcal{B}_{su}$.
 \end{lemma}
 \vspace{6pt}
  Based on the {\em commutator rule} $[{\cal W},\mathcal{B}_{su}]=0$,
  the subspace ${\cal W}\subset{su(2^p)}$ spanned by
  generators commuting with a bi-subalgebra $\mathcal{B}_{su}$ is 
  defined to be {\em the commutator subspace determined by} $\mathcal{B}_{su}$
  and denoted as ${\cal W}(\mathcal{B}_{su})$ whenever necessary.
  Any two such subspaces are disjoint.
 \vspace{6pt}
 \begin{lemma}\label{lemdisjointComm}
  Two commutator subspaces ${\cal W}_1$ and ${\cal W}_2\subset{su(2^p)}$
  respectively determined by two maximal bi-subalgebras $\mathcal{B}_{su,1}$
  and $\mathcal{B}_{su,2}\in\mathcal{G}(\mathcal{B}^{[r]}_{su})$
  of an $r$-th maximal bi-subalgebra $\mathcal{B}^{[r]}_{su}$ of
  $su(2^p)$, i.e., $[{\cal W}_1,\mathcal{B}_{su,1}]=0$
  and $[{\cal W}_2,\mathcal{B}_{su,2}]=0$,
  share the null intersection ${\cal W}_1\cap{\cal W}_2=\{0\}$.
 \end{lemma}
 \vspace{6pt}
  Thanks to Lemmas from~\ref{lemabeGinsu} to~\ref{lemdisjointComm},
  a partition is affirmed in $su(2^p)$.
 \vspace{6pt}
 \begin{thm}\label{thmCommPar}
  The group $\mathcal{G}(\mathcal{B}^{[r]}_{su})$ comprising
  maximal bi-subalgebras of an $r$-th maximal bi-subalgebra $\mathcal{B}^{[r]}_{su}$
  in the Lie algebra $su(2^p)$ by the commutator rule decides a partition of $su(2^p)$.
 \end{thm}
 \vspace{6pt}
  This partition, consisting of $2^{2p-r}$ commutator subspaces in total,
  is termed
  {\em the commutator partition of order $r$ generated by} $\mathcal{B}^{[r]}_{su}$
  and
  denoted as $\{{\cal P}_{\mathcal{C}}(\mathcal{B}^{[r]}_{su})\}$.
  The partition has a group structure too.
 \vspace{6pt}
 \begin{lemma}\label{lemcommsubclose}
  For two commutator subspaces ${\cal W}_1$ and ${\cal W}_2$
  respectively determined by the maximal bi-subalgebras $\mathcal{B}_{su,1}$
  and $\mathcal{B}_{su,2}\in\mathcal{G}(\mathcal{B}^{[r]}_{su})$
  of an $r$-th maximal bi-subalgebra $\mathcal{B}^{[r]}_{su}$
  of $su(2^p)$, the closure holds that,
  $\forall\hspace{2pt}{\cal S}^{\zeta}_{\alpha}\in{\cal W}_1$ and ${\cal S}^{\eta}_{\beta}\in{\cal W}_2$,
  the bi-additive generator ${\cal S}^{\zeta+\eta}_{\alpha+\beta}$
  belongs to the commutator subspace ${\cal W}=[{\cal W}_1,{\cal W}_2]$
  determined by the bi-subalgebra
  $\mathcal{B}_{su,1}\sqcap\mathcal{B}_{su,2}\in\mathcal{G}(\mathcal{B}^{[r]}_{su})$,
  i.e., $[{\cal S}^{\zeta+\eta}_{\alpha+\beta},\mathcal{B}_{su,1}\sqcap\mathcal{B}_{su,2}]=0$
  and thus $[{\cal W},\mathcal{B}_{su,1}\sqcap\mathcal{B}_{su,2}]=0$.
 \end{lemma}
 \vspace{6pt}

 Generated by an $r$-th maximal bi-subalgebra, group isomorphisms analogous to
 those of Corollary~2 in~\cite{SuTsai1} are preserved.
 \vspace{6pt}
 \begin{cor}\label{corisoBrsu}
  Given an $r$-th maximal bi-subalgebra $\mathcal{B}^{[r]}_{su}$
  of the Lie algebra $su(2^p)$, there exists the isomorphism relation
  $\{\mathcal{P}_{\mathcal{C}}(\mathcal{B}^{[r]}_{su})\}
  \cong\mathcal{G}(\mathcal{B}^{[r]}_{su})
  \cong\{{\cal S}^{\zeta}_{\alpha}:{\cal S}^{\zeta}_{\alpha}\in\mathcal{B}^{[r]}_{su}\}
  \cong{Z^{2p-r}_2}$ for the sets of all commutator subspaces in $su(2^p)$,
  of all maximal bi-subalgebras, of all spinor generators in $\mathcal{B}^{[r]}_{su}$
  and of all $(2p-r)$-digit strings under the commutator, the
  $\sqcap$, the bi-addition $\diamond$ operations and the bit-wise
  addition respectively.
 \end{cor}
 \vspace{6pt}
  Due to this isomorphism relation,
  it is favorable to label every maximal bi-subalgebra
  in $\mathcal{G}(\mathcal{B}^{[r]}_{su})$
  by a $(2p-r)$-digit binary string such that
  $\mathcal{B}_{su,i}\sqcap\mathcal{B}_{su,j}=\mathcal{B}_{su,i+j}$
  for $i,j\in{Z^{2p-r}_2}$ with the designation
  $\mathcal{B}^{[r]}_{su}=\mathcal{B}_{su,\mathbf{0}}$.
  To every commutator partition, there corresponds a
  {\em dual partition} yielded by a coset rule. 
 \vspace{6pt}
 \begin{thm}\label{thmbisubpar}
  An $r$-th maximal bi-subalgebra $\mathcal{B}^{[r]}_{su}\equiv\mathcal{B}^{[r,\mathbf{0}]}_{su}\subset{su(2^p)}$
  can generate a partition in $su(2^p)$ consisting of $2^r$ disjoint subspaces $\mathcal{B}^{[r,i]}_{su}$
  via the coset rule of partition,
  namely $su(2^p)=\bigcup_{i\in{Z^r_2}}\mathcal{B}^{[r,i]}_{su}$
  complying with the condition $\forall\hspace{2pt}{\cal S}^{\zeta}_{\alpha}\in\mathcal{B}^{[r,i]}_{su}$,
  ${\cal S}^{\eta}_{\beta}\in\mathcal{B}^{[r,j]}_{su}$,
  $\exists !\hspace{3pt}\mathcal{B}^{[r,l]}_{su}$  such that
  ${\cal S}^{\zeta+\eta}_{\alpha+\beta}\in\mathcal{B}^{[r,l]}_{su}$
  with $i+j=l$, $\forall\hspace{2pt}i,j,l\in{Z^r_2}$.
 \end{thm}
 \vspace{6pt}
  Under the operation of bi-addition every subspace $\mathcal{B}^{[r,i]}_{su}$ is
  a coset of the ``subgroup" $\mathcal{B}^{[r]}_{su}$ in $su(2^p)$
  and is thus named {\em a coset subspace of} $\mathcal{B}^{[r]}_{su}$.
  Denoted as $\{\mathcal{P}_{\mathcal{B}}(\mathcal{B}^{[r]}_{su})\}$,
  the partition comprising these $2^r$ cosets is known as
  {\em the bi-subalgebra partition of order $r$ generated by} $\mathcal{B}^{[r]}_{su}$.

  \section{Partition Duality\label{secPARDUAL}}
  The partitions over the Lie algebra $su(2^p)$ generated by the commutator
  and the coset rules introduced in last section are in general distinct.
  However, as immediately to be shown, the two partitions
  have an elegant duality relation.
 \vspace{6pt}
 \begin{lemma}\label{lemcommBiofBrsu}
  The subspace spanned by spinors commuting with an $r$-th maximal
  bi-subalgebra $\mathcal{B}^{[r]}_{su}$ of $su(2^p)$, $0\leq r\leq 2p$,
  forms a $(2p-r)$-th maximal bi-subalgebra of $su(2^p)$.
 \end{lemma}
 \vspace{3pt}
 \begin{proof}
  Suppose $V\subset{su(2^p)}$ is the subspace spanned by spinors
  commuting with $\mathcal{B}^{[r]}_{su}$.
  For any spinors ${\cal S}^{\zeta_1}_{\alpha_1},{\cal S}^{\zeta_2}_{\alpha_2}\in{V}$,
  the bi-additive  ${\cal S}^{\zeta_1+\zeta_2}_{\alpha_1+\alpha_2}$
  commutes with $\mathcal{B}^{[r]}_{su}$ owing to the vanishing
  commutators
  $[{\cal S}^{\zeta_1}_{\alpha_1},\mathcal{B}^{[r]}_{su}]=[{\cal S}^{\zeta_2}_{\alpha_2},\mathcal{B}^{[r]}_{su}]=0$.
  Thus ${\cal S}^{\zeta_1+\zeta_2}_{\alpha_1+\alpha_2}\in{V}$
  and the subspace $V\equiv\mathcal{B}$ is a bi-subalgebra.
  By Lemma~\ref{lemnumbi-subinsu}, the fact of $\mathcal{B}$
  being a $(2p-r)$-th maximal bi-subalgebra of $su(2^p)$
  can be affirmed by counting the number of
  spinors commuting with $\mathcal{B}^{[r]}_{su}$.
  Let the subalgebra $\mathcal{B}^{[r]}_{su}$ be
  spanned by $2p-r$ independent generators
  $\{{\cal S}^{\eta_1}_{\beta_1},{\cal S}^{\eta_2}_{\beta_2},\cdots,{\cal S}^{\eta_{2p-r}}_{\beta_{2p-r}}\}$.
  It is easy to derive a number $2^r$ of spinors ${\cal S}^{\zeta}_{\alpha}\in{su(2^p)}$
  satisfying the identities
  $\zeta\cdot\beta_t+\eta_t\cdot\alpha=0$ for all $1\leq t\leq 2p-r$.
  Hence $\mathcal{B}$ is composed of in total $2^r$ spinor generators and
  is a $(2p-r)$-th maximal bi-subalgebra of $su(2^p)$.
 \end{proof}
 \vspace{6pt}
  That is to say, there exists a {\em unique}
  $(2p-r)$-th maximal bi-subalgebra $\mathcal{B}^{[2p-r]}_{su}$ 
  commuting with a given $r$-th maximal bi-subalgebra
  $\mathcal{B}^{[r]}_{su}\subset{su(2^p)}$.
  The following two lemmas pave the way to the duality relation.
 \vspace{6pt}
 \begin{lemma}\label{lemcommsubcoset}
  A commutator subspace of the commutator partition
  $\{{\cal P}_{\mathcal{C}}(\mathcal{B}^{[r]}_{su})\}$
  generated by an $r$-th maximal bi-subalgebra $\mathcal{B}^{[r]}_{su}\subset{su(2^p)}$
  is a coset subspace of the bi-subalgebra partition
  $\{{\cal P}_{\mathcal{B}}(\mathcal{B}^{[2p-r]}_{su})\}$
  generated by a $(2p-r)$-th maximal bi-subalgebra
  $\mathcal{B}^{[2p-r]}_{su}\subset{su(2^p)}$
  as long as
  $[\mathcal{B}^{[r]}_{su},\mathcal{B}^{[2p-r]}_{su}]=0$, $0\leq r\leq 2p$.
 \end{lemma}
 \vspace{3pt}
 \begin{proof}
  This lemma is an implication of Lemma~\ref{lemcommsubclose}
  by relating the commutator subspaces of
  $\{{\cal P}_{\mathcal{C}}(\mathcal{B}^{[r]}_{su})\}$
  in terms of bitwise addition of index strings. 
  Specifically for two arbitrary commutator subspaces
  ${\cal W}(\mathcal{B}_{su,i})$ and ${\cal W}(\mathcal{B}_{su,j})$
  determined by two maximal bi-subalgebras
  $\mathcal{B}_{su,i}$ and $\mathcal{B}_{su,j}\in\mathcal{G}(\mathcal{B}^{[r]}_{su})$
  respectively,
  the bi-additive ${\cal S}^{\zeta+\eta}_{\alpha+\beta}$
  of any two spinors
  ${\cal S}^{\zeta}_{\alpha}\in{\cal W}(\mathcal{B}_{su,i})$
  and
  ${\cal S}^{\eta}_{\beta}\in{\cal W}(\mathcal{B}_{su,j})$
  belongs to ${\cal W}(\mathcal{B}_{su,i+j}=\mathcal{B}_{su,i}\sqcap\mathcal{B}_{su,j})$,
  $i,j\in{Z^{2p-r}_2}$.
  By this index rule, 
  the subalgebra
  $\mathcal{B}^{[2p-r]}_{su}=\mathcal{B}_{su,\mathbf{0}}$
  is designated to be ${\cal W}(\mathcal{B}_{su,\mathbf{0}})$.
 \end{proof}
 \vspace{6pt}
 \vspace{6pt}
 \begin{lemma}\label{lemcosetcommwith}
  A coset subspace of the bi-subalgebra partition $\{{\cal P}_{\mathcal{B}}(\mathcal{B}^{[r]}_{su})\}$
  generated by an $r$-th maximal bi-subalgebra $\mathcal{B}^{[r]}_{su}\subset{su(2^p)}$
  is a commutator subspace of the commutator partition $\{{\cal P}_{\mathcal{C}}(\mathcal{B}^{[2p-r]}_{su})\}$
  generated by a $(2p-r)$-th maximal bi-subalgebra
  $\mathcal{B}^{[2p-r]}_{su}\subset{su(2^p)}$
  as long as
  $[\mathcal{B}^{[r]}_{su},\mathcal{B}^{[2p-r]}_{su}]=0$,
  $0\leq r \leq 2p$.
 \end{lemma}
 \vspace{3pt}
 \begin{proof}
  This lemma is asserted by showing the fact that,
  for any coset subspace
  $\mathcal{B}^{[r,i]}_{su}\in\{{\cal P}_{\mathcal{B}}(\mathcal{B}^{[r]}_{su})\}$,
  there exists a unique maximal bi-subalgebra
  $\mathcal{B}_{su,i}$ 
  of $\mathcal{B}^{[2p-r]}_{su}$ such that
  $\mathcal{B}^{[r,i]}_{su}$ commutes with $\mathcal{B}_{su,i}$
  but with no part of the complement
  ${\mathcal{B}}^c_{su,i}
  =\mathcal{B}^{[2p-r]}_{su}-\mathcal{B}_{su,i}$,
  here $\mathcal{B}^{[r]}_{su}=\mathcal{B}^{[r,\mathbf{0}]}_{su}$
  and  $i\in{Z^r_2}$.
  With an arbitrary choice ${\cal S}^{\zeta_0}_{\alpha_0}\in\mathcal{B}^{[r,i]}_{su}$,
  every generator of $\mathcal{B}^{[r,i]}_{su}$ can be written as the bi-additive
  ${\cal S}^{\zeta_0+\xi}_{\alpha_0+\gamma}$ of ${\cal S}^{\zeta_0}_{\alpha_0}$
  and a spinor ${\cal S}^{\xi}_{\gamma}\in\mathcal{B}^{[r]}_{su}$,
  {\em cf}. Theorem~\ref{thmbisubpar}.
  Due to the vanishing commutator
  $[\mathcal{B}^{[r]}_{su},\mathcal{B}^{[2p-r]}_{su}]=0$,
  a spinor ${\cal S}^{\eta}_{\beta}\in\mathcal{B}^{[2p-r]}_{su}$
  commutes with the whole bi-subalgebra $\mathcal{B}^{[r,i]}_{su}$
  if $[{\cal S}^{\zeta_0}_{\alpha_0},{\cal S}^{\eta}_{\beta}]=0$
  or with none subset of it if $[{\cal S}^{\zeta_0}_{\alpha_0},{\cal S}^{\eta}_{\beta}]\neq 0$.
  There exists at least one generator of $\mathcal{B}^{[2p-r]}_{su}$
  commuting with $\mathcal{B}^{[r,i]}_{su}$, otherwise
  contradictions occur.
  Let $\mathcal{V}\subset\mathcal{B}^{[2p-r]}_{su}$
  be the subspace spanned by all spinors
  commuting with $\mathcal{B}^{[r,i]}_{su}$.
  The bi-additive ${\cal S}^{\eta_1+\eta_2}_{\beta_1+\beta_2}$
  of any ${\cal S}^{\eta_1}_{\beta_1}$ and ${\cal S}^{\eta_2}_{\beta_2}\in\mathcal{V}$
  must commute with $\mathcal{B}^{[r,i]}_{su}$ because of 
  the vanishing commutators
  $[{\cal S}^{\eta_1}_{\beta_1},\mathcal{B}^{[r,i]}_{su}]=[{\cal S}^{\eta_2}_{\beta_2},\mathcal{B}^{[r,i]}_{su}]=0$.
  Further by the nonvanishing commutators
  $[{\cal S}^{\zeta_0}_{\alpha_0},{\cal S}^{\hat{\eta}_1}_{\hat{\beta}_1}]\neq 0$
  and $[{\cal S}^{\zeta_0}_{\alpha_0},{\cal S}^{\hat{\eta}_2}_{\hat{\beta}_2}]\neq 0$
  for any pair ${\cal S}^{\hat{\eta}_1}_{\hat{\beta}_1}$
  and
  ${\cal S}^{\hat{\eta}_2}_{\hat{\beta}_2}\in\mathcal{V}^c=\mathcal{B}^{[2p-r]}_{su}-\mathcal{V}$,
  it derives
  the commuting of
  the bi-additive ${\cal S}^{\hat{\eta}_1+\hat{\eta}_2}_{\hat{\beta}_1+\hat{\beta}_2}$
  and ${\cal S}^{\zeta_0}_{\alpha_0}$ and then
  $[{\cal S}^{\hat{\eta}_1+\hat{\eta}_2}_{\hat{\beta}_1+\hat{\beta}_2},\mathcal{B}^{[r,i]}_{su}]=0$.
  This validates $\mathcal{V}\equiv\mathcal{B}_{su,i}$ to be a maximal bi-subalgebra
  of $\mathcal{B}^{[2p-r]}_{su}$.
  The uniqueness of $\mathcal{B}_{su,i}$ is shown by contradiction.
  Suppose there have two maximal bi-subalgebras
  $\mathcal{B}_{su,1}$ and $\mathcal{B}_{su,2}$ both commuting with $\mathcal{B}^{[r,i]}_{su}$
  and at least one generator
  ${\cal S}^{\eta_3}_{\beta_3}\in\mathcal{B}_{su,1}$
  and
  ${\cal S}^{\eta_3}_{\beta_3}\in\mathcal{B}^{[2p-r]}_{su}-\mathcal{B}_{su,2}$.
  An inconsistency arises that
  $[{\cal S}^{\eta_3}_{\beta_3},\mathcal{B}^{[r,i]}_{su}]=0$
  as well as
  $[{\cal S}^{\eta_3}_{\beta_3},\mathcal{B}^{[r,i]}_{su}]\neq 0$.
  The lemma is thus affirmed.
 \end{proof}
 \vspace{6pt}
  A {\em partition duality} hidden in the Lie algebra $su(2^p)$
  is thus disclosed.
 \vspace{6pt}
 \begin{thm}\label{thmBiCommBrsu}
  In the Lie algebra $su(2^p)$,
  the commutator partition of order $r$
  $\{{\cal P}_{\mathcal{C}}(\mathcal{B}^{[r]}_{su})\}$
  generated by an $r$-th maximal bi-subalgebra
  $\mathcal{B}^{[r]}_{su}\subset{su(2^p)}$
  is the bi-subalgebra partition of order $2p-r$
  $\{{\cal P}_{\mathcal{B}}(\mathcal{B}^{[2p-r]}_{su})\}$
  generated by a $(2p-r)$-th maximal bi-subalgebra
  $\mathcal{B}^{[2p-r]}_{su}\subset{su(2^p)}$
  provided the two bi-subalgebras
  $\mathcal{B}^{[r]}_{su}$ and $\mathcal{B}^{[2p-r]}_{su}$
  commute, $0\leq r\leq 2p$.
 \end{thm}
 \vspace{6pt}
  The partitions
  $\{{\cal P}_{\mathcal{C}}(\mathcal{B}^{[r]}_{su})\}$ and $\{{\cal P}_{\mathcal{B}}(\mathcal{B}^{[r]}_{su})\}$
  become identical when
  $\mathcal{B}^{[r]}_{su}$ is a Cartan subalgebra. 
  For a Cartan subalgebra 
  is a $p$-th maximal bi-subalgebra of $su(2^p)$
  and is the maximal subspace commuting with itself.
 \vspace{6pt}
 \begin{cor}\label{coroBiComCartan}
  The commutator partition and the bi-subalgebra partition generated by a same
  Cartan subalgebra are identical.
 \end{cor}
 \vspace{6pt}

  An important linkage is to be examined that
  both a bi-subalgebra partition
  $\{{\cal P}_{\mathcal{B}}(\mathcal{B}^{[r]}_{su})\}$
  and a commutator partition
  $\{{\cal P}_{\mathcal{C}}(\mathcal{B}^{[r]}_{su})\}$
  can return to be a {\em Quotient Algebra Partition (QAP)}
  if the generating bi-subalgebra $\mathcal{B}^{[r]}_{su}$
  is abelian.
  The partition conversions mainly reply upon the fact that
  an abelian bi-subalgebra of order $r$ $\mathcal{B}^{[r]}_{su}$ is
  an $(r-p)$-th maximal bi-subalgebra
  of a Cartan subalgebra $\mathfrak{C}\subset{su(2^p)}$, $p\leq r\leq 2p$.
  While $\mathcal{B}^{[r]}_{su}=\mathfrak{B}^{[r-p]}$ being a bi-subalgebra of $\mathfrak{C}$,
  each coset subspace of $\{{\cal P}_{\mathcal{B}}(\mathcal{B}^{[r]}_{su})\}$
  is also a {\em partitioned conjugate-pair subspace}
  ${\cal W}(\mathfrak{B},\mathfrak{B}^{[r-p]};s)$
  of the doublet $(\mathfrak{B},\mathfrak{B}^{[r-p]})$,
  for $s\in{Z^{r-p}_2}$ 
  and $\mathfrak{B}\in\mathcal{G}(\mathfrak{C})$ being a maximal bi-subalgebra of $\mathfrak{C}$,
  {\em cf.} Lemma~5 of~\cite{SuTsai2}.
  This equivalence 
  requires the existence
  of a unique maximal bi-subalgebra $\mathfrak{B}$ 
  commuting with 
  the former subspace.
  \vspace{6pt}
   \begin{lemma}\label{lemcosubconjupair}
    A coset subspace of the bi-subalgebra partition
    $\{{\cal P}_{\mathcal{B}}(\mathcal{B}^{[r]}_{su})\}$
    generated by an abelian $r$-th maximal bi-subalgebra
    $\mathcal{B}^{[r]}_{su}\subset{su(2^p)}$, $p\leq r\leq 2p$,
    uniquely commutes with a maximal bi-subalgebra
    of a Cartan subalgebra $\mathfrak{C}\supset\mathcal{B}^{[r]}_{su}$.
   \end{lemma}
   \vspace{2pt}
   \begin{proof}
    Every spinor generator of $su(2^p)$ must commute with a unique maximal bi-subalgebra
    of $\mathfrak{C}$ by Lemma~3 in~\cite{SuTsai1}.
    Let ${\cal S}^{\hspace{.5pt}\zeta}_{\alpha}$
    be an arbitrary generator in a coset subspace
    $\mathcal{B}^{[r,i]}_{su}\in\{{\cal P}_{\mathcal{B}}(\mathcal{B}^{[r]}_{su})\}$
    and commute with a maximal bi-subalgebra
    $\mathfrak{B}\in\mathcal{G}(\mathfrak{C})$ of $\mathfrak{C}$,
    here $i\in{Z^r_2}$ and $\mathcal{B}^{[r]}_{su}=\mathcal{B}^{[r,\mathbf{0}]}_{su}$.
    According to Theorem~\ref{thmbisubpar},
    each generator in $\mathcal{B}^{[r,i]}_{su}$ else from ${\cal S}^{\hspace{.5pt}\zeta}_{\alpha}$
    can be written as a bi-additive ${\cal S}^{\hspace{.5pt}\zeta+\eta}_{\alpha+\beta}$
    with a generator ${\cal S}^{\hspace{.2pt}\eta}_{\beta}\in\mathcal{B}^{[r]}_{su}$,
    ${\cal S}^{\hspace{.2pt}\eta}_{\beta}\neq{\cal S}^{\mathbf{0}}_{\hspace{.4pt}\mathbf{0}}$.
    Owing to the vanishing commutator $[\mathcal{B}^{[r]}_{su},\mathfrak{B}]=0$
    as $\mathcal{B}^{[r]}_{su}\subset\mathfrak{C}$
    and by Lemma~1 in~\cite{SuTsai1}, it derives the commuting of
    ${\cal S}^{\hspace{.5pt}\zeta+\eta}_{\alpha+\beta}$ and $\mathfrak{B}$.
    The uniqueness of a such maximal bi-subalgebra $\mathfrak{B}$
    can be confirmed by Lemma~3 in~\cite{SuTsai2}. 
   \end{proof}
   \vspace{6pt}
  With the application of the {\em coset rule of bisection}~\cite{SuTsai1,SuTsai2},
  the subspace ${\cal W}(\mathfrak{B},\mathfrak{B}^{[r-p]};s)$ further divides into
  two {\em conditioned subspaces} $W(\mathfrak{B},\mathfrak{B}^{[r-p]};s)$
  and $\hat{W}(\mathfrak{B},\mathfrak{B}^{[r-p]};s)$,
  which are respectively abelian and a member subspace of
  the partition $\{{\cal P}_{\mathcal{Q}}(\mathcal{B}^{[r]}_{su})\}$.
  Returning to a quotient-algebra partition from a bi-subalgebra partition is then arrived.
 \vspace{6pt}
 \begin{cor}\label{coroBiPartoQA}
  Imposed with the coset rule of bisection, the bi-subalgebra partition $\{{\cal P}_{\mathcal{B}}(\mathcal{B}^{[r]}_{su})\}$
  generated by an $r$-th maximal bi-subalgebra
  $\mathcal{B}^{[r]}_{su}\subset su(2^p)$  
  recovers to be a quotient algebra partition
  $\{{\cal P}_{\mathcal{Q}}(\mathcal{B}^{[r]}_{su})\}$ 
  when $\mathcal{B}^{[r]}_{su}$ being abelian.
 \end{cor}
 \vspace{6pt}

  In advance of converting to a quotient-algebra partition, 
  a commutator partition
  $\{{\cal P}_{\mathcal{C}}(\mathcal{B}^{[r]}_{su})\}$
  transits to the bi-subalgebra partition
  $\{{\cal P}_{\mathcal{B}}(\mathcal{B}^{[r]}_{su})\}$
  resorting to the {\em coset rule of partition}~\cite{SuTsai2}.
  \vspace{6pt}
 \begin{lemma}\label{lemcommsubdivide}
  Refined by applying the coset rule of partition,
  the commutator partition $\{{\cal P}_{\mathcal{C}}(\mathcal{B}^{[r]}_{su})\}$
  generated by an $r$-th maximal bi-subalgebra
  $\mathcal{B}^{[r]}_{su}\subset su(2^p)$ 
  changes into the bi-subalgebra partition
  $\{{\cal P}_{\mathcal{B}}(\mathcal{B}^{[r]}_{su})\}$.
 \end{lemma}
 \vspace{3pt}
 \begin{proof}
  Specifically,
  every commutator subspace ${\cal W}(\mathcal{B}_{su,i})$ of
  $\{{\cal P}_{\mathcal{C}}(\mathcal{B}^{[r]}_{su})\}$
  can divide into a number $2^{2r-2p}$ of
  subspaces ${\cal W}(\mathcal{B}_{su,i};s)$
  respecting the coset rule of partition:
  ${\cal S}^{\zeta+\eta}_{\alpha+\beta}\in{\cal W}(\mathcal{B}_{su,i+j};s+t)$
  for every pair
  ${\cal S}^{\zeta}_{\alpha}\in{\cal W}(\mathcal{B}_{su,i};s)$ and
  ${\cal S}^{\eta}_{\beta}\in{\cal W}(\mathcal{B}_{su,j};t)$
  with ${\cal W}(\mathcal{B}_{su,\mathbf{0}};\mathbf{0})=\mathcal{B}^{[r]}_{su}$,
  here $\mathcal{B}_{su,i},\mathcal{B}_{su,j}\in\mathcal{G}(\mathcal{B}^{[r]}_{su})$,
  $i,j\in{Z^{2p-r}_2}$, and $s,t\in{Z^{2r-2p}_2}$.
  This division is validated following the same procedures in Lemmas~1, 5 and~15
  of~\cite{SuTsai2}
  except replacing
  the bi-subalgebra $\mathfrak{B}^{[r]}$ of $\mathfrak{C}$ by
  $\mathcal{B}^{[r]}_{su}$ and
  the Cartan subalgebra
  $\mathfrak{C}\subset{su(2^p)}$ by a $(2p-r)$-th maximal bi-subalgebra
  $\mathcal{B}^{[2p-r]}_{su}\subset{su(2^p)}$.
  Here, $\mathcal{B}^{[2p-r]}_{su}$ is the unique bi-subalgebra
  commuting with $\mathcal{B}^{[r]}_{su}$, {\em cf.} Lemma~\ref{lemcommBiofBrsu}, and
  the latter is a $(2r-2p)$-th maximal bi-subalgebra of the former. 
  With the above substitution,
  Lemmas~1 and~5 of~\cite{SuTsai2} affirm
  the coset rule on partitioned subspaces
  ${\cal W}(\mathcal{B}_{su,i};s)$, $s\in{Z^{2r-2p}_2}$, of a same maximal bi-subalgebra
  $\mathcal{B}_{su,i}\in\mathcal{G}(\mathcal{B}^{[r]}_{su})$.
  While Lemma~15 of~\cite{SuTsai2} accounts for
  the rule relating partitioned subspaces
  associated to different maximal bi-subalgebras, 
  recalling the labelling choice
  $\mathcal{B}_{su,i}\sqcap\mathcal{B}_{su,j}=\mathcal{B}_{su,i+j}$. 
 \end{proof}
 \vspace{6pt}
  Notice that this lemma is valid whether or not the bi-subalgebra
  $\mathcal{B}^{[r]}_{su}$ is abelian.
  After a required bisection on each divided subspace, the conversion completes.
 \vspace{6pt}
 \begin{cor}\label{coroCommPartoQA}
  Imposed with the coset rules of partition and bisection,
  the commutator partition $\{{\cal P}_{\mathcal{C}}(\mathcal{B}^{[r]}_{su})\}$
  generated by an $r$-th maximal bi-subalgebra
  $\mathcal{B}^{[r]}_{su}\subset su(2^p)$ 
  recovers to be a quotient algebra partition
  $\{{\cal P}_{\mathcal{Q}}(\mathcal{B}^{[r]}_{su})\}$
  when $\mathcal{B}^{[r]}_{su}$ being abelian.
 \end{cor}
 \vspace{6pt}
  Remind that the imposition of the two coset rules is orderless~\cite{SuTsai2}.
  It is concluded that a bi-subalgebra partition and
  a commutator partition are two notons sharing a duality relation, and both
  can respectively return to a quotient-algebra partition as the generating
  bi-subalgebra is abelian.

  A quotient-algebra partition
  is endowed with an {\em abelian group structure}.
  Recall Theorem~3 in~\cite{SuTsai2} that,
  within the quotient-algebra partition of rank $r$
  $\{{\cal P}_{\mathcal{Q}}(\mathfrak{B}^{[r]})\}$ generated by an $r$-th maximal bi-subalgebra
  $\mathfrak{B}^{[r]}$ of a Cartan subalgebra
  $\mathfrak{C}\subset{su(N)}$,
  $0\leq r\leq p$,
  the conditioned subspaces ${W}^{\epsilon}(\mathfrak{B},\mathfrak{B}^{[r]};i)$
  satisfy the {\em quaternion condition of closure of rank $r$},
  for all $\epsilon,\sigma\in{Z_2}$, $i,j\in{Z^r_2}$
  and $\mathfrak{B},\mathfrak{B}'\in\mathcal{G}(\mathfrak{C})$,
 \begin{align}\label{eqgenWcommrankr}
  [{W}^{\epsilon}(\mathfrak{B},\mathfrak{B}^{[r]};i), {W}^{\sigma}(\mathfrak{B}',\mathfrak{B}^{[r]};j)]\subset
  {W}^{\epsilon+\sigma}(\mathfrak{B}\sqcap\mathfrak{B}',\mathfrak{B}^{[r]};i+j),
  \end{align}
  here ${W}^{0}(\mathfrak{C},\mathfrak{B}^{[r]};\mathbf{0})=\{0\}$
  and
  ${W}^{1}(\mathfrak{C},\mathfrak{B}^{[r]};\mathbf{0})=\mathfrak{C}$.
  Moreover,
  an operation $\circledcirc$ called {\em tri-addition} is then defined in
  $\{{\cal P}_{\mathcal{Q}}(\mathfrak{B}^{[r]})\}$,
  {\em cf.} Corollary~2 in~\cite{SuTsai2}:
  ${W}^{\epsilon}(\mathfrak{B},\mathfrak{B}^{[r]};i)\circledcirc{W}^{\sigma}(\mathfrak{B}',\mathfrak{B}^{[r]};j)
  ={W}^{\epsilon+\sigma}(\mathfrak{B}\sqcap\mathfrak{B}',\mathfrak{B}^{[r]};i+j)\in\{\mathcal{P}_{\mathcal{Q}}(\mathfrak{B}^{[r]})\}$
  for every pair
  ${W}^{\epsilon}(\mathfrak{B},\mathfrak{B}^{[r]};i)$
  and
  ${W}^{\sigma}(\mathfrak{B}',\mathfrak{B}^{[r]};j)
  \in\{\mathcal{P}_{\mathcal{Q}}(\mathfrak{B}^{[r]})\}$.
  Under this operation, 
  the set of subspaces in
  $\{{\cal P}_{\mathcal{Q}}(\mathfrak{B}^{[r]})\}$ is isomorphic
  to the additive group $Z^{p+r+1}_2$.
  Expositions in the subsequent sections will make use of
  this group structure of importance.

 \section{Merging and Detaching a Co-Quotient Algebra\label{secmergedetach}}
  One of major subjects concerned in this serial of articles~\cite{Su,SuTsai1,SuTsai2}
  is to establish the scheme comprehensively producing admissible decompositions of unitary Lie algebras.
  Relying upon the framework of quaotient-algebra partition, such a scheme
  makes easy the systematic generating of Cartan decompositions.
  A Cartan decomposition
  $su(N)=\mathfrak{t}\oplus\mathfrak{p}$ is the direct sum 
  of a subalgebra $\mathfrak{t}$ and a vector subspace $\mathfrak{p}$
  satisfying the decomposition condition
  $[\mathfrak{t},\mathfrak{t}]\subset\mathfrak{t}$,
  $[\mathfrak{t},\mathfrak{p}]\subset\mathfrak{p}$,
  $[\mathfrak{p},\mathfrak{p}]\subset\mathfrak{t}$,
  and
  $\text{Tr}(\mathfrak{t}\hspace{.5pt}\mathfrak{p})=0$.
  The success of the decomposition production is attributed to
  a merit of the scheme recognizing the truth that
  the subalgebra $\mathfrak{t}$ of a decomposition
  $\mathfrak{t}\oplus\mathfrak{p}$ is a proper maximal subgroup
  of a quotient-algebra partition under the operation of tri-addition~\cite{SuTsai2}.
  Thanks to the group-structured nature of quotient-algebra partitions,
  a terse law to decide the type of a decomposition has been concluded~\cite{SuTsai2}:
  a Cartan decomposition $su(N)=\mathfrak{t}\oplus\mathfrak{p}$
  is of type {\bf AI} if the maximal
  abelian subalgebra of $\mathfrak{p}$ is a Cartan subalgebra $\mathfrak{C}\subset{su(N)}$,
  a decomposition of type {\bf AII} if the maximal
  abelian subalgebra of $\mathfrak{p}$ is a proper maximal bi-subalgebra $\mathfrak{B}$ of $\mathfrak{C}$,
  or a type {\bf AIII} if the maximal
  abelian subalgebra of $\mathfrak{p}$ is a complement
  $\mathfrak{B}^c=\mathfrak{C}-\mathfrak{B}$.

  The arrangements of quotient and co-quotient algebras help
  schematic manipulations of a quotient-algebra partition~\cite{Su,SuTsai1,SuTsai2}.
  In the quotient-algebra partition of rank $r$
  $\{\mathcal{P}_{\mathcal{Q}}(\mathfrak{B}^{[r]})\}$
  generated by an $r$-th maximal bi-subalgebra
  $\mathfrak{B}^{[r]}$ of a Cartan subalgebra $\mathfrak{C}\subset{su(N)}$,
  for $0\leq r\leq p$ and $2^{p-1}<N\leq 2^p$,
  there determine the quotient algebra $\{{\cal Q}(\mathfrak{B}^{[r]})\}$
  given by the center subalgebra $\mathfrak{B}^{[r]}$ and
  a set of co-quotient algebras, each of which
  $\{\mathcal{Q}({W}^{\epsilon}(\mathfrak{B},\mathfrak{B}^{[r]};i))\}$
  is given by a non-null conditioned subspace
  ${W}^{\epsilon}(\mathfrak{B},\mathfrak{B}^{[r]};i)\in\{\mathcal{P}_{\mathcal{Q}}(\mathfrak{B}^{[r]})\}$
  other than $\mathfrak{B}^{[r]}$,
  here $\epsilon\in{Z_2}$, $i\in{Z^r_2}$ and
  $\mathfrak{B}\in\mathcal{G}(\mathfrak{C})$ being a maximal bi-subalgebra of $\mathfrak{C}$.
  Yet, two kinds of co-quotient algebras are admitted
  according to the two occasions that the center subalgebra
  ${W}^{\epsilon}(\mathfrak{B},\mathfrak{B}^{[r]};i)$ is a
  {\em degrade} conditioned subspace as
  $\mathfrak{B}\supset\mathfrak{B}^{[r]}$
  or a {\em regular} one as $\mathfrak{B}\nsupseteq\mathfrak{B}^{[r]}$~\cite{SuTsai2}.
  The two kinds exhibit distinct properties as detailed in this section:
  the former allows a procedure of  {\em merging} conjugate pairs and
  the latter can convert to a refined version
  by {\em detaching} the pairs.

 \subsection{Merging a Co-Quotient Algebra\label{subsecmerge}}
  To illustrate the merging procedure, the quotient algebra of rank
  $r$  $\{{\cal Q}(\mathfrak{B}^{[r]};2^{p+r}-1)\}$
  given by $\mathfrak{B}^{[r]}\subset\mathfrak{C}$
  and the co-quotient algebra of rank $r$
  $\{{\cal Q}(\mathfrak{B}^{[r,l]};2^{p+r}-2^{2r-2})\}$ given by
  a coset $\mathfrak{B}^{[r,l]}$ of $\mathfrak{B}^{[r]}$ in $\mathfrak{C}$,
  $l\in{Z^r_2-\{\mathbf{0}\}}$,
  are respectively displayed in Figs.~\ref{figmergQA}(a) and~\ref{figmergQA}(b).
  Here it is no lack of universality to take the coset $\mathfrak{B}^{[r,l]}$
  as the center subalgebra of the co-quotient algebra,
  for a non-null degrade conditioned subspace
  ${W}^{\epsilon}(\mathfrak{B},\mathfrak{B}^{[r]};i)$ with $\mathfrak{B}\supset\mathfrak{B}^{[r]}$
  is a coset of $\mathfrak{B}^{[r]}$ in a Cartan subalgebra
  $\mathfrak{C}^{\ast}$ being a superset of both
  $\mathfrak{B}^{[r]}$ and
  ${W}^{\epsilon}(\mathfrak{B},\mathfrak{B}^{[r]};i)$.
  As depicted in Fig.~\ref{figmergQA},
  the maximal bi-subalgebras $\mathfrak{B}_m$ and $\mathfrak{B}_n\in\mathcal{G}(\mathfrak{C})$
  keep the closure identity 
  $\mathfrak{B}_1=\mathfrak{B}_m\sqcap\mathfrak{B}_n$, $1<m,n<2^p$,
  and the relations hold for subspace indices
  $s+\hat{s}=i+\hat{i}=j+\hat{j}=l$ and $s=i+j=\hat{i}+\hat{j}$,
  $i,\hat{i},j,\hat{j},s,\hat{s}\in{Z^r_2}$.
  In addition, $\mathfrak{B}_1$ is a superset of $\mathfrak{B}^{[r]}$
  but $\mathfrak{B}_1\nsupseteq\mathfrak{B}^{[r,l]}$.

  The merging procedure is conducted in $\{{\cal Q}(\mathfrak{B}^{[r,l]};2^{p+r}-2^{2r-2})\}$
  by pairwise combining conditioned subspaces that commute.
  Two options of merged co-quotient algebras
  are then rendered, the one via {\em parallel} merging as shown
  in Fig.~\ref{figmergQA}(c) and
  the other via {\em crossing} merging in~\ref{figmergQA}(d).
  Based on the original structure in Figs.~\ref{figmergQA}(a) or~\ref{figmergQA}(b),
  it is plain to verify
  the commutation relation of Eq.~\ref{eqgenWcommrankr} preserved in these two algebras.
  As a consequence of the merging, there yield two choices of $(r-1)$-th maximal bi-subalgebras
  $\mathfrak{B}^{[r-1]}=\mathfrak{B}^{[r]}\cup{W}(\mathfrak{B}_1,\mathfrak{B}^{[1]};s)$
  of a Cartan subalgebra $\mathfrak{C}_{merg}$
  and
  $\hat{\mathfrak{B}}^{[r-1]}=\mathfrak{B}^{[r]}\cup\hat{W}(\mathfrak{B}_1,\mathfrak{B}^{[1]};\hat{s})$
  of another Cartan subalgebra $\hat{\mathfrak{C}}_{merg}$,
  for ${W}(\mathfrak{B}_1,\mathfrak{B}^{[r]};s)$ and
  $\hat{W}(\mathfrak{B}_1,\mathfrak{B}^{[r]};\hat{s})$ being a conditioned
  subspace of the doublet $(\mathfrak{B}_1,\mathfrak{B}^{[1]})$.
  The next crucial step is to assert the fact that each abelian subspace thus merged is a conditioned subspace
  in the quotient-algebra of rank $r-1$
  $\{\mathcal{P}_{\mathcal{Q}}(\widetilde{\mathfrak{B}}^{[r-1]})\}$
  given by
  $\widetilde{\mathfrak{B}}^{[r-1]}=\mathfrak{B}^{[r-1]}\subset\mathfrak{C}^{\hspace{0.5pt}\ast}=\mathfrak{C}_{merg}$
  or by
  $\widetilde{\mathfrak{B}}^{[r-1]}=\hat{\mathfrak{B}}^{[r-1]}\subset\mathfrak{C}^{\hspace{0.5pt}\ast}=\hat{\mathfrak{C}}_{merg}$.

   Without loss of generality, let the merged subspace
   $W=W(\mathfrak{B}_m,\mathfrak{B}^{[r]};i)\cup W(\mathfrak{B}_n,\mathfrak{B}^{[r]};j)$
   in Fig.~\ref{figmergQA}(c) be  an example.
   There have two requirements for $W$ to be a conditioned subspace in the quotient-algebra
   partition $\{\mathcal{P}_{\mathcal{Q}}(\mathfrak{B}^{[r-1]})\}$,
   specifically, $W$ commuting
   with a maximal bi-subalgebra $\mathfrak{B}$ of $\mathfrak{C}_{merg}$ and
   $\forall\hspace{2pt}{\cal S}^{\zeta}_{\alpha},{\cal S}^{\eta}_{\beta}\in{W}$,
   ${\cal S}^{\zeta+\eta}_{\alpha+\beta}\in\mathfrak{B}\cap\mathfrak{B}^{[r-1]}$,
   {\em cf.} Definition~2 in~\cite{SuTsai2}.
   The subspace $W$ is abelian due to the commutation relation
   of Eq.~\ref{eqgenWcommrankr}
   $[{W}(\mathfrak{B}_m,\mathfrak{B}^{[r]};i), {W}(\mathfrak{B}_n,\mathfrak{B}^{[r]};j)]\subset
    \hat{W}(\mathfrak{B}_1,\mathfrak{B}^{[r]};s)=\{0\}$.
   For any two generators ${\cal S}^{\zeta}_{\alpha}$ and ${\cal S}^{\eta}_{\beta}\in{W}$,
   the inclusion holds that either
   ${\cal S}^{\zeta}_{\alpha},{\cal S}^{\eta}_{\beta}\in W(\mathfrak{B}_m,\mathfrak{B}^{[r]};i)$
   ($\in W(\mathfrak{B}_n,\mathfrak{B}^{[r]};j)$) or
   ${\cal S}^{\zeta}_{\alpha}\in W(\mathfrak{B}_m,\mathfrak{B}^{[r]};i)$ and
   ${\cal S}^{\eta}_{\beta}\in W(\mathfrak{B}_n,\mathfrak{B}^{[r]};j)$.
   By definition, the bi-additive generator ${\cal S}^{\zeta+\eta}_{\alpha+\beta}$ is in
   $\mathfrak{B}^{[r]}\subset\mathfrak{C}_{merg}$ for the $1$st case.
   In the $2$nd case, the generator ${\cal S}^{\zeta+\eta}_{\alpha+\beta}$
   is contained in the subspace
   ${W}(\mathfrak{B}_1,\mathfrak{B}^{[r]};s)\subset\mathfrak{C}_{merg}$ owing to
   Corollary~3 in~\cite{SuTsai2}.
   Thus the generator ${\cal S}^{\zeta+\eta}_{\alpha+\beta}$ must
   belong to $\mathfrak{B}^{[r-1]}$.
    Since the generator ${\cal S}^{\zeta+\eta}_{\alpha+\beta}\in\mathfrak{C}_{merg}$ commutes
    with not only ${\cal S}^{\zeta}_{\alpha}$ and ${\cal S}^{\eta}_{\beta}\in W$
    but also the whole $W$ of being abelian,
    it deduces the existence of a non-null subspace $V$ in $\mathfrak{C}_{merg}$
    commuting with $W$.
    Further endorsed by Lemma~4 in~\cite{SuTsai1}, such a non-null subspace ought to be a maximal
   bi-subalgebra  $V\equiv\mathfrak{B}$, {\em i.e.},  $[W,\mathfrak{B}]=0$,
   for $\mathfrak{C}_{merg}$ being a Cartan subalgebra.
   Meanwhile, ${\cal S}^{\zeta+\eta}_{\alpha+\beta}$ is in $\mathfrak{B}$
   by Lemma~14 in~\cite{SuTsai1}.
   The assertion is then arrived that $W$
   is a conditioned subspace of the doublet $(\mathfrak{B},\mathfrak{B}^{[r-1]})$
   in the quotient-algebra partition $\{\mathcal{P}_{\mathcal{Q}}(\mathfrak{B}^{[r-1]})\}$.
   Similarly, the other merged subspaces in Fig.~\ref{figmergQA}(c)
   are respectively a conditioned subspace in $\{\mathcal{P}_{\mathcal{Q}}(\mathfrak{B}^{[r-1]})\}$.
   By the same token on the other hand, each subspace in Fig.~\ref{figmergQA}(d) is a
   conditioned subspace in $\{\mathcal{P}_{\mathcal{Q}}(\mathfrak{B}^{[r-1]})\}$.
   That is to say, the partition $\{\mathcal{P}_{\mathcal{Q}}(\mathfrak{B}^{[r-1]})\}$
   admits co-quotient algebras of both ranks $r$ and $r-1$.

 \subsection{Detaching a Co-Quotient Algebra\label{subsecmerge}}
  The exposition of the detaching starts with,
  respectively in Figs.~\ref{figdetQA}(a) and~\ref{figdetQA}(b),
  the quotient algebra of rank
  $r$  $\{{\cal Q}(\mathfrak{B}^{[r]};2^{p+r}-1)\}$ given by $\mathfrak{B}^{[r]}\subset\mathfrak{C}$
  and the co-quotient algebra of rank $r$
  $\{{\cal Q}({W}(\mathfrak{B}_1,\mathfrak{B}^{[r]};s);2^{p+r}-1)\}$ given by
  a regular conditioned subspace ${W}(\mathfrak{B}_1,\mathfrak{B}^{[r]};s)$,
  $\mathfrak{B}_1\nsupseteq\mathfrak{B}^{[r]}$ and $s\in{Z^r_2}$.
  The maximal bi-subalgebras $\mathfrak{B}_m$ and $\mathfrak{B}_n\in\mathcal{G}(\mathfrak{C})$
  in Fig.~\ref{figdetQA} 
  holds the closure
  $\mathfrak{B}_1=\mathfrak{B}_m\sqcap\mathfrak{B}_n$, $1<m,n<2^p$,
  and the identity $s=i+j$ relates the indices $i,j,s\in{Z^r_2}$.
  Notice that the intersection
  $\mathfrak{B}^{[r+1]}=\mathfrak{B}_1\cap\mathfrak{B}^{[r]}$ is an
  $(r+1)$-th maximal bi-subalgebra of $\mathfrak{C}$ for $\mathfrak{B}_1\nsupseteq\mathfrak{B}^{[r]}$.
  The detaching procedure is performed in the co-quotient algebra
  $\{{\cal Q}({W}(\mathfrak{B}_1,\mathfrak{B}^{[r]};s))\}$
  by bisecting each conditioned subspace
  in Fig.~\ref{figdetQA}(a) or~\ref{figdetQA}(b). 
  In order to validate the two refined versions of co-quotient algebras,
  via {\em parallel} and {\em crossing} detachings
  as in Figs.~\ref{figdetQA}(c) and (d),
  it requires the affirmation that subspaces in these two
  structures respect the commutation relation of Eq.~\ref{eqgenWcommrankr}
  and are respectively a conditioned subspace in the
  quotient-algebra partition of rank $r+1$
  $\{\mathcal{P}_{\mathcal{Q}}(\mathfrak{B}^{[r+1]})\}$ given by $\mathfrak{B}^{[r+1]}$.

  Take the subspace ${W}(\mathfrak{B}_m,\mathfrak{B}^{[r]};i)$
  in Fig.~\ref{figdetQA}(b) as an example.
  This subspace is divided into two halves 
  ${W}(\mathfrak{B}_m,\mathfrak{B}^{[r+1]};0\circ i)$
  and ${W}(\mathfrak{B}_m,\mathfrak{B}^{[r+1]};1\circ i)$
  following the coset rule:
  $\forall\hspace{2pt}{\cal S}^{\zeta}_{\alpha},{\cal S}^{\eta}_{\beta}
  \in{W}(\mathfrak{B}_m,\mathfrak{B}^{[r+1]};0\circ i)$
  ($\in{W}(\mathfrak{B}_m,\mathfrak{B}^{[r+1]};1\circ i)$),
  ${\cal S}^{\zeta+\eta}_{\alpha+\beta}\in\mathfrak{B}_m\cap\mathfrak{B}^{[r+1]}$;
  here the subspace index $0\circ s$
  is the concatenation of the digit $0$ and string $s\in{Z^r_2}$
  and $\hat{W}(\mathfrak{B}_1,\mathfrak{B}^{[r+1]};0\circ s)=\{0\}$.
  Neither of the subspaces is null if
  $\mathfrak{B}_m\nsupseteq\mathfrak{B}^{[r+1]}$
  or either of them is so if
  $\mathfrak{B}_m\supset\mathfrak{B}^{[r+1]}$.
  Apparently both the subspaces commute with
  $\mathfrak{B}_m\subset\mathfrak{C}$ and are separately a conditioned subspace of
  the doublet $(\mathfrak{B}_m,\mathfrak{B}^{[r+1]})$ by Definition~2 in~\cite{SuTsai2}.
  Hence each of the two refined subspaces is a conditioned subspace of
  the quotient-algebra partition of rank $r+1$
  $\{\mathcal{P}_{\mathcal{Q}}(\mathfrak{B}^{[r+1]})\}$.
  Likewise, every of the other subspaces in Figs.~\ref{figdetQA}(c) and (d)
  is as well a conditioned subspace of $\{\mathcal{P}_{\mathcal{Q}}(\mathfrak{B}^{[r+1]})\}$.
  Remark that the center subalgebra {\em splits} into a non-null
  ${W}(\mathfrak{B}_1,\mathfrak{B}^{[r+1]};0\circ s)$
  and an empty subspace
  $\hat{W}(\mathfrak{B}_1,\mathfrak{B}^{[r+1]};0\circ s)=\{0\}$
  due to the inclusion $\mathfrak{B}_1\supset\mathfrak{B}^{[r+1]}$.
  Therefore, besides one rank $r$,
  a regular conditioned subspace
  ${W}(\mathfrak{B}_1,\mathfrak{B}^{[r]};s)$
  can generate a co-quotient algebra of rank $r+1$ through the detaching procedure,
  $1\leq r<p$.

 \section{More on Cartan Decompositions of Type AIII\label{secgenAIII}}
 \renewcommand{\theequation}{\arabic{section}.\arabic{equation}}
\setcounter{equation}{0} \noindent
  Supplementary to early discussions of Cartan decompositions of the three types~\cite{SuTsai2},
  the attention of this section will focus on the type {\bf AIII} of $su(m+n)$ as $m\neq n$.
  The {\em intrinsic} Cartan decomposition
  $su(m+n)=\hat{\mathfrak{t}}_{\hspace{1pt}\rm III}\oplus\hat{\mathfrak{p}}_{\hspace{1pt}\rm III}$
  of type {\bf AIII} for $m\geq n$, as designated in~\cite{Helgason,SuTsai1},
  is composed of the subalgebra
  $\hat{\mathfrak{t}}_{\hspace{1pt}{\rm III}}=c\otimes su(m)\otimes su(n)
  =\{(\begin{array}{cc}
  A&0\\
  0&B
  \end{array}):A\in{u(m)},B\in{u(n)}\text{ and }{\text{Tr}(A+B)=0}\}$ and
  the subspace
  $\hat{\mathfrak{p}}_{\hspace{1pt}{\rm III}}=
  \{(\begin{array}{cc}
  0&{\cal Z}\\
  -{\cal Z}^{\dagger}&0
  \end{array}):{\cal Z} \in M_{m\times n}(\mathbb{C})\}$;
  note $[c,\hat{\mathfrak{t}}_{\hspace{1pt}{\rm III}}]=0$.
  The subalgebra $\hat{\mathfrak{t}}_{\hspace{1pt}{\rm III}}$ is spanned by a total
  number $m^2+n^2-1$ of generators and
  the set $\hat{\mathfrak{a}}_{\hspace{1pt}{\rm III}}=\{i\ket{k}\bra{k+m}-i\ket{k+m}\bra{k}:1\leq k\leq n\}$
  is optionally a maximal abelian subalgebra in $\hat{\mathfrak{p}}_{\hspace{1pt}{\rm III}}$.
  Since those of $su(N)$ for $2^{p-1}<N<2^p$
  are acquirable by applying the removing process~\cite{Su,SuTsai1,SuTsai2},
  as always only the decompositions of $su(2^p)$ are considered. 
  The decompositions of $su(m+n)$ with $m=n$
  have been addressed in Theorem~3 in~\cite{SuTsai2}.
  In examining other instances of $m>n$, the $\lambda$-generators
  are a better choice as the generating set of the algebra. 
  The set consists of the off-diagonal matrices $\lambda_{kl}=\ket{k}\bra{l}+\ket{l}\bra{k}$ and
  $\hat{\lambda}_{kl}=-i\ket{k}\bra{l}+i\ket{l}\bra{k}$ and
  the diagonal ones $d_{kl}=\ket{k}\bra{k}-\ket{l}\bra{l}$, $1\leq k,l\leq N$,
  which satisfy the commutator relations as listed in from Eqs.~A.1 to A.6 in Appendix~A of~\cite{Su}.

  To generate decompositions of type {\bf AIII} in $su(m+n=2^p)$ for $m>n$,
  it is essentail to perform an appropriate 
  {\em division} over the conditioned subspaces of a quotient-algebra
  partition of {\em rank zero}.
  The exposition is confined to 
  the {\em intrinsic} quotient-algebra partition of this rank 
  $\{\mathcal{P}_{\mathcal{Q}}(\mathfrak{C}_{[\mathbf{0}]})\}$
  given by the {\em intrinsic} Cartan subalgebra
  $\mathfrak{C}_{[\mathbf{0}]}=\{{\cal S}^{\nu_0}_{\hspace{.5pt}\mathbf{0}}:\forall\hspace{2pt}\nu_0\in{Z^p_2}\}
  \subset{su(2^p)}$. 
  For the other decomposition of the same type can be mapped to a decomposition determined
  in $\{\mathcal{P}_{\mathcal{Q}}(\mathfrak{C}_{[\mathbf{0}]})\}$
  via a conjugate transformation~\cite{Su,SuTsai1,SuTsai2}.
  The partition $\{\mathcal{P}_{\mathcal{Q}}(\mathfrak{C}_{[\mathbf{0}]})\}$
  comprises the following conditioned subspaces, $\gamma\in{Z^p_2}-\{\mathbf{0}\}$,
  \begin{align}\label{eqintrW}
    &{W}^0(\mathfrak{B}_{\mathbf{0}})=\{0\},\notag\\
    &{W}^1(\mathfrak{B}_{\mathbf{0}})=\mathfrak{C}_{[\mathbf{0}]}
    =\{{\cal S}^{\nu}_{\mathbf{0}}:\forall\hspace{2pt}\nu\in{Z^p_2},\}
    =\{d_{ij}:\forall\hspace{2pt}0<i,j\leq 2^p\},\notag\\
    &{W}^0(\mathfrak{B}_{\gamma})
     =\{{\cal S}^{\hat{\xi}}_{\gamma}:\forall\hspace{2pt}\hat{\xi}\in{Z^p_2},\hat{\xi}\cdot\gamma=1\}\notag\\
    &=\{\hat{\lambda}_{ij}:1\leq i,j\leq 2^p,
   \hspace{2pt}i-1=\omega,j-1=\tau,\omega+\tau=\gamma\text{ for }\omega,\tau\in{Z^p_2}\},\text{ and}\notag\\
   &{W}^1(\mathfrak{B}_{\gamma})
    =\{{\cal S}^{\xi}_{\gamma}:\forall\hspace{2pt}\xi\in{Z^p_2},\xi\cdot\gamma=0\}\notag\\
   &=\{\lambda_{ij}:1\leq i,j\leq 2^p,
   \hspace{2pt}i-1=\omega,j-1=\tau,\omega+\tau=\gamma\text{ for }\omega,\tau\in{Z^p_2}\},
  \end{align}
  where $\mathfrak{B}_{\gamma}=\{{\cal S}^{\mu}_{\mathbf{0}}:\forall\hspace{2pt}\mu\in{Z^p_2}\text{ and }\mu\cdot\gamma=0\}$
  is a maximal bi-subalgebra of
  $\mathfrak{C}_{[\mathbf{0}]}=\mathfrak{B}_{\mathbf{0}}$.
  Thanks to the commutation relation of Eq.~\ref{eqgenWcommrankr} aforesaid
  and by Corollary~2 in~\cite{SuTsai2},
  these conditioned subspaces form an {\em abelian group}
  obeying the closure under the {\em tri-addition}
  $\circledcirc$: for all
  $\alpha,\beta\in{Z^p_2}$ and $\epsilon,\sigma\in{Z_2}$,
  \begin{align}\label{eqtriaddinEp4}
   {W}^{\epsilon}(\mathfrak{B}_{\alpha})\circledcirc{W}^{\sigma}(\mathfrak{B}_{\beta})
   ={W}^{\epsilon+\sigma}(\mathfrak{B}_{\alpha+\beta}).
  \end{align}
  The division cuts each conditioned subspace
  ${W}^{\epsilon}(\mathfrak{B}_{\alpha})$ into two subspaces
  ${W}^{\epsilon}(\mathfrak{B}_{\alpha};\kappa)$ additionally tagged with
  an index $\kappa\in{Z_2}$.
  Let $\{\mathcal{P}_{\mathcal{Q}}(\mathfrak{C}_{[\mathbf{0}]})\}_{div}$
  denote the partition constituted by these
  divided conditioned subspaces. 
  The key of the subspace division is to preserve the same group structure in this 
  version of partition.

  In 
  $\{\mathcal{P}_{\mathcal{Q}}(\mathfrak{C}_{[\mathbf{0}]})\}_{div}$,
  the subspace
  ${W}^0(\mathfrak{B}_{\mathbf{0}})={W}^0(\mathfrak{B}_{\mathbf{0}};0)\cup{W}^0(\mathfrak{B}_{\mathbf{0}};1)$
  is divided into two null subspaces and
  the subspace ${W}^1(\mathfrak{B}_{\mathbf{0}})$
  into the null ${W}^1(\mathfrak{B}_{\mathbf{0}};1)=\{0\}$
  and ${W}^1(\mathfrak{B}_{\mathbf{0}};0)=\mathfrak{C}_{[\mathbf{0}]}$.
  As to regular conditioned subspaces, each ${W}^\epsilon(\mathfrak{B}_{\gamma\neq\mathbf{0}})$
  splits into two non-null subspaces ${W}^\epsilon(\mathfrak{B}_{\gamma\neq\mathbf{0}};0)$
  and ${W}^\epsilon(\mathfrak{B}_{\gamma\neq\mathbf{0}};1)$
  by a cut over the generating set depending on the integer $m$ of $su(m+n)$.
  Specifically, the divided subspace ${W}^0(\mathfrak{B}_{\gamma\neq\mathbf{0}};0)$
  (or ${W}^1(\mathfrak{B}_{\gamma\neq\mathbf{0}};0)$)
  contains the generators $\hat{\lambda}_{kl}$ (or $\lambda_{kl}$)
  with subscripts ranging in $0<k<l\leq m$,
  while ${W}^0(\mathfrak{B}_{\gamma\neq\mathbf{0}};1)$ 
  (or ${W}^1(\mathfrak{B}_{\gamma\neq\mathbf{0}};1)$)
  carries $\hat{\lambda}_{k'l'}$ (or $\lambda_{k'l'}$)
  with $0<k'<m<l'<2^p$. 
  Thus, the general forms of the divided subspaces can be written as follows,
  $\gamma\in{Z^p_2}-\{\mathbf{0}\}$,
  \begin{align}\label{eqQAPrefined}
   &{W}^0(\mathfrak{B}_{\mathbf{0}};0)
   ={W}^0(\mathfrak{B}_{\mathbf{0}};1)=\{0\},\notag\\
   &{W}^1(\mathfrak{B}_{\mathbf{0}};0)=\mathfrak{C}_{[\mathbf{0}]},\hspace{2pt}
   {W}^1(\mathfrak{B}_{\mathbf{0}};1)=\{0\},\notag\\
   &{W}^0(\mathfrak{B}_{\gamma};0)
   =\{\hat{\lambda}_{kl}:\forall\hspace{2pt}0<k<l\leq m,\omega=k-1,\tau=l-1\text{ and }\omega+\tau=\gamma,\omega,\tau\in{Z^p_2}\},\notag\\
   &{W}^0(\mathfrak{B}_{\gamma};1)
   =\{\hat{\lambda}_{k'l'}:\forall\hspace{2pt}0<k'\leq m<l'\leq 2^p,\omega'=k'-1,\tau'=l'-1\text{ and }\omega'+\tau'=\gamma,\omega',\tau'\in{Z^p_2}\},\notag\\
   &{W}^1(\mathfrak{B}_{\gamma};0)
   =\{\lambda_{kl}:\forall\hspace{2pt}0<k<l\leq m,\omega=k-1,\tau=l-1\text{ and }\omega+\tau=\gamma,\omega,\tau\in{Z^p_2}\},\text{ and}\notag\\
   &{W}^1(\mathfrak{B}_{\gamma};1)
   =\{\lambda_{k'l'}:\forall\hspace{2pt}0<k'\leq m<l'\leq 2^p,\omega'=k'-1,\tau'=l'-1\text{ and }\omega'+\tau'=\gamma,\omega',\tau'\in{Z^p_2}\}.
  \end{align}
  It is easy to validate the tri-addition closure for these subspaces in terms of the identity
  \begin{align}\label{eqtriaddrefsub}
   {W}^{\epsilon}(\mathfrak{B}_{\alpha};\kappa)\circledcirc{W}^{\sigma}(\mathfrak{B}_{\beta};\kappa')
   ={W}^{\epsilon+\sigma}(\mathfrak{B}_{\alpha+\beta};\kappa+\kappa'),
  \end{align}
  for $\epsilon,\sigma,\kappa,\kappa'\in{Z_2}$ and
  $\alpha,\beta\in{Z^p_2}$,
  which establishes the demanded structure of an abelian group
  in $\{\mathcal{P}_{\mathcal{Q}}(\mathfrak{C}_{[\mathbf{0}]})\}_{div}$.
  Based on this group structure,
  generating Cartan decompositions
  becomes routine taking advantage of the assertion
  in Lemma~20 of~\cite{SuTsai2}.
  That is, the subalgebra $\mathfrak{t}$ of a decomposition $\mathfrak{t}\oplus\mathfrak{p}$
  is a {\em proper maximal subgroup} of $\{\mathcal{P}_{\mathcal{Q}}(\mathfrak{C}_{[\mathbf{0}]})\}_{div}$
  under the tri-addition.

  Owing to the truth of the Cartan subalgebra $\mathfrak{C}_{[\mathbf{0}]}={W}^1(\mathfrak{B}_{\mathbf{0}};0)$
  being either in $\mathfrak{t}$ or $\mathfrak{p}$
  of a decomposition $\mathfrak{t}\oplus\mathfrak{p}$,
  there breed two types of Cartan decompositions in the partition
  $\{\mathcal{P}_{\mathcal{Q}}(\mathfrak{C}_{[\mathbf{0}]})\}_{div}$, {\em cf.} Corollary~5 in~\cite{SuTsai2}:
  a type {\bf AI} as $\mathfrak{C}_{[\mathbf{0}]}\subset\mathfrak{p}$
  or a type {\bf AIII} as
  $\mathfrak{C}_{[\mathbf{0}]}\subset\mathfrak{t}$.
  In the former case, the subalgebra $\mathfrak{t}_{\rm I}$ of
  a type-{\bf AI} decomposition $\mathfrak{t}_{\rm I}\oplus\mathfrak{p}_{\rm I}$
  coincides with the algebra $so(2^p)$ up to a conjugate transformation.
  Being a proper maximal subgroup of $\{\mathcal{P}_{\mathcal{Q}}(\mathfrak{C}_{[\mathbf{0}]})\}_{div}$,
  the set $\hat{\mathfrak{t}}_{\hspace{1pt}{\rm III}}$ embracing
  all divided subspaces with the subspace index $\kappa=0$,
  {\em i.e.},
  \begin{align}\label{eqA3tp}
   \hat{\mathfrak{t}}_{\hspace{1pt}{\rm III}}=\{{W}^{\epsilon}(\mathfrak{B}_{\alpha};0):
   \forall\hspace{2pt}\epsilon\in{Z_2}\text{ and }\alpha\in{Z^p_2}\},
  \end{align}
  is the subalgebra $\hat{\mathfrak{t}}_{\hspace{1pt}{\rm III}}=c\oplus{su(m)}\oplus{su(n)}$
  of the {\em intrinsic} type-{\bf AIII} decomposition
  $\hat{\mathfrak{t}}_{\hspace{1pt}{\rm III}}\oplus\hat{\mathfrak{p}}_{\hspace{1pt}{\rm III}}$. 
  The subspace $\hat{\mathfrak{a}}_{\hspace{1pt}{\rm III}}={W}^0(\mathfrak{B}_{10\cdots 0};1)$
  is an option of a maximal abelian subalgebra of the complement
  $\hat{\mathfrak{p}}_{\hspace{1pt}{\rm III}}
  =\{\mathcal{P}_{\mathcal{Q}}(\mathfrak{C}_{[\mathbf{0}]})\}_{div}-\hat{\mathfrak{t}}_{\hspace{1pt}{\rm III}}$.
  As a final remark, besides the intrinsic, non-intrinsic decompositions of type {\bf AIII} of $su(m'+n')$
  with integer pairs $(m',n')=(m-2l,n+2l)$, $0\leq l\leq m-2^{p-1}$,
  are also attainable in $\{\mathcal{P}_{\mathcal{Q}}(\mathfrak{C}_{[\mathbf{0}]})\}_{div}$.
  Refer to Figs.~\ref{figsu8tp5+3}, \ref{figsu8tp6+2} and \ref{figsu8tp7+1} for
  example demonstrations of $su(8)$.

\section{Recursive $\mathfrak{t}$-$\mathfrak{p}$ Decomposition\label{secRecurtpD}}
 As stated in~\cite{SuTsai2},
 a quotient algebra partition
 $\{ \mathcal{P}_{{\cal Q}}(\mathfrak{B}^{[r]}) \}$
 of rank $r$ admits a $\mathfrak{t}$-$\mathfrak{p}$ decomposition
 of $1$st level obeying the criterion that, under the tri-addition,
 the subalgebra $\mathfrak{t}$ is a proper maximal subgroup of
 $\{ \mathcal{P}_{{\cal Q}}(\mathfrak{B}^{[r]}) \}$
 and the subspace $\mathfrak{p}$
 the coset of $\mathfrak{t}$.
 This criterion is also satisfied by decompositions of higher levels.
 \vspace{6pt}
 \begin{defn}\label{defnl-thleveltpD}
  In a quotient algebra partition $\{ \mathcal{P}_{{\cal Q}}(\mathfrak{B}^{[r]}) \}$,
  a $\mathfrak{t}$-$\mathfrak{p}$ decomposition of the $l$-th-level
  $\mathfrak{t}_{[l]}\oplus\mathfrak{p}_{[l]}$
  is conducted on the subalgebra $\mathfrak{t}_{[l-1]}=\mathfrak{t}_{[l]}\oplus\mathfrak{p}_{[l]}$
  of a $\mathfrak{t}$-$\mathfrak{p}$ decomposition of the $(l-1)$-th level
  $\mathfrak{t}_{[l-1]}\oplus\mathfrak{p}_{[l-1]}$, where
  the subalgebra $\mathfrak{t}_{[l]}$ is a proper maximal subgroup of $\mathfrak{t}_{[l-1]}$
  and the subspace
  $\mathfrak{p}_{[l]}=\mathfrak{t}_{[l-1]}-\mathfrak{t}_{[l]}$
  the coset of $\mathfrak{t}_{[l]}$
  in $\mathfrak{t}_{[l-1]}$,
  the partition $\{ \mathcal{P}_{{\cal Q}}(\mathfrak{B}^{[r]}) \}=\mathfrak{t}_{[0]}\oplus\mathfrak{p}_{[0]}$
  being the decomposition of $0$th level with
  $\mathfrak{p}_{[0]}=\{0\}$.
 \end{defn}
 \vspace{6pt}
   The following covering relation is essential to
  the type decision of a decomposition.
 \vspace{6pt}
 \begin{lemma}\label{lemLvlthin1st}
  For a $\mathfrak{t}$-$\mathfrak{p}$ decomposition of the $l$-th
  level $\mathfrak{t}_{[l]}\oplus\mathfrak{p}_{[l]}$ in $su(N)$,
  $2^{p-1}<N\leq 2^p$ and $1< l\leq p+r+1$,
  there exists a $\mathfrak{t}$-$\mathfrak{p}$ decomposition of the
  $1$st level $\mathfrak{t}_{[1]}\oplus\mathfrak{p}_{[1]}$
  endowed with the coverings 
  $\mathfrak{t}_{[1]}\supset\mathfrak{t}_{[l]}$
  and $\mathfrak{p}_{[1]}\supset\mathfrak{p}_{[l]}$.
 \end{lemma}
 \vspace{2pt}
 \begin{proof}
  Assume the subalgebra $\mathfrak{t}_{[l]}$
  is an $l$-th maximal subgroup of the quotient-algebra partition
  of rank $r$
  $\{\mathcal{P}_{\mathcal{Q}}(\mathfrak{B}^{[r]})\}
  =\{{W}^{\epsilon}(\mathfrak{B},\mathfrak{B}^{[r]};i):\forall\hspace{2pt}
  \mathfrak{B}\in\mathcal{G}(\mathfrak{C}),\epsilon\in{Z_2},i\in{Z^r_2}\}$
  generated by an $r$-th maximal bi-subalgebra $\mathfrak{B}^{[r]}$
  of a Cartan subalgebra $\mathfrak{C}\subset{su(N)}$ under the tri-addition,
  referring to Definition~\ref{defnl-thleveltpD}.
  The construction of a required $1$st-level $\mathfrak{t}_{[1]}\oplus\mathfrak{p}_{[1]}$
  decomposition relies upon 
  a chosen set 
  ${\cal M}_{l-1}=\{{W}^{\epsilon_s}(\mathfrak{B}_s,\mathfrak{B}^{[r]};i_s):1\leq s\leq l-1\}$
  comprising a number $l-1$ of conditioned subspaces in
  $\{\mathcal{P}_{\mathcal{Q}}(\mathfrak{B}^{[r]})\}-\mathfrak{t}_{[l]}\oplus\mathfrak{p}_{[l]}$.
  This set is independent in the sense that 
  no subspace in ${\cal M}_{l-1}$ is the tri-additive of any two other members in the set. 
  In addition, the subspaces in ${\cal M}_{l-1}$ meet the disjoint criterion that the two tri-additives
  ${W}^{\epsilon_s+\epsilon_{s'}}(\mathfrak{B}_s\sqcap\mathfrak{B}_{s'},\mathfrak{B}^{[r]};i_s+i_{s'})$
  and
  ${W}^{\epsilon_t+\epsilon_{t'}}(\mathfrak{B}_t\sqcap\mathfrak{B}_{t'},\mathfrak{B}^{[r]};i_t+i_{t'})$
  belong to two distinct cosets of $\mathfrak{t}_{[l]}$
  for four arbitrary members
  ${W}^{\epsilon_s}(\mathfrak{B}_s,\mathfrak{B}^{[r]};i_s)$,
  ${W}^{\epsilon_{s'}}(\mathfrak{B}_{s'},\mathfrak{B}^{[r]};i_{s'})$,
  ${W}^{\epsilon_t}(\mathfrak{B}_t,\mathfrak{B}^{[r]};i_t)$ and
  ${W}^{\epsilon_{t'}}(\mathfrak{B}_{t'},\mathfrak{B}^{[r]};i_{t'})\in{\cal M}_{l-1}$.
  Let the subalgebra $\mathfrak{t}_{[1]}=Span\{\mathfrak{t}_{[l]},{\cal M}_{l-1}\}$
  be spanned by $\mathfrak{t}_{[l]}$ and ${\cal M}_{l-1}$ and
  the subspace $\mathfrak{p}_{[1]}=Span\{\mathfrak{p}_{[l]},{\cal M}_{l-1}\}$ 
  by $\mathfrak{p}_{[l]}$ and ${\cal M}_{l-1}$.
  Thereupon the composition $\mathfrak{t}_{[1]}\oplus\mathfrak{p}_{[1]}$
  is a $1$st-level $\mathfrak{t}$-$\mathfrak{p}$ decomposition as demanded
  possessing the inclusions $\mathfrak{t}_{[1]}\supset\mathfrak{t}_{[l]}$
  and $\mathfrak{p}_{[1]}\supset\mathfrak{p}_{[l]}$.
  The explicit construction is recursive and delivered level by level as follows.

  Initiated with a set
  ${\cal M}_1=\{{W}^{\epsilon_1}(\mathfrak{B}_1,\mathfrak{B}^{[r]};i_1)\}$
  of one arbitrary conditioned subspace 
  in $\{\mathcal{P}_{\mathcal{Q}}(\mathfrak{B}^{[r]})\}-\mathfrak{t}_{[l]}\oplus\mathfrak{p}_{[l]}$,
  the composition $\mathfrak{t}_{[l-1]}\oplus\mathfrak{p}_{[l-1]}$
  of the subalgebra
  $\mathfrak{t}_{[l-1]}=Span\{\mathfrak{t}_{[l]},{\cal M}_1\}
  =\{{W}^{\epsilon+\epsilon_1}(\mathfrak{B}\sqcap\mathfrak{B}_1,\mathfrak{B}^{[r]};i+i_1):
  \forall\hspace{2pt}{W}^{\epsilon}(\mathfrak{B},\mathfrak{B}^{[r]};i)\subset\mathfrak{t}_{[l]}\}$
  and the coset
  $\mathfrak{p}_{[l-1]}=Span\{\mathfrak{p}_{[l]},{\cal M}_1\}
  =\{{W}^{\epsilon'+\epsilon_1}(\mathfrak{B}'\sqcap\mathfrak{B}_1,\mathfrak{B}^{[r]};i'+i_1):
  \forall\hspace{2pt}{W}^{\epsilon'}(\mathfrak{B}',\mathfrak{B}^{[r]};i')\subset\mathfrak{p}_{[l]}\}$
  of $\mathfrak{t}_{[l-1]}$
  under the tri-addition is an $(l-1)$-th level $\mathfrak{t}$-$\mathfrak{p}$
  decomposition satisfying $\mathfrak{t}_{[l-1]}\supset\mathfrak{t}_{[l]}$
  and $\mathfrak{p}_{[l-1]}\supset\mathfrak{p}_{[l]}$.
  With another choice of conditioned subspace ${W}^{\epsilon_2}(\mathfrak{B}_2,\mathfrak{B}^{[r]};i_2)$
  in $\{\mathcal{P}_{\mathcal{Q}}(\mathfrak{B}^{[r]})\}-\mathfrak{t}_{[l-1]}\oplus\mathfrak{p}_{[l-1]}$
  to form the set
  ${\cal M}_2=\{{W}^{\epsilon_s}(\mathfrak{B}_s,\mathfrak{B}^{[r]};i_s):s=1,2\}$,
  the composition $\mathfrak{t}_{[l-2]}\oplus\mathfrak{p}_{[l-2]}$ of
  the subalgebra
  $\mathfrak{t}_{[l-2]}=Span\{\mathfrak{t}_{[l]},{\cal M}_2\}
  =\{{W}^{\epsilon+\epsilon_2}(\mathfrak{B}\sqcap\mathfrak{B}_2,\mathfrak{B}^{[r]};i+i_2):
  \forall\hspace{2pt}{W}^{\epsilon}(\mathfrak{B},\mathfrak{B}^{[r]};i)\subset\mathfrak{t}_{[l-1]}\}$
  and the coset
  $\mathfrak{p}_{[l-2]}=Span\{\mathfrak{p}_{[l]},{\cal M}_2\}
  =\{{W}^{\epsilon'+\epsilon_2}(\mathfrak{B}'\sqcap\mathfrak{B}_2,\mathfrak{B}^{[r]};i'+i_2):
  \forall\hspace{2pt}{W}^{\epsilon'}(\mathfrak{B}',\mathfrak{B}^{[r]};i')\subset\mathfrak{p}_{[l-1]}\}$
  of $\mathfrak{t}_{[l-2]}$
  is an $(l-2)$-th-level decomposition realizing the inclusions
  $\mathfrak{t}_{[l-2]}\supset\mathfrak{t}_{[l]}$
  and $\mathfrak{p}_{[l-2]}\supset\mathfrak{p}_{[l]}$.
  The procedure of constructing such $\mathfrak{t}$-$\mathfrak{p}$ decompositions
  completes when a set ${\cal M}_{l-1}$ is yielt
  and a required $1$st-level decomposition is obtained.
  Notice that it is plain to verify the compliance of a $\mathfrak{t}$-$\mathfrak{p}$ decomposition
  $\mathfrak{t}_{[k]}\oplus\mathfrak{p}_{[k]}$ so achieved at every level, $1\leq k<l$,
  with the decomposition condition
  $[\mathfrak{t}_{[k]},\mathfrak{t}_{[k]}]\subset\mathfrak{t}_{[k]}$,
  $[\mathfrak{t}_{[k]},\mathfrak{p}_{[k]}]\subset\mathfrak{p}_{[k]}$,
  $[\mathfrak{p}_{[k]},\mathfrak{p}_{[k]}]\subset\mathfrak{t}_{[k]}$
  and
  ${\rm Tr}\{\mathfrak{t}_{[k]}\hspace{.5pt}\mathfrak{p}_{[k]}\}=0$.
  The proof ends.
 \end{proof}
 \vspace{6pt}

  \vspace{6pt}
 \begin{lemma}\label{LvlsupLvlMaxAbelSubAlg}
  For a 1st-level $\mathfrak{t}$-$\mathfrak{p}$ decomposition
  $\mathfrak{t}_{[1]}\oplus\mathfrak{p}_{[1]}$ and an $l$-th-level decomposition
  $\mathfrak{t}_{[l]}\oplus\mathfrak{p}_{[l]}$ having the inclusions
  $\mathfrak{t}_{[1]}\supset\mathfrak{t}_{[l]}$ and $\mathfrak{p}_{[1]}\supset\mathfrak{p}_{[l]}$,
  the four occasions hold for the two maximal abelian subalgebras
  ${\cal A}_{[1]}$ of $\mathfrak{p}_{[1]}$ and ${\cal A}_{[l]}$ of $\mathfrak{p}_{[l]}$
  that
  ${\cal A}_{[1]}$ is a Cartan subalgebra $\mathfrak{C}$ if so is ${\cal A}_{[l]}$,
  ${\cal A}_{[1]}$ is $\mathfrak{C}$ or a $1$st maximal
  bi-subalgebra $\mathfrak{B}^{[1]}$ of $\mathfrak{C}$ if ${\cal A}_{[l]}$ is
  an $r$-th maximal bi-subalgebra of $\mathfrak{B}^{[r]}\subset\mathfrak{B}^{[1]}\subset\mathfrak{C}$ with $r\geq 1$,
  ${\cal A}_{[1]}$ is $\mathfrak{C}$ or the coset $\mathfrak{B}^{[1,1]}$ of $\mathfrak{B}^{[1]}$ in $\mathfrak{C}$
  if ${\cal A}_{[l]}$ is $\mathfrak{B}^{[1,1]}$,
  and finally ${\cal A}_{[1]}$ is $\mathfrak{C}$, $\mathfrak{B}^{[1]}$ or $\mathfrak{B}^{[1,1]}$ if
  ${\cal A}_{[l]}$ is a coset $\mathfrak{B}^{[r',i]}$ of an
  $r'$-th maximal bi-subalgebra of $\mathfrak{B}^{[r]}\subset\mathfrak{B}^{[1]}\subset\mathfrak{C}$
  with $r'\geq 2$ and $i\neq \mathbf{0}$.
 \end{lemma}
 \vspace{2pt}
 \begin{proof}
  Suppose that
  $\mathfrak{t}_{[l]}$ is a subgroup and $\mathfrak{p}_{[l]}$ a
  coset of $\mathfrak{t}_{[l]}$ of the rank-$s$ quotient-algebra partition
  $\{{\cal P}_{\mathcal{Q}}(\mathfrak{B}^{[s]})\}$ generated by
  an $s$-th maximal bi-subalgebra $\mathfrak{B}^{[s]}$ of a Cartan subalgebra $\mathfrak{C}'$
  under the tri-addition, $0\leq s\leq p$,
  {\em cf.} Definition~\ref{defnl-thleveltpD}.
  Based on the proof of Lemma~\ref{lemLvlthin1st},
  a set of $l-1$ indepedent conditioned subspaces ${\cal M}=\{W^{\epsilon_t}(\mathfrak{B}_t,\mathfrak{B}^{[s]};j_t):1\leq t\leq l-1\}$
  can be chosen from $\{{\cal P}_{\mathcal{Q}}(\mathfrak{B}^{[s]})\}-\mathfrak{t}_{[l]}\oplus\mathfrak{p}_{[l]}$
  to form the $1$st-level decomposition
  $\mathfrak{t}_{[1]}\oplus\mathfrak{p}_{[1]}$ with
  $\mathfrak{t}_{[1]}=Span\{\mathfrak{t}_{[l]},{\cal M}\}$ and
  $\mathfrak{p}_{[1]}=Span\{\mathfrak{p}_{[l]},{\cal M}\}$.
  The proof will divided into two occasions that ${\cal A}_{[l]}$
  is a bi-subalgebra $\mathfrak{C}$ or $\mathfrak{B}^{[r]}$
  with $r\geq 1$, or ${\cal A}_{[l]}$
  is the coset $\mathfrak{B}^{[1,1]}$ or $\mathfrak{B}^{[r',i]}$
  with $r'\geq 2$ and $i\neq \mathbf{0}$.

  On the first occasion, if ${\cal A}_{[l]}$ is the Cartan subalgebra $\mathfrak{C}$,
  the $l-1$ subspaces of ${\cal M}$ are chosen arbitrarily from
  $\{{\cal P}_{\mathcal{Q}}(\mathfrak{B}^{[s]})\}-\mathfrak{t}_{[l]}\oplus\mathfrak{p}_{[l]}$
  to form $\mathfrak{p}_{[1]}$.
  Here, the maximal abelian subalgebra ${\cal A}_{[1]}$ must be a
  Cartan subalgebra $\mathfrak{C}'$ identical to $\mathfrak{C}$
  or to another Cartan subalgebra
  by composing $\mathfrak{B}^{[s]}$ and other
  conditioned subspaces of $\mathfrak{p}_{[1]}$ as in the proof of Lemmas~28 and~29
  in~\cite{SuTsai2}.
  If ${\cal A}_{[l]}$ is the $r$-th maximal bi-subalgebra $\mathfrak{B}^{[r]}$,
  two cases for the independent set ${\cal M}$ should be considered.
  One case is to choose ${\cal M}$ consisting of the $p-s$ subspaces ${W}^{\epsilon_m}(\mathfrak{B}_m,\mathfrak{B}^{[s]};j_m)$
  obeying the condition
  $\mathfrak{B}_m\cap\mathfrak{B}^{[r]}=\mathfrak{B}^{[r]}$
  for $1\leq m\leq p-s$ and the remaining subspaces following
  $\mathfrak{B}_n\cap\mathfrak{B}^{[r]}\neq\mathfrak{B}^{[r]}$ for $p-s<n\leq l-1$.
  The other case is to choose ${\cal M}$ composed of the $t=p-s-1$ subspaces of ${\cal M}$
  satisfying $\mathfrak{B}_t\cap\mathfrak{B}^{[r]}=\mathfrak{B}^{[r]}$
  for $1\leq t\leq p-s-1$
  and the remaining having $\mathfrak{B}_t\cap\mathfrak{B}^{[r]}\neq\mathfrak{B}^{[r]}$.
  One thus derives that the maximal abelian subalgebra ${\cal A}_{[1]}\subset\mathfrak{p}_{[1]}$ is a
  Cartan subalgebra $\mathfrak{C}'$ in the first case and is a 1st
  maximal bi-subalgebra $\mathfrak{B}'^{[1]}$ of $\mathfrak{C}'$
  in the latter case.
  Likewise, $\mathfrak{C}'$ (or $\mathfrak{B}'^{[1]}$)
  is equal to $\mathfrak{C}$ (or $\mathfrak{B}^{[1]}$), or to
  another Cartan subalgebra (or another 1st maximal bi-subalgebra)
  by composing $\mathfrak{B}^{[s]}$ with other
  conditioned subspaces in $\mathfrak{p}_{[1]}$, referring to the proofs of Lemmas~28 and~29 in~\cite{SuTsai2}.

  On the other occasion, if ${\cal A}_{[l]}$ is the coset $\mathfrak{B}^{[1,1]}$
  of $\mathfrak{B}^{[1]}$, the maximal abelian subalgebra ${\cal A}_{[1]}$
  is also the coset $\mathfrak{B}^{[1,1]}$ via a similar argument on the 1st occasion of ${\cal A}_{[l]}=\mathfrak{B}^{[r]}$
  except the bi-subalgebra
  $B^{[s]}=W^1(\mathfrak{B}^{[1]},\mathfrak{B}^{[s]};\mathbf{0})$
  being obligated to be in ${\cal M}$ and $B^{[s]}\subset\mathfrak{t}_{[1]}$.
  While ${\cal A}_{[1]}$
  is the Cartan subalgebra $\mathfrak{C}'$ as
  $B^{[s]}$ is not in ${\cal M}$ but in $\mathfrak{p}_{[1]}$ to form $\mathfrak{C}'$
  through a similar argument on the 1st occasion.
  Finally, if ${\cal A}_{[l]}$ is a coset $\mathfrak{B}^{[r',i]}$
  of $\mathfrak{B}^{[r']}$ with $r'\geq 2$ and $i\neq\mathbf{0}$,
  the maximal abelian subalgebra ${\cal A}_{[1]}$ is the coset $\mathfrak{B}^{[1,1]}$
  when the bi-subalgebra
  $B^{[s]}=W^{\epsilon_1}(\mathfrak{B}_1,\mathfrak{B}^{[s]};j_1)$
  is in ${\cal M}$ and $B^{[s]}\subset\mathfrak{t}_{[1]}$ together with
  a similar argument on the 1st occasion.
  Yet, ${\cal A}_{[1]}$
  is a Cartan subalgebra $\mathfrak{C}'$ (or a 1st maximal bi-subalgebra $\mathfrak{B}'^{[1]}$) as
  $B^{[s]}$ is not in ${\cal M}$ but in $\mathfrak{p}_{[1]}$ to form $\mathfrak{C}'$ (or $\mathfrak{B}'^{[1]}$)
  by a similar argument on the 1st occasion.
 \end{proof}
 \vspace{6pt}
  This consequence will be applied to decide the type of an
  $l$-th-level decomposition.

  To factorize a unitary action $U\in{SU(N)}$,
  $2^{p-1}<N\leq 2^p$,
  it is necessary to prepare level by level a series of Cartan decompositions
  and to decide the {\em type} of the decomposition at each level.
  The type of a $1$st-level $\mathfrak{t}$-$\mathfrak{p}$
  $su(N)=\mathfrak{t}_{[1]}\oplus\mathfrak{p}_{[1]}$,
  as stated in Theorem~4 of~\cite{SuTsai2},
  is decided by the maximal abelian subalgebra $\mathfrak{a}$ of $\mathfrak{p}_{[1]}$
  according to a concise law:
  the decomposition $su(N)=\mathfrak{t}_{[1]}\oplus\mathfrak{p}_{[1]}$
  is a Cartan decomposition of type {\bf AI} if
  $\mathfrak{a}$ is a Cartan subalgebra $\mathfrak{C}\subset{su(N)}$,
  of type {\bf AII} if
  $\mathfrak{a}$ is a proper maximal bi-subalgebra $\mathfrak{B}$ of $\mathfrak{C}$,
  or of type {\bf AIII} if
  $\mathfrak{a}$ is a complement
  $\mathfrak{B}^c=\mathfrak{C}-\mathfrak{B}$.
  An extended version of this law applicable to succeeding levels is the foucs of the
  continued discussion.

  Granted by the {\em KAK theorem}~\cite{Knapp},
  a unitary action $U\in{SU(2^p)}$ can be factorized into the form
  $U=K_{[1],0}AK_{[1],1}$ due to a Cartan decomposition
  $su(2^p)=\mathfrak{t}_{[1]}\oplus\mathfrak{p}_{[1]}$.
  Herein the factor $A=e^{i\mathbf{a}_{[0]}}$ is the exponential mapping of a vector
  $\mathbf{a}_{[0]}$ of a maximal abelian subalgebra $\mathfrak{a}$ in the subspace $\mathfrak{p}_{[1]}$
  and  $K_{[1],s}=e^{i\mathfrak{t}_s}$ is contributed by a vector $\mathbf{t}_s$
  of the subalgebra $\mathfrak{t}_{[1]}$ for $s\in{Z_2}$.
  The calculation of the factorization is in practice realized only on the occasion
  of the {\em intrinsic} Cartan decomposition
  $\hat{\mathfrak{t}}_{[1]}\oplus\hat{\mathfrak{p}}_{[1]}$~\cite{Su,SuTsai1,SuTsai2}.
  For other occasions,
  the decomposition $\mathfrak{t}_{[1]}\oplus\mathfrak{p}_{[1]}$
  must first be mapped to the intrinsic by a conjugate transformation
  fulfilling
  $\hat{\mathfrak{t}}_{[1]}=Q\mathfrak{t}_{[1]}Q^\dag$
  and
  $\hat{\mathfrak{p}}_{[1]}=Q\mathfrak{p}_{[1]}Q^\dag$ via a operator
  $Q\in{SU(2^p)}$, which is explicitly written in Appendix~2 of~\cite{SuTsai2}.
  Let the calculation proceed on the modified factorization
  $\hat{U}=QUQ^{\dag}=\hat{K}_{[1],0}\hat{A}\hat{K}_{[1],1}$
  and then the desired factorization is arrived through the reverse mapping $U=Q^{\dag}\hat{U}Q$
  with $\hat{A}=QAQ^{\dag}$ and
  $\hat{K}_{[1],s}=QK_{[1],s}Q^{\dag}$, $s\in{Z_2}$.
  The matrix computation of the factorization
  $\hat{U}=\hat{K}_{[1],0}\hat{A}\hat{K}_{[1],1}$ due to the
  intrinsic decomposition
  $\hat{\mathfrak{t}}_{[1]}\oplus\hat{\mathfrak{p}}_{[1]}$
  is carried out within the {\em bilinear constraint}
  $\hat{K}_{[1],s}M\hat{K}^t_{[1],s}=M$ (or $\hat{K}_{[1],s}M\hat{K}^{\dag}_{[1],s}=M$),
  where the {\em metric} $M$ is
  a $2^p\times 2^p$ {\em invertible real symmetric} matrix
  and the factor $\hat{A}$ is also {\em real} and {\em symmetric}~\cite{Helgason,Knapp,RG,GVL}.

   One type of intrinsic Cartan decomposition
   $\hat{\mathfrak{t}}_{[1]}\oplus\hat{\mathfrak{p}}_{[1]}$ is associated with
   one distinct metric $M$.
   When the metric $M=M_{\hspace{.5pt}\mathbf{I}}=I={\cal S}^{\mathbf{0}}_{\mathbf{0}}$ is the identity matrix,
   the factorization $\hat{U}=\hat{K}_{[1],0}\hat{A}\hat{K}_{[1],1}$ is
   the well-known {\em Singular Value Decomposition (SVD)} requiring
   computation over the relations
   $\hat{K}_{[1],s}\hat{K}^t_{[1],s}=I$ for $s\in{Z_2}$, whose factor $\hat{A}$
   is the exponential mapping of a vector in the maximal abelian subalgebra
   $\hat{\mathfrak{a}}=\{{{\cal S}^{\zeta}_{\mathbf{0}}:\forall\hspace{2pt}\zeta\in{Z^p_2}}\}\subset\hat{\mathfrak{p}}_{[1]}$
   consisting of all diagonal generators of $su(2^p)$.
   The intrinsic Cartan decomposition $\hat{\mathfrak{t}}_{[1]}\oplus\hat{\mathfrak{p}}_{[1]}$
   causing SVD
   is of type {\bf AI} with the subalgebra $\hat{\mathfrak{t}}_{[1]}=\hat{\mathfrak{t}}_{\mathbf{I}}=so(2^p)$~\cite{SuTsai2}.
   The KAK factorization 
   is called the {\em Symplectic Decomposition (SpD)}
   as $M=M_{\hspace{.5pt}\mathbf{II}}=J_{2^{p-1}}={\cal S}^{\zeta_0}_{\alpha_0}$,
   here the strings $\zeta_0=\alpha_0=10\cdots 0\in{Z^p_2}$
   having only one single nonzero bit at the leftmost digit.
   The components $\hat{K}_{[1],s}$ comply with the constraint
   $\hat{K}_{[1],s}J_{2^{p-1}}\hat{K}^t_{[1],s}J^t_{2^{p-1}}=I$
   and the factor $\hat{A}$ is evolved by a vector of the maximal abelian subalgebra
   $\hat{\mathfrak{a}}=\{{\cal S}^{\eta}_{\mathbf{0}}:
   \forall\hspace{2pt}\eta\in{Z^p_2},\eta\cdot\alpha_0=0\}\subset\hat{\mathfrak{p}}_{[1]}$.
   The corresponding intrinsic decomposition 
   is a type {\bf AII} with the subalgebra $\hat{\mathfrak{t}}_{\mathbf{II}}=sp(2^{p-1})$.
   Finally, the intrinsic type-{\bf AIII} decomposition, 
   bearing the subalgebra $\hat{\mathfrak{t}}_{\mathbf{III}}=c\otimes{su(2^{p-1})}\otimes{su(2^{p-1})}$
   of the center $c=\{{\cal S}^{\zeta_0}_{\mathbf{0}}\}$,
   leads to the factorization known as the {\em Cosine-Sine Decomposition (CSD)}, which is
   associated with
   the metric $M=M_{\hspace{.5pt}\mathbf{III}}=I_{2^{p-1},2^{p-1}}={\cal S}^{\zeta_0}_{\mathbf{0}}$.
   Its component computation is guided by the identity
   $\hat{K}_{[1],s}I_{2^{p-1},2^{p-1}}\hat{K}^\dag_{[1],s}I_{2^{p-1},2^{p-1}}=I$
   and the factor $\hat{A}$ is the contribution of a vector in the maximal abelian subalgebra
   $\hat{\mathfrak{a}}=\{{\cal S}^{\zeta}_{\alpha_0}:\forall\hspace{2pt}\zeta\in{Z^p_2}\}\subset\hat{\mathfrak{p}}_{[1]}$.
   For the convenience of below expositions, the 
            letters {\bf A$\Omega$} will stand for the type index {\bf AI}, {\bf AII} or {\bf AIII}.

   The action $U$ admits further factorizations owing to decompositions of higher levels.
   As a result of decomposing the subalgebra
   $\mathfrak{t}_{[1]}=\mathfrak{t}_{[2]}\oplus\mathfrak{p}_{[2]}$
   into a $\mathfrak{t}$-$\mathfrak{p}$ decomposition of the $2$nd level
   by the {\em KAK} theorem again,
   the component $K_{[1],s}$ is factorized into the form
   $K_{[1],s}=K_{[2],s_0}A_{[1],s}K_{[2],s_1}$,
   here $s\in{Z_2}$ and $s_\epsilon=s\circ\epsilon$ being a
   concatenation of $s$ and $\epsilon\in{Z_2}$.
   Likewise, the factor $A_{[1],s}$ is the exponential evolution of a vector
   of a maximal abelian subalgebra in the subspace $\mathfrak{p}_{[2]}$
   and $K_{[2],s_\epsilon}$ contributed by a vector of the subalgebra $\mathfrak{t}_{[2]}$.
   By Lemma~\ref{lemLvlthin1st} there must exist a $1$st-level
   $\mathfrak{t}$-$\mathfrak{p}$ decomposition $\mathfrak{t}_{\Omega_2}\oplus\mathfrak{p}_{\Omega_2}$
   such that
   $\mathfrak{t}_{\Omega_2}\supset\mathfrak{t}_{[2]}$
   and
   $\mathfrak{p}_{\Omega_2}\supset\mathfrak{p}_{[2]}$.
   Suppose
   $\mathfrak{t}_{\Omega_2}\oplus\mathfrak{p}_{\Omega_2}$ is a type {\bf A$\Omega_2$}.
   Similar to the case at the $1$st level,
   the key step of the KAK factorization is practiced only
   according to the {\em intrinsic} Cartan decomposition.
   The calculation is conducted on the modified version
   $\hat{K}_{[1],s}=\hat{K}_{[2],s_0}\hat{A}_{[1],s}\hat{K}_{[2],s_1}$
   under the constraint
   $\hat{K}_{[2],s_0}M_{\Omega_2}\hat{K}^t_{[2],s_1}=M_{\Omega_2}$
   (or $\hat{K}_{[2],s_0}M_{\Omega_2}\hat{K}^{\dag}_{[2],s_1}=M_{\Omega_2}$)
   with the associated metric $M_{\Omega_2}$.
   After the calculation completes, each factorization component at the 2nd level is delivered by the reverse mapping
   $K_{[1],s}=Q^{\dag}_{\Omega_2}\hat{K}_{[1],s}Q_{\Omega_2}$;
   here $Q_{\Omega_2}$ is the operator
   mapping $\mathfrak{t}_{\Omega_2}\oplus\mathfrak{p}_{\Omega_2}$ into
   the intrinsic
   $\hat{\mathfrak{t}}_{\Omega_2}\oplus\hat{\mathfrak{p}}_{\Omega_2}$~\cite{SuTsai2}
   via the conjugate transformation
   $\hat{A}_{[1],s}=Q_{\Omega_2}A_{[1],s}Q^{\dag}_{\Omega_2}$
   and
   $\hat{K}_{[2],s_\epsilon}=Q_{\Omega_2}K_{[2],s_\epsilon}Q^{\dag}_{\Omega_2}$.
   Based on the allowed factorization of this choice, it is legitimate
   to label the $2$nd-level decomposition
   $\mathfrak{t}_{[2]}\oplus\mathfrak{p}_{[2]}$ as a type {\bf A$\Omega_2$}.
   When the decomposition goes forward to the $l$-th level,
   $l>2$,
   the major task to accomplish is the recursive factorization
   $K_{[l-1],s}=K_{[l],s_0}A_{[l-1],s}K_{[l],s_1}$
   owing to a $\mathfrak{t}$-$\mathfrak{p}$ decomposition
   $\mathfrak{t}_{[l]}\oplus\mathfrak{p}_{[l]}$ of this level,
   $s\in{Z^{l-1}_2}$ and $s_\epsilon=s\circ \epsilon$ for $\epsilon\in{Z_2}$.
   Similarly, there exists a $1$st-level decomposition
   $\mathfrak{t}_{\Omega_l}\oplus\mathfrak{p}_{\Omega_l}$
   supporting the coverings
   $\mathfrak{t}_{\Omega_l}\supset\mathfrak{t}_{[l]}$
   and
   $\mathfrak{p}_{\Omega_l}\supset\mathfrak{p}_{[l]}$. 
   Suppose $\mathfrak{t}_{\Omega_l}\oplus\mathfrak{p}_{\Omega_l}$
   is a type {\bf A$\Omega_l$} and
   let the factorization proceed 
   exactly follow the same recipe employed at the $2$nd level.
   The decomposition
   $\mathfrak{t}_{[l]}\oplus\mathfrak{p}_{[l]}$ is then
   considered of type {\bf A$\Omega_l$},
   although the choice is not unique.

   Furnished with the factorization recipe as above, 
   a decomposition can be 
   identified in the type with a $1$st level supporting subspace coverings.
 \vspace{6pt}
 \begin{defn}\label{defnl-thCDtype}
   A $\mathfrak{t}$-$\mathfrak{p}$ decomposition of the $l$-th level $\mathfrak{t}_{[l]}\oplus\mathfrak{p}_{[l]}$
   in the Lie algebra ${su(N)}$, $2^{p-1}<N\leq 2^p$ and $1< l\leq p$,
   is a Cartan decomposition of type {\bf A$\Omega$},
   if there exists a $1$st-level $\mathfrak{t}$-$\mathfrak{p}$ decomposition
   $su(N)=\mathfrak{t}_{\Omega}\oplus\mathfrak{p}_{\Omega}$ of the same type {\bf A$\Omega$}
   bearing the inclusions $\mathfrak{t}_{[l]}\subseteq\mathfrak{t}_{\Omega}$
   and $\mathfrak{p}_{[l]}\subseteq\mathfrak{p}_{\Omega}$; 
   the type {\bf A$\Omega$} is referring to the type {\bf AI}, {\bf AII} or {\bf AIII}.
 \end{defn}
 \vspace{6pt}
   In other words, the type assigned to the subalgebra $\mathfrak{t}_{[l]}$
   is offered by 
   a choice of a $1$st maximal subgroup $\mathfrak{t}_{\Omega}$
   respecting the condition that the the former subalgebra is
   an $(l-1)$-th maximal subgroup of the latter under the tri-addition.
   When necessary, 
   a type once chosen can be preserved throughout a serial of decompositions.
  \vspace{6pt}
  \begin{lemma}\label{lemlto1stlevelCDtype}
   For every $l$-th-level $\mathfrak{t}$-$\mathfrak{p}$
   decomposition $\mathfrak{t}_{[l]}\oplus\mathfrak{p}_{[l]}$ of type {\bf A$\Omega$}
   in $su(N)$, $l>1$,
   there exist a serial of $\mathfrak{t}$-$\mathfrak{p}$ decompositions
   $\mathfrak{t}_{[k]}\oplus\mathfrak{p}_{[k]}$ of the same type {\bf A$\Omega$},
   from the $1$st to the $l$-th level,  
   respecting the condition that every
   $\mathfrak{t}_{[k]}$ is a proper maximal subgroup of $\mathfrak{t}_{[k-1]}$
   under the tri-addition
   and $\mathfrak{p}_{[k]}\subset\mathfrak{p}_{[k-1]}$
   for all $1<k\leq l$. 
  \end{lemma}
  \vspace{3pt}
  \begin{proof}
   Since $\mathfrak{t}_{[l]}\oplus\mathfrak{p}_{[l]}$ is a type {\bf A$\Omega$},
   there exists a $1$st-level decomposition
   $\mathfrak{t}_{\Omega}\oplus\mathfrak{p}_{\Omega}$ of the same type
   holding the coverings 
   $\mathfrak{t}_{\Omega}\supset\mathfrak{t}_{[l]}$ and $\mathfrak{p}_{\Omega}\supset\mathfrak{p}_{[l]}$
   according to Definition~\ref{defnl-thCDtype}.
   A such serial of $\mathfrak{t}$-$\mathfrak{p}$
   decompositions $\mathfrak{t}_{[k]}\oplus\mathfrak{p}_{[k]}$,
   $1<k\leq l$, can be constructed
   following the similar procedure of recursion as shown in the proof
   of Lemma~\ref{lemLvlthin1st}.
   By taking an arbitrary conditioned subspace
   ${W}^{\epsilon_1}(\mathfrak{B}_1,\mathfrak{B}^{[r]};i_1)$ in
   $\mathfrak{t}_{\Omega}-\mathfrak{t}_{[l]}$ for
   a coset
   $\mathfrak{p}'_{[l]}=\{{W}^{\epsilon_1+\epsilon}(\mathfrak{B}_1\sqcap\mathfrak{B},\mathfrak{B}^{[r]};i_1+i):
   \forall\hspace{2pt}{W}^\epsilon(\mathfrak{B},\mathfrak{B}^{[r]};i)\in\mathfrak{t}_{[l]}\}$
   of $\mathfrak{t}_{[l]}$ under the tri-addition,
   the composition $\mathfrak{t}_{[l-1]}=\mathfrak{t}_{[l]}\oplus\mathfrak{p}'_{[l]}$
   forms an $(l-1)$-th maximal subgroup of the given quotient-algebra partition 
   and belongs to $\mathfrak{t}_{\Omega}$.
   In addition, with the subspace
   $\mathfrak{p}''_{[l]}=\{{W}^{\epsilon_1+\sigma}(\mathfrak{B}_1\sqcap\mathfrak{B}^\dag,\mathfrak{B}^{[r]};i_1+j):
   \forall\hspace{2pt}{W}^\sigma(\mathfrak{B}^\dag,\mathfrak{B}^{[r]};j)\in\mathfrak{p}_{[l]}\}$
   being another coset of $\mathfrak{t}_{[l]}$ under the tri-addition,
   the composition $\mathfrak{p}_{[l-1]}=\mathfrak{p}_{[l]}\oplus\mathfrak{p}''_{[l]}$
   is a coset of $\mathfrak{t}_{[l-1]}$ and
   a subset of $\mathfrak{p}_{\Omega}$, recalling
   a rule of decomposition condition
   $[\mathfrak{t}_{\Omega},\mathfrak{p}_{\Omega}]\subset\mathfrak{p}_{\Omega}$.
   Thus, the composition $\mathfrak{t}_{[l-1]}\oplus\mathfrak{p}_{[l-1]}$ is of type {\bf A$\Omega$}. 
   Let this procedure be recursively performed, which then produces
   a serial of decompositions
   $\mathfrak{t}_{[k]}\oplus\mathfrak{p}_{[k]}$ of the same type 
   as demanded for $1<k\leq l$.
  \end{proof}
  \vspace{6pt}

   By Lemmas~\ref{lemLvlthin1st} and~\ref{LvlsupLvlMaxAbelSubAlg},
   when the maximal abelian subalgebras of the subspace $\mathfrak{p}_{[l]}$
   of an $l$-th-level $\mathfrak{t}$-$\mathfrak{p}$ decomposition
   $\mathfrak{t}_{[l]}\oplus\mathfrak{p}_{[l]}$ is given,
   there exists a $1$st-level decomposition
   $\mathfrak{t}_{[1]}\oplus\mathfrak{p}_{[1]}$ to have the
   inclusions
   $\mathfrak{t}_{[1]}\supset\mathfrak{t}_{[l]}$ and
   $\mathfrak{p}_{[1]}\supset\mathfrak{p}_{[l]}$.
   Thus, together with Definition~\ref{defnl-thCDtype},
   it is no difficult to extend the rule of type decision to 
   decompositions of levels beyond the $1$st.
  \vspace{6pt}
  \begin{thm}\label{thm3typesoflthtp}
   For an $l$-th-level $\mathfrak{t}$-$\mathfrak{p}$ decomposition
   $\mathfrak{t}_{[l]}\oplus\mathfrak{p}_{[l]}$ of a rank-$r$ quotient-algebra partition over $su(N)$,
   $2^{p-1}<N\leq 2^p$, $0\leq r\leq p$ and $1< l\leq p+r$, 
   the decomposition is a Cartan decomposition of type {\bf AI} if the maximal abelian subalgebra ${\cal A}$
   of the subspace $\mathfrak{p}_{[l]}$
   is a Cartan subalgebra $\mathfrak{C}\subset{su(N)}$,
   is a type {\bf AI} or {\bf AIII} if ${\cal A}$ is the coset $\mathfrak{B}^{[1,1]}$
   of a $1$st maximal bi-subalgebra $\mathfrak{B}^{[1]}$ in $\mathfrak{C}$,
   is a type {\bf AI} or {\bf AII} if ${\cal A}$ is an $r'$-th
   maximal bi-subalgebra $\mathfrak{B}^{[r']}$ of $\mathfrak{C}$ for $1\leq r'\leq p$,
   or is one of the three types {\bf AI}, {\bf AII} and {\bf AIII} if ${\cal A}$ is
   a coset of an $r''$-th maximal bi-subalgebra $\mathfrak{B}^{[r'']}$ in $\mathfrak{C}$ for $1<r''\leq p$.
  \end{thm}
  \vspace{2pt}
  \begin{proof}
   This theorem is a direct consequence of
   Lemmas~\ref{lemLvlthin1st} and~\ref{LvlsupLvlMaxAbelSubAlg} as well as
   Definition~\ref{defnl-thCDtype}.
 \end{proof}
 \vspace{6pt}
  There hence implies the nonunique freedom of types choosing
  at each decomposition level during factorizing a unitary action.

  A series of $\mathfrak{t}$-$\mathfrak{p}$ decompositions in strict accord with
  the abelian-group structure of a quotient algebra partition
  and halting at an abelian subalgebra establish 
  a {\em $\mathfrak{t}$-$\mathfrak{p}$ decomposition sequence}.
 \vspace{6pt}
 \begin{defn}\label{defntpseq}
   Denoted as
   $seq_{\mathfrak{t}\mathfrak{p}}(\mathfrak{t}_{[M]})$,
   a $\mathfrak{t}$-$\mathfrak{p}$ decomposition sequence of length $M$ 
   consists of a series of $\mathfrak{t}$-$\mathfrak{p}$ decompositions
   $\mathfrak{t}_{[l]}\oplus\mathfrak{p}_{[l]}$ from the $1$st to the $M$-th level
   in a quotient-algebra partition of the Lie algebra $su(N)$,
   $2^{p-1}<N\leq 2^p$ and $1\leq l\leq M$,
   where $su(N)=\mathfrak{t}_{[0]}=\mathfrak{t}_{[1]}\oplus\mathfrak{p}_{[1]}$ and
   $\mathfrak{p}_{[0]}=\{0\}$,
   the subalgebra $\mathfrak{t}_{[l]}$ is a proper maximal subgroup of
   $\mathfrak{t}_{[l-1]}$ under the tri-addition
   with the complement
   $\mathfrak{p}_{[l]}=\mathfrak{t}_{[l-1]}-\mathfrak{t}_{[l]}$,
   every $\mathfrak{t}_{[l']}$ for $1\leq l'<M$ is nonabelian
   and only the subalgebra $\mathfrak{t}_{[M]}$ at the final level is abelian.
 \end{defn}
 \vspace{6pt}
  The length of a decomposition sequence is bounded between
  the rank of the partition and the power of the algebra's dimension.
  \vspace{6pt}
  \begin{lemma}\label{lemtplength}
   A quotient-algebra partition of rank $r$ of the Lie algebra $su(N)$,
   $2^{p-1}<N \leq 2^p$ and $0\leq r\leq p$,
   admits $\mathfrak{t}$-$\mathfrak{p}$ decomposition
   sequences of lengths ranging from $p$ to $p+r$.
  \end{lemma}
  \vspace{3pt}
  \begin{proof}
   It will be shown first there has no decomposition sequence of length shorter than
   $p$; namely,
   every subalgebra $\mathfrak{t}_{[k]}$ of a decomposition
   $\mathfrak{t}_{[k]}\oplus\mathfrak{p}_{[k]}$
   at the $k$-th level for $1\leq k<p$
   is nonabelian. 
   In the quotient-algebra partition of rank $r$
   $\{\mathcal{P}_{\mathcal{Q}}(\mathfrak{B}^{[r]})\}
   =\{{W}^\epsilon(\mathfrak{B},\mathfrak{B}^{[r]};i):\forall\hspace{2pt}
   \mathfrak{B}\in\mathcal{G}(\mathfrak{C}),\epsilon\in{Z_2}\text{ and }i\in{Z^r_2}\}$
   generated by an $r$-th maximal bi-subalgebra $\mathfrak{B}^{[r]}$ of a Cartan
   subalgebra $\mathfrak{C}\subset{su(N)}$, consider
   a maximal abelian subalgebra
   $\mathfrak{C'}=\bigcup_{i\in{Z^r_2}}{W}^1(\mathfrak{C},\mathfrak{B}^{[r]};i)$
   formed by 
   a number $2^{r}$ of non-null conditioned subspaces.
   A fact is revealed that
   the subalgebra
   $\mathfrak{t}_{[p]}
   =\{{W}^0(\mathfrak{C},\mathfrak{B}^{[r]};i),{W}^1(\mathfrak{C},\mathfrak{B}^{[r]};i):\forall\hspace{2pt}i\in{Z^r_2}\}$
   is the smallest subgroup of $\{\mathcal{P}_{\mathcal{Q}}(\mathfrak{B}^{[r]})\}$
   under the tri-addition which is abelian as well as a superset of $\mathfrak{C'}$.
   Thus, every subalgebra $\mathfrak{t}_{[k]}$ of a sequence
   in $\{\mathcal{P}_{\mathcal{Q}}(\mathfrak{B}^{[r]})\}$
   as $1\leq k<p$ is nonabelian
   and per Definition~\ref{defntpseq} there is no
   decomposition sequence of length shorter than $p$.

   However, there always exist decomposition sequences
   of lengths ranging from $p$ to $p+r$.
   Let the construction begin with a sequence of length $p+r$.
   In $\{\mathcal{P}_{\mathcal{Q}}(\mathfrak{B}^{[r]})\}$,
   the subalgebra
   $\mathfrak{t}_{[p+r]}
   =\{{W}^0(\mathfrak{C},\mathfrak{B}^{[r]};\mathbf{0}),{W}^\epsilon(\mathfrak{B},\mathfrak{B}^{[r]};i)\}$
   comprising the group identity ${W}^0(\mathfrak{C},\mathfrak{B}^{[r]};\mathbf{0})=\{0\}$
   and a non-null subspace
   ${W}^\epsilon(\mathfrak{B},\mathfrak{B}^{[r]};i)$
   is a smallest abelian subgroup under the tri-addition.
   With another choice of non-null conditioned subspace
   ${W}^\sigma(\mathfrak{B}_1,\mathfrak{B}^{[r]};j)$,
   the subspace
   $\mathfrak{p}_{[p+r]}=
   \{{W}^{\sigma+\tau}(\mathfrak{B}_1\sqcap\mathfrak{B}',\mathfrak{B}^{[r]};j+s):
   \forall\hspace{2pt}{W}^{\tau}(\mathfrak{B}',\mathfrak{B}^{[r]};s)\in\mathfrak{t}_{[p+r]}\}$
   corresponds to a coset of $\mathfrak{t}_{[p+r]}$.
   Their composition
   $\mathfrak{t}_{[p+r-1]}=\mathfrak{t}_{[p+r]}\oplus\mathfrak{p}_{[p+r]}$
   is nonabelian provided the two subspaces ${W}^\sigma(\mathfrak{B}_1,\mathfrak{B}^{[r]};j)$
   and $\mathfrak{t}_{[p+r]}$ do not commute.
   Consequently every subalgebra $\mathfrak{t}_{[K_{[1],1}]}\supset\mathfrak{t}_{[p+r-1]}$
   for $1\leq K_{[1],1}<p+r-1$ is nonabelian too. 
   Hereby, one
   $\mathfrak{t}$-$\mathfrak{p}$ decomposition sequence of length
   $p+r$ with $\mathfrak{t}_{[p+r]}\oplus\mathfrak{p}_{[p+r]}$ at
   the final level is obtained. 

   A decomposition sequence of length $p+r-1$ is created undertaking similar steps.
   Given an abelian subalgebra $\mathfrak{t}_{[p+r]}$,
   the composition
   $\mathfrak{t}_{[p+r-1]}=\mathfrak{t}_{[p+r]}\oplus\mathfrak{p}^{\dag}_{[p+r]}$
   of the subalgebra $\mathfrak{t}_{[p+r]}$
   and its coset
   $\mathfrak{p}^{\dag}_{[p+r]}=
   \{{W}^{\nu+\tau}(\mathfrak{B}_2\sqcap\mathfrak{B}',\mathfrak{B}^{[r]};m+s):
   \forall\hspace{2pt}{W}^{\tau}(\mathfrak{B}',\mathfrak{B}^{[r]};s)\in\mathfrak{t}_{[p+r]}\}$
   is abelian by picking some conditioned subspace
   ${W}^{\nu}(\mathfrak{B}_2,\mathfrak{B}^{[r]};m)$ commuting with $\mathfrak{t}_{[p+r]}$.
   Then, a nonabelian subalgebra
   $\mathfrak{t}_{[p+r-2]}=\mathfrak{t}_{[p+r-1]}\oplus\mathfrak{p}_{[p+r-1]}$
   with $\mathfrak{p}_{[p+r-1]}$ being a coset of $\mathfrak{t}_{[p+r-1]}$
   is rendered as long as
   $[\mathfrak{t}_{[p+r-1]},\mathfrak{p}_{[p+r-1]}]\neq 0$.
   By the same token, due to the nonabelianness of every subalgebra $\mathfrak{t}_{[k_2]}\supset\mathfrak{t}_{[p+r-2]}$
   at the $k_2$-th level as $1\leq k_2<p+r-2$, 
   a decomposition sequence of length $p+r-1$ is thus built.
   Recursively this procedure can generates
   decomposition sequences of lengths from $p+r$ to $p$.
   The construction terminates at the length
   $p$, because as aforesaid there always have abelian subalgebras $\mathfrak{t}_{[p]}$
   but every subalgebra $\mathfrak{t}_{[k]}$ in a sequence for $1\leq k<p$ is nonabelian.
  \end{proof}
  \vspace{6pt}
   Thus, the Lie algebra $su(2^p)$ has decomposition sequences
   from the shortest of length $p$ to the longest $2p$.

   Based on Theorem~5 in~\cite{SuTsai2},
   a $1$st-level $\mathfrak{t}$-$\mathfrak{p}$ decomposition determined in a quotient-algebra of rank $r$
   is reconstructed in that of rank higher than $r$.
   A such reconstruction is applicable to a $\mathfrak{t}$-$\mathfrak{p}$ decomposition of the $l$-th level
   since a quotient-algebra partition can be partitioned into a quotient-algebra partition of a higher rank.
    \vspace{6pt}
  \begin{lemma}\label{leminclul-thCD}
   A $\mathfrak{t}$-$\mathfrak{p}$ decomposition of the $l$-th level decided in the
   quotient-algebra partition of rank $r$
   generated by an $r$-th maximal bi-subalgebra $\mathfrak{B}^{[r]}$
   of a Cartan subalgebra
   $\mathfrak{C}\subset{su(N)}$, $2^{p-1}<N\leq 2^p$,
   is a decomposition of the same level
   in the partition of a higher rank
   generated by an $r'$-th maximal bi-subalgebra
   $\mathfrak{B}^{[r']}\subset\mathfrak{B}^{[r]}$ of $\mathfrak{C}$, $0\leq r<r'\leq p$ and $1\leq l\leq p+r+1$.
  \end{lemma}
  \vspace{2pt}
  \begin{proof}
   The quotient algebra partition
   $\{\mathcal{P}_{\mathcal{Q}}(\mathfrak{B}^{[r]})\}$
   generated by
   $\mathfrak{B}^{[r]}$
   allows a further partitioning into the partition of a higher rank $\{\mathcal{P}_{\mathcal{Q}}(\mathfrak{B}^{[r']})\}$
   generated by
   $\mathfrak{B}^{[r']}$.
   The partitioning is obtained by the fact
   that every conditioned subspace ${W}^\epsilon(\mathfrak{B},\mathfrak{B}^{[r]};i)$
   in $\{\mathcal{P}_{\mathcal{Q}}(\mathfrak{B}^{[r]})\}$
   can split into
   $2^{r'-r}$ conditioned subspaces ${W}^\epsilon(\mathfrak{B},\mathfrak{B}^{[r']};i')$
   in $\{\mathcal{P}_{\mathcal{Q}}(\mathfrak{B}^{[r']})\}$,
   here $\mathfrak{B}\in\mathcal{G}(\mathfrak{C})$, $i\in{Z^r_2}$ and $i'\in{Z^{r'}_2}$.
   Since the subalgebra $\mathfrak{t}_{[l]}$ and the subspace $\mathfrak{p}_{[l]}$
   of an $l$-th-level decomposition
   $\mathfrak{t}_{[l]}\oplus\mathfrak{p}_{[l]}$
   both comprise the conditioned subspaces in
   $\{\mathcal{P}_{\mathcal{Q}}(\mathfrak{B}^{[r]})\}$,
   the decomposition $\mathfrak{t}_{[l]}\oplus\mathfrak{p}_{[l]}$
   is simply a $\mathfrak{t}$-$\mathfrak{p}$ decomposition of the same level consisting
   of the refined conditioned subspaces in
   $\{\mathcal{P}_{\mathcal{Q}}(\mathfrak{B}^{[r']})\}$.
  \end{proof}
  \vspace{6pt}
   As an implication of Lemma~\ref{leminclul-thCD},
   a $\mathfrak{t}$-$\mathfrak{p}$ decomposition sequence decided in
   a quotient-algebra partition of rank $r$ is reconstructed in that of a higher rank.
 \vspace{6pt}
 \begin{thm}\label{thmincluRecCD}
   A $\mathfrak{t}$-$\mathfrak{p}$ decomposition sequence
   in the quotient-algebra partition generated by an $r$-th maximal bi-subalgebra $\mathfrak{B}^{[r]}$
   of a Cartan subalgebra $\mathfrak{C}$ in the Lie algebra ${su(N)}$,
   $2^{p-1}<N\leq 2^p$,
   can also be 
   constructed in the partition of a higher rank
   generated by an $r'$-th maximal bi-subalgebra
   $\mathfrak{B}^{[r']}\subset\mathfrak{B}^{[r]}$ of $\mathfrak{C}$, $0\leq r<r'\leq p$.
 \end{thm}
 \vspace{3pt}
 \begin{proof}
  Let $seq_{\mathfrak{t}\mathfrak{p}}(\mathfrak{t}_{[M]})=\{\mathfrak{t}_{[l]}\oplus\mathfrak{p}_{[l]};l=1,2,\cdots,M\}$
  be a $\mathfrak{t}$-$\mathfrak{p}$ decomposition sequence of the length $M$
  decided in the quotient algebra partition generated by $\mathfrak{B}^{[r]}$.
  Since the decomposition of the $l$-th level
  $\mathfrak{t}_{[l]}\oplus\mathfrak{p}_{[l]}\in{seq_{\mathfrak{t}\mathfrak{p}}(\mathfrak{t}_{[M]})}$
  is a $\mathfrak{t}$-$\mathfrak{p}$ decomposition of the same level in the partition
  generated by $\mathfrak{B}^{[r']}\subset\mathfrak{B}^{[r]}$
  according to Lemma~\ref{leminclul-thCD},
  the sequence ${seq_{\mathfrak{t}\mathfrak{p}}(\mathfrak{t}_{[M]})}$
  is rebuildable in a more refined partition. 
 \end{proof}
 \vspace{6pt}
   Recovering a decomposition sequence in a higher-rank quotient-algebra partition
   is a direct consequence of the above theorem.
 \vspace{6pt}
 \begin{cor}\label{corincluSeqDec}
  A decomposition sequence determined in the quotient-algebra partition generated by an $r$-th maximal
  bi-subalgebra $\mathfrak{B}^{[r]}$ of a Cartan subalgebra
  $\mathfrak{C}\subset{su(N)}$, $2^{p-1}<N\leq 2^p$,
  is recoverable in the partition generated by an
  $r'$-th maximal bi-subalgebras
  $\mathfrak{B}^{[r']}\subset\mathfrak{B}^{[r]}$ of $\mathfrak{C}$,
  $0\leq r<r'\leq p$.
 \end{cor}
 \vspace{6pt}
  It is apparent that all decomposition sequences
  are acquirable in the quotient algebra partition of the highest rank.
 \vspace{6pt}
 \begin{cor}\label{corcomplte}
  The complete set of decomposition sequences of the Lie algebra
  $su(N)$, $2^{p-1}<N\leq 2^p$, 
  is produced in the
  quotient-algebra partition of the highest rank $p$
  $\{\mathcal{P}_{\mathcal{Q}}(\mathfrak{B}^{[p]})\}$ generated
  by the $p$-th maximal bi-subalgebra $\mathfrak{B}^{[p]}=\{{\cal S}^{\bf 0}_{\hspace{.01cm}{\bf 0}}\}$
  of a Cartan subalgebra $\mathfrak{C}\subset{su(N)}$.
 \end{cor}
 \vspace{6pt}
  Therefore, all admissible factorizations of a unitary action are derivable from the complete set of decomposition sequences.

\nonumsection{References}

 \appendix{~~Universality of the QAP Structure in Lie Algebras\label{appUniversalQAP}}
 It is known that semisimple Lie algebras are classified into
 {\em classical} and {\em exceptional} Lie algebras
 according to the so-called scheme of {\em root system}.
 As shown in~\cite{RG},
 for a semisimple Lie algebra $\mathfrak{g}$ of {\em rank} \hspace{1pt}$l$,
 there exists a Cartan subalgebra
 $\mathfrak{C}=span\{ H_k:k=1,2,\cdots,l \}\subset\mathfrak{g}$
 spanned by a number $l$ of independent commuting generators (vectors) $H_k$,
 such that every generator $E_{\mathbf{r}}\in\mathfrak{g}$ is associated with a unique
 $l$-dimensional real vector
 $\mathbf{r}=(r_1,r_2,\cdots,r_l)\in\mathbb{R}_l$ obeying $[H_s,E_{\mathbf{r}}]=r_sE_{\mathbf{r}}$,
 termed as a {\em root} of $\mathfrak{g}$.
 These roots have to satisfy two criteria.
 One is that a negative root $-\mathbf{r}$ is also a root,
 which is associated with the generator
 $E_{-\mathbf{r}}$
 fulfilling $[E_{\mathbf{r}},E_{-\mathbf{r}}]\neq 0$ and $[E_{\mathbf{r}},E_{-\mathbf{r}}]\in\mathfrak{C}$.
 The other is that the addition $\mathbf{r}_1+\mathbf{r}_2$ of two roots $\mathbf{r}_1$ and $\mathbf{r}_2$
 is still a root if
 $[E_{\mathbf{r}_1},E_{\mathbf{r}_2}]\neq 0$
 and is not a root if otherwise,
 $E_{\mathbf{r}_1}$ and $E_{\mathbf{r}_2}$
 being the associated generators of $\mathbf{r}_1$ and $\mathbf{r}_2$.

 Let the two criteria be rephrased in the QAP structure
 of the partition $\{{\cal P}_{{\cal Q}}(\mathfrak{C})\}$
 generated by $\mathfrak{C}$ over $\mathfrak{g}$.
 To satisfy the 1st criterion,
 associated to a pair of roots $\mathbf{r}$ and $-\mathbf{r}$,
 the generators $E_{\mathbf{r}}$ and $E_{-\mathbf{r}}$
 are respectively in conditioned subspaces
 $W$ and $\hat{W}$ of a conjugate pair ${\cal W}=W\cup \hat{W}$
 in $\{{\cal P}_{{\cal Q}}(\mathfrak{C})\}$.
 For the 2nd criterion,
 assume the generators $E_{\mathbf{r}_1}$ and $E_{\mathbf{r}_2}$
 of two roots in conditioned subspaces
 $W^{\epsilon}_1\subset{\cal W}_1$ and
 $W^{\sigma}_2\subset{\cal W}_2$,
 the parity indices $\epsilon$
 and $\sigma\in{Z_2}$.
 If the addition $\mathbf{r}_1+\mathbf{r}_2$ is a root,
 its associated generator $E_{\mathbf{r}_1+\mathbf{r}_2}$
 belongs to the conditioned subspace $W^{\epsilon+\sigma}_3\subset{\cal W}_3$
 by the closure of conditioned subspaces
 $[W^{\epsilon}_1,W^{\sigma}_2]\subset W^{\epsilon+\sigma}_3$.

 In the classification of semisimple Lie algebras~\cite{RG},
 there have four types of root systems of classical Lie algebras
 named $A$, $B$, $C$ and $D$.
 Respectively written by unit vectors
 $\mathbf{e}_i=(r_{ih})$ in $\mathbb{R}_l$,
 $r_{ih}=\delta_{ih}$ and $1\leq i,h\leq l$,
 the root systems of the four types are
 \begin{align}\label{rootsTypesClassical}
             &A_{l-1}, \hspace{20pt} \mathbf{e}_i-\mathbf{e}_j,  \hspace{50pt}1\leq i\neq j\leq l;\notag\\
             &D_l,     \hspace{22pt}\pm \mathbf{e}_i \pm \mathbf{e}_j,          \hspace{50pt} 1\leq i\neq j\leq l,\hspace{4pt}l> 3;\notag\\
             &B_l,     \hspace{23pt}\pm \mathbf{e}_i \pm \mathbf{e}_j,\hspace{4pt}\pm \mathbf{e}_i,  \hspace{25pt} 1\leq i\neq j\leq l,\hspace{4pt}l>2;\notag\\
             &C_l,     \hspace{23pt}\pm \mathbf{e}_i \pm \mathbf{e}_j,\hspace{4pt}\pm 2\mathbf{e}_i,  \hspace{20pt} 1\leq i\neq j\leq l,\hspace{4pt}l> 1,
 \end{align}
 where $A_{l-1}$ is the root system for type-$A$ Lie algebra of
 rank $l-1$,
 and $B_l$, $C_l$ and $D_l$ are those for types $B$, $C$
 and $D$ of rank $l$.
 In the following,
 examples of the four types are derive from a unitary Lie algebra with dimension
 of a power of $2$~\cite{Su,SuTsai1},
 and each QAP complies with the two aforesaid criteria.
 This demonstrates a general methodology to construct a QAP for
 a classical and exceptional Lie algebra.
   The QAP structure of $A_4$ is isomorphic to that of $su(4)$ as in~\cite{Su}
   and is illustrated in Fig.~\ref{figA_3QAP}.
   The examples of remaining types $D_4$, $B_3$ and $C_4$
   in Figs.~\ref{figD4QAP},~\ref{figB3QAP} and~\ref{figC4QAP}
   are respectively acquired from the QAP of $su(8)$ through the {\em removing process}~\cite{Su,SuTsai1}.
   Note that only conjugate pairs are listed in the last two figures.
   \begin{figure}[h]
 \begin{center}
   \[\begin{array}{c}
             \begin{array}{ccccccc}
             &&& \mathfrak{C}_{A_3} &&&
             \\
             W_1&E_{\mathbf{e}_1-\mathbf{e}_2}\hspace{8pt}E_{\mathbf{e}_3-\mathbf{e}_4}&& &&
             E_{-\mathbf{e}_1+\mathbf{e}_2}\hspace{8pt}E_{-\mathbf{e}_3+\mathbf{e}_4}&\hat{W}_1
             \\
             W_2&E_{\mathbf{e}_1-\mathbf{e}_3}\hspace{8pt}E_{\mathbf{e}_2-\mathbf{e}_4} && &&
             E_{-\mathbf{e}_1+\mathbf{e}_3}\hspace{8pt}E_{-\mathbf{e}_2+\mathbf{e}_4}&\hat{W}_2
             \\
             W_3&E_{\mathbf{e}_1-\mathbf{e}_4}\hspace{8pt}E_{\mathbf{e}_2-\mathbf{e}_3} && &&
             E_{-\mathbf{e}_1+\mathbf{e}_4}\hspace{8pt}E_{-\mathbf{e}_2+\mathbf{e}_3}&\hat{W}_3
             \end{array}
    \end{array}\]
 \end{center}
 \fcaption{ The QAP structure of $A_3$;
 each $\mathbf{e}_i$ is a $4$-dimensional real vector in the Euclidean space $\mathbb{R}_4$
 with the scalar $1$ on the $i$-th component and $0$ on the others,
 and this QAP is isomorphic to that of $su(4)$.~\label{figA_3QAP}}
 \end{figure}
   \begin{figure}[h!]
 \begin{center}
   \[\begin{array}{c}
             \begin{array}{ccccccc}
             &&& \mathfrak{C}_{D_4} &&&
             \\
             W_1&E_{\mathbf{e}_1\pm\mathbf{e}_2}\hspace{8pt}E_{\mathbf{e}_3\pm\mathbf{e}_4}&& &&
             E_{-(\mathbf{e}_1\pm\mathbf{e}_2)}\hspace{8pt}E_{-(\mathbf{e}_3\pm\mathbf{e}_4)}&\hat{W}_1
             \\
             W_2&E_{\mathbf{e}_1\pm\mathbf{e}_3}\hspace{8pt}E_{\mathbf{e}_2\pm\mathbf{e}_4} && &&
             E_{-(\mathbf{e}_1\pm\mathbf{e}_3)}\hspace{8pt}E_{-(\mathbf{e}_2\pm\mathbf{e}_4)}&\hat{W}_2
             \\
             W_3&E_{\mathbf{e}_1\pm\mathbf{e}_4}\hspace{8pt}E_{\mathbf{e}_2\pm\mathbf{e}_3} && &&
             E_{-(\mathbf{e}_1\pm\mathbf{e}_4)}\hspace{8pt}E_{-(\mathbf{e}_2\pm\mathbf{e}_3)}&\hat{W}_3
             \end{array}
    \end{array}\]
 \end{center}
 \fcaption{ The QAP structure of $D_4$, which is derived from the QAP of $su(8)$ through the removing process.~\label{figD4QAP}}
 \end{figure}
   \begin{figure}[h!]
 \begin{center}
   \[\begin{array}{c}
             \begin{array}{ccccc}
             \hspace{4pt}\mathfrak{C}_{B_3}&&&
             \\
                {\cal W}_1
               &&&&
                 E_{\pm(\mathbf{e}_1\pm\mathbf{e}_2)}\hspace{16pt}E_{\pm(\mathbf{e}_3\pm\mathbf{e}_4)}
             \\
                {\cal W}_2
               &&&&
                 E_{\pm(\mathbf{e}_1\pm\mathbf{e}_3)}\hspace{16pt}E_{\pm(\mathbf{e}_2\pm\mathbf{e}_4)}
             \\
                {\cal W}_3
                &&&&
                 E_{\pm(\mathbf{e}_1\pm\mathbf{e}_4)}\hspace{16pt}E_{\pm(\mathbf{e}_2\pm\mathbf{e}_3)}
             \end{array}
    \end{array}\]
 \end{center}
 \fcaption{ The partition of conjugate-pairs for the root system of $B_3$;
 each conjugate-pair ${\cal W}_q$ can further divide into two conditioned subspaces
 $W_q$ and $\hat{W}_q$ via appropriately linear superposing the generator in this pair,
 $q=1,2,3$.~\label{figB3QAP}}
 \end{figure}
   \begin{figure}[h!]
 \begin{center}
   \[\begin{array}{c}
             \begin{array}{ccccccc}
             \hspace{4pt}\mathfrak{C}_{C_4}&
             \\
                {\cal W}_1
               &&&&
                 E_{\mathbf{e}_1\pm\mathbf{e}_2} & &E_{\mathbf{e}_3\pm\mathbf{e}_4}
             \\
                {\cal W}_2
               &&&&
                 E_{\mathbf{e}_1\pm\mathbf{e}_3} & &E_{\mathbf{e}_2\pm\mathbf{e}_4}
             \\
                {\cal W}_3
                &&&&
                 E_{\mathbf{e}_1\pm\mathbf{e}_4} & & E_{\mathbf{e}_2\pm\mathbf{e}_3}
             \\
                {\cal W}_4
                &&&&
                 E_{\pm 2\mathbf{e}_1} \hspace{16pt}
                 E_{\pm 2\mathbf{e}_2}
                 &&
                 E_{\pm 2\mathbf{e}_3}
                 \hspace{16pt}
                 E_{\pm 2\mathbf{e}_4}
             \\
                {\cal W}_5
                &&&&
                 \hspace{14pt}E_{-(\mathbf{e}_1\pm\mathbf{e}_2)}
                 &&
                 \hspace{14pt}E_{-(\mathbf{e}_3\pm\mathbf{e}_4)}
             \\
                {\cal W}_6
                &&&&
                \hspace{14pt} E_{-(\mathbf{e}_1\pm\mathbf{e}_3)}
                 &&
                \hspace{14pt} E_{-(\mathbf{e}_2\pm\mathbf{e}_4)}
             \\
                {\cal W}_7
                &&&&
                \hspace{14pt} E_{-(\mathbf{e}_1\pm\mathbf{e}_4)}
                 &&
                \hspace{14pt} E_{-(\mathbf{e}_2\pm\mathbf{e}_3)}
             \end{array}
    \end{array}\]
 \end{center}
 \fcaption{ The partition of conjugate-pairs for the root system of $C_4$;
 each conjugate-pair ${\cal W}_q$ can further divide into two conditioned subspaces
 $W_q$ and $\hat{W}_q$ via appropriately linear superposing the roots in this pair,
 $q=1,2,\cdots,7$.~\label{figC4QAP}}
 \end{figure}

 The QAP structure is also applicable to exceptional Lie algebras.
 For example,
 consider the root system of the exceptional Lie algebra $G_2$:
 \begin{align}\label{rootsExpG2}
 G_2,\hspace{20pt}\mathbf{e}_i-\mathbf{e}_j,\hspace{4pt}\pm(\mathbf{e}_i+\mathbf{e}_j-2\mathbf{e}_k),
      \hspace{20pt}1\leq i\neq j\neq k\leq 3.
 \end{align}
 There derives the QAP of $G_2$ from that of $su(8)$
  via the removing process,
  {\em cf.} Fig.~\ref{figG2QAP} with a list of its conjugate pairs.
 The QAP structures of remaining exceptional Lie algebras $F_4$, $E_6$, $E_7$ and
  $E_8$~\cite{RG}
  are derived via the same procedure.
   \begin{figure}[h!]
 \begin{center}
   \[\begin{array}{c}
             \begin{array}{ccc}
             \hspace{4pt}\mathfrak{C}_{G_2}&
             \\
                {\cal W}_1
               &&
                 E_{\mathbf{e}_1-\mathbf{e}_2}\hspace{8pt}E_{-\mathbf{e}_1+\mathbf{e}_2}
                 \hspace{8pt}E_{\mathbf{e}_1+\mathbf{e}_2-2\mathbf{e}_3}\hspace{8pt}E_{-\mathbf{e}_1-\mathbf{e}_2+2\mathbf{e}_3}
             \\
                {\cal W}_2
               &&
                \hspace{0pt}
                E_{\mathbf{e}_1-\mathbf{e}_3}\hspace{8pt}E_{-\mathbf{e}_1+\mathbf{e}_3}
                  \hspace{8pt}
                E_{\mathbf{e}_1-2\mathbf{e}_2+\mathbf{e}_3}\hspace{8pt}
                E_{-\mathbf{e}_1+2\mathbf{e}_2-\mathbf{e}_3}
             \\
                {\cal W}_3
                &&
                 \hspace{0pt}E_{\mathbf{e}_2-\mathbf{e}_3}\hspace{8pt}E_{-\mathbf{e}_2+\mathbf{e}_3}
                  \hspace{8pt}
                  E_{2\mathbf{e}_1-\mathbf{e}_2-\mathbf{e}_3}\hspace{8pt}E_{-2\mathbf{e}_1+\mathbf{e}_2+\mathbf{e}_3}
             \end{array}
    \end{array}\]
 \end{center}
 \fcaption{ The partition of conjugate-pairs for the root system of $G_2$;
 each conjugate-pair ${\cal W}_q$ can further divide into two conditioned subspaces
 $W_q$ and $\hat{W}_q$ via appropriately linear superposing the roots in this pair,
 $q=1,2,3$.~\label{figG2QAP}}
 \end{figure}

 Based on Ado's theorem~\cite{Knapp,RG},
 there is a faithful finite dimensional representation of  $\mathfrak{g}$
 if $\mathfrak{g}$ is a finite dimensional Lie algebra over a field $\mathbb{F}=\mathbb{R}$ or
 $\mathbb{C}$.
 That is, there is an imbedding $\mathfrak{g}\rightarrow sl(N, \mathbb{F})$
 in the Lie algebra $sl(N, \mathbb{F})$ of the general Lie group $SL(N, \mathbb{F})$ of a dimension $N$.
 The QAP structure is also applicable to $sl(N)$.
  \vspace{6pt}
  \begin{lem}\label{QAPonSlN}
  The Lie algebra $sl(N)$, $2^{p-1}<N\leq 2^p$, admits the
  structure of quotient algebra partition of rank $r$ generated by
  an $r$-th maximal bi-subalgebra of a Cartan subalgebra in
  ${sl(N)}$, $0\leq r\leq p$.
  \end{lem}
  \vspace{2pt}
  \begin{proof}
   This lemma is obvious because, similar to the unitary Lie algebra $su(N)$,
   the generators of $sl(N)$ with $N=2^p$
   is expressible in the $s$-representation.
  \end{proof}
  \vspace{6pt}
  \vspace{6pt}
  \begin{thm}\label{slNtpD}
  The Lie algebra $sl(N)=\mathfrak{t}\oplus\mathfrak{p}$ admits a
  $\mathfrak{t}$-$\mathfrak{p}$ decomposition
  composed of a compact subalgebra $\mathfrak{t}$
  and noncompact  subspace $\mathfrak{p}$,
  $2^{p-1}<N\leq 2^p$.
  \end{thm}
  \vspace{2pt}
  \begin{proof}
   According to the assertions in~\cite{Su,SuTsai1,SuTsai2},
   the unitary Lie algebra $su(N)$ permits $\mathfrak{t}$-$\mathfrak{p}$ decompositions
   $su(N)=\mathfrak{t}\oplus\hat{\mathfrak{p}}$ in which both
   the subalgebra $\mathfrak{t}$ and the subspace $\hat{\mathfrak{p}}$
   are compact.
   By Lemma~\ref{QAPonSlN} and the Weyl's unitarity trick~\cite{Knapp,RG},
   the decomposition $\mathfrak{t}\oplus i\hat{\mathfrak{p}}$
   with $\mathfrak{p}=i\hat{\mathfrak{p}}$ of $\mathfrak{t}\oplus\hat{\mathfrak{p}}$
   is a required $\mathfrak{t}$-$\mathfrak{p}$ decomposition of $sl(N)$.
  \end{proof}
  \vspace{6pt}

 \newpage
 \appendix{~~Figures of Merged and Detached Co-quotient Algebras\label{appMDCo-QAFigs}}
 \begin{figure}
 \begin{center}
      \[\begin{array}{cc}
             \begin{array}{c}
             \hspace{-175pt}(a)\\
             \begin{array}{c}
             \mathfrak{B}^{[r]}
             \end{array}\\
                        \\
             \begin{array}{ccc}
             \mathfrak{B}^{[r,l]} && \{0\}\\
             &\hspace{10pt}&\\
             W(\mathfrak{B}_1,\mathfrak{B}^{[r]};s)&&\{0\} \\
             &&\\
             \{0\}&&\hat{W}(\mathfrak{B}_1,\mathfrak{B}^{[r]};\hat{s})\\
             &&\\
             W(\mathfrak{B}_m,\mathfrak{B}^{[r]};i)&&\hat{W}(\mathfrak{B}_m,\mathfrak{B}^{[r]};i)\\
             &&\\
             W(\mathfrak{B}_m,\mathfrak{B}^{[r]};\hat{i})&&\hat{W}(\mathfrak{B}_m,\mathfrak{B}^{[r]};\hat{i})\\
             &&\\
             W(\mathfrak{B}_n,\mathfrak{B}^{[r]};j)&&\hat{W}(\mathfrak{B}_n,\mathfrak{B}^{[r]};j)\\
             &&\\
             W(\mathfrak{B}_n,\mathfrak{B}^{[r]};\hat{j})&&\hat{W}(\mathfrak{B}_n,\mathfrak{B}^{[r]};\hat{j})
             \end{array}
             \end{array}
       &\hspace{30pt}\begin{array}{c}
              \hspace{-175pt}(b)\\
             \begin{array}{c}
             \mathfrak{B}^{[r,l]}
             \end{array}\\
                        \\
             \begin{array}{ccc}
             \mathfrak{B}^{[r]} && \{0\}\\
             &\hspace{10pt}&\\
             W(\mathfrak{B}_1,\mathfrak{B}^{[r]};s)&& \hat{W}(\mathfrak{B}_1,\mathfrak{B}^{[r]};\hat{s})\\
             &&\\
             \{0\}&&\{0\}\\
             &&\\
             {W}(\mathfrak{B}_m,\mathfrak{B}^{[r]};\hat{i})&&\hat{W}(\mathfrak{B}_m,\mathfrak{B}^{[r]};i)\\
             &&\\
             {W}(\mathfrak{B}_m,\mathfrak{B}^{[r]};i)&&\hat{W}(\mathfrak{B}_m,\mathfrak{B}^{[r]};\hat{i})\\
             &&\\
             {W}(\mathfrak{B}_n,\mathfrak{B}^{[r]};\hat{j})&&\hat{W}(\mathfrak{B}_n,\mathfrak{B}^{[r]};j)\\
             &&\\
             {W}(\mathfrak{B}_n,\mathfrak{B}^{[r]};j)&&\hat{W}(\mathfrak{B}_n,\mathfrak{B}^{[r]};\hat{j})
             \end{array}
             \end{array}
       \end{array}\]
\[\begin{array}{c}
             \hspace{-395pt}(c)\\
             \begin{array}{c}
             \hspace{-5pt}\mathfrak{B}^{[r,l]}
             \end{array}\\
                        \\
             \begin{array}{ccc}
             \hspace{0pt}\mathfrak{B}^{[r-1]}=\mathfrak{B}^{[r]}\cup {W}(\mathfrak{B}_1,\mathfrak{B}^{[r]};s)&&\hat{W}(\mathfrak{B}_1,\mathfrak{B}^{[r]};\hat{s})\\
             &\hspace{20pt}&\\
             {W}(\mathfrak{B}_m,\mathfrak{B}^{[r]};\hspace{1pt}i)\cup {W}(\mathfrak{B}_n,\mathfrak{B}^{[r]};\hspace{1pt}j)
             &&
             \hat{W}(\mathfrak{B}_m,\mathfrak{B}^{[r]};\hat{i})\cup\hat{W}(\mathfrak{B}_n,\mathfrak{B}^{[r]};\hat{j})\\
             &&\\
             \hspace{-2pt}{W}(\mathfrak{B}_m,\mathfrak{B}^{[r]};\hat{i})\cup {W}(\mathfrak{B}_n,\mathfrak{B}^{[r]};\hat{j})
             &&
             \hat{W}(\mathfrak{B}_m,\mathfrak{B}^{[r]};\hspace{1pt}{i})\cup\hat{W}(\mathfrak{B}_n,\mathfrak{B}^{[r]};\hspace{1pt}{j})
             \end{array}
    \end{array}\]
   \[\begin{array}{c}
             \hspace{-395pt}(d)\\
             \begin{array}{c}
             \hspace{-5pt}\mathfrak{B}^{[r,l]}
             \end{array}\\
                        \\
             \begin{array}{ccc}
             \hspace{0pt}\hat{\mathfrak{B}}^{[r-1]}=\mathfrak{B}^{[r]}\cup\hat{W}(\mathfrak{B}_1,\mathfrak{B}^{[r]};\hat{s})&&{W}(\mathfrak{B}_1,\mathfrak{B}^{[r]};s)\\
             &\hspace{20pt}&\\
             {W}(\mathfrak{B}_m,\mathfrak{B}^{[r]};\hspace{1pt}i)\cup\hat{W}(\mathfrak{B}_n,\mathfrak{B}^{[r]};\hspace{1pt}\hat{j})
             &&
             \hat{W}(\mathfrak{B}_m,\mathfrak{B}^{[r]};\hat{i})\cup {W}(\mathfrak{B}_n,\mathfrak{B}^{[r]};{j})\\
             &&\\
             \hspace{-1pt}{W}(\mathfrak{B}_m,\mathfrak{B}^{[r]};\hat{i)}\cup \hat{W}(\mathfrak{B}_n,\mathfrak{B}^{[r]};{j})
             &&
             \hspace{1pt}\hat{W}(\mathfrak{B}_m,\mathfrak{B}^{[r]};\hspace{1pt}{i})\cup W(\mathfrak{B}_n,\mathfrak{B}^{[r]};\hspace{1pt}\hat{j})
             \end{array}
    \end{array}\]
 \end{center}
 \fcaption{The quotient algebra of rank $r$ $\{{\cal Q}(\mathfrak{B}^{[r]};2^{p+r}-1)\}$ in (a)
  and the co-quotient algebra of rank $r$ $\{{\cal Q}(\mathfrak{B}^{[r,l]};2^{p+r}-2^{2r-2})\}$ in (b)
  respectively given by an $r$-th maximal bi-subalgebra $\mathfrak{B}^{[r]}$
  of a Cartan subalgebra $\mathfrak{C}$
  and the coset $\mathfrak{B}^{[r,l]}$, followed by
  two merged co-quotient algebras of rank $r$
  acquired respectively by parallel merging in (c) and by crossing merging in (d);
  here $\mathfrak{B}_1=\mathfrak{B}_m\sqcap\mathfrak{B}_n\supset\mathfrak{B}^{[r]}$,
  $\mathfrak{B}_1\nsupseteq\mathfrak{B}^{[r,l]}$, $1< m,n<2^p$, $s+\hat{s}=i+\hat{i}=j+\hat{j}=l$
  and $s=i+j=\hat{i}+\hat{j}$.  \label{figmergQA}}
 \end{figure}
 \begin{figure}
 \begin{center}
 \[\begin{array}{c}
  \begin{array}{cc}
      \hspace{-15pt}
      \begin{array}{c}
             \hspace{-180pt}(a)\\
             \\
             \begin{array}{c}
             \mathfrak{B}^{[r]}={W}(\mathfrak{C},\mathfrak{B}^{[r]};\mathbf{0})
             \end{array}\\
                        \\
             \begin{array}{cc}
             \hspace{0pt}{W}(\mathfrak{B}_1,\hspace{0pt}\mathfrak{B}^{[r]};\hspace{2pt}s)& \hspace{0pt}\hat{W}(\mathfrak{B}_1,\hspace{0pt}\mathfrak{B}^{[r]};\hspace{2pt}s)\\
             &\\
             {W}(\mathfrak{B}_m,\mathfrak{B}^{[r]};i)& \hat{W}(\mathfrak{B}_m,\mathfrak{B}^{[r]};i)\\
             &\\
             \hspace{0pt}{W}(\mathfrak{B}_n,\hspace{2pt}\mathfrak{B}^{[r]};\hspace{0pt}j)&\hspace{0pt}\hat{W}(\mathfrak{B}_n,\hspace{2pt}\mathfrak{B}^{[r]};\hspace{0pt}j)
             \end{array}
      \end{array}
&\hspace{5pt}
     \hspace{-15pt}
      \begin{array}{c}
             \hspace{-180pt}(b)\\
             \\
             \begin{array}{c}
              {W}(\mathfrak{B}_1,\hspace{0pt}\mathfrak{B}^{[r]};\hspace{0pt}s)
             \end{array}\\
                        \\
             \begin{array}{cc}
             \mathfrak{B}^{[r]}={W}(\mathfrak{C},\mathfrak{B}^{[r]};\mathbf{0})& \hspace{0pt}\hat{W}(\mathfrak{B}_1,\hspace{0pt}\mathfrak{B}^{[r]};\hspace{2pt}s)\\
             &\\
             {W}(\mathfrak{B}_m,\mathfrak{B}^{[r]};i)& \hat{W}(\mathfrak{B}_n,\hspace{0pt}\mathfrak{B}^{[r]};\hspace{0pt}j)\\
             &\\
             \hspace{0pt}{W}(\mathfrak{B}_n,\hspace{0pt}\mathfrak{B}^{[r]};\hspace{0pt}j)&\hspace{0pt}\hat{W}(\mathfrak{B}_m,\hspace{0pt}\mathfrak{B}^{[r]};\hspace{0pt}i)
             \end{array}
      \end{array}
  \end{array}
  \\
  \\
      \begin{array}{c}
             \hspace{-380pt}(c)\\
             \\
             \begin{array}{c}
              {W}(\mathfrak{B}_1,\hspace{0pt}\mathfrak{B}^{[r+1]};\hspace{0pt}0\circ\mathbf{0})
             \end{array}\\
                        \\
             \begin{array}{ccc}
             \mathfrak{B}^{[r+1]}={W}(\mathfrak{C},\mathfrak{B}^{[r+1]};0\circ\mathbf{0})&& \hspace{0pt}\hat{W}(\mathfrak{B}_1,\hspace{0pt}\mathfrak{B}^{[r+1]};\hspace{2pt}0\circ s)\\
             \\
             {W}(\hspace{2pt}\mathfrak{C},\hspace{2pt}\mathfrak{B}^{[r+1]};\hspace{2pt}1\circ\mathbf{0})&& \hspace{0pt}\hat{W}(\mathfrak{B}_1,\hspace{0pt}\mathfrak{B}^{[r+1]};\hspace{2pt}1\circ s)\\
             \\
             {W}(\mathfrak{B}_m,\mathfrak{B}^{[r+1]};0\circ i)&& \hat{W}(\mathfrak{B}_n,\hspace{0pt}\mathfrak{B}^{[r+1]};\hspace{2pt}0\circ j)\\
             \\
             {W}(\mathfrak{B}_m,\mathfrak{B}^{[r+1]};1\circ i)&& \hat{W}(\mathfrak{B}_n,\hspace{0pt}\mathfrak{B}^{[r+1]};\hspace{2pt}1\circ j)\\
             \\
             \hspace{0pt}{W}(\mathfrak{B}_n,\hspace{0pt}\mathfrak{B}^{[r+1]};\hspace{0pt}0\circ j)&&\hspace{0pt}\hat{W}(\mathfrak{B}_m,\hspace{0pt}\mathfrak{B}^{[r+1]};\hspace{0pt}0\circ i)\\
             \\
             \hspace{0pt}{W}(\mathfrak{B}_n,\hspace{0pt}\mathfrak{B}^{[r+1]};\hspace{0pt}1\circ j)&&\hspace{0pt}\hat{W}(\mathfrak{B}_m,\hspace{0pt}\mathfrak{B}^{[r+1]};\hspace{0pt}1\circ i)
             \end{array}
      \end{array}
      \\
      \\
      \begin{array}{c}
             \hspace{-380pt}(d)\\
             \\
             \begin{array}{c}
              {W}(\mathfrak{B}_1,\hspace{0pt}\mathfrak{B}^{[r+1]};\hspace{0pt}0\circ\mathbf{0})
             \end{array}\\
                        \\
             \begin{array}{ccc}
             \mathfrak{B}^{[r+1]}={W}(\mathfrak{C},\mathfrak{B}^{[r+1]};0\circ\mathbf{0})&& \hspace{0pt}\hat{W}(\mathfrak{B}_1,\hspace{0pt}\mathfrak{B}^{[r+1]};\hspace{2pt}0\circ s)\\
             \\
             \hat{W}(\mathfrak{B}_1,\hspace{0pt}\mathfrak{B}^{[r+1]};\hspace{2pt}1\circ s)&& \hspace{0pt}{W}(\hspace{2pt}\mathfrak{C},\hspace{2pt}\mathfrak{B}^{[r+1]};\hspace{2pt}1\circ\mathbf{0})\\
             \\
             {W}(\mathfrak{B}_m,\mathfrak{B}^{[r+1]};0\circ i)&& \hat{W}(\mathfrak{B}_n,\hspace{0pt}\mathfrak{B}^{[r+1]};\hspace{2pt}0\circ j)\\
             \\
             \hat{W}(\mathfrak{B}_n,\hspace{0pt}\mathfrak{B}^{[r+1]};\hspace{2pt}1\circ j)&&{W}(\mathfrak{B}_m,\mathfrak{B}^{[r+1]};1\circ i) \\
             \\
             \hspace{0pt}{W}(\mathfrak{B}_n,\hspace{0pt}\mathfrak{B}^{[r+1]};\hspace{0pt}0\circ j)&&\hspace{0pt}\hat{W}(\mathfrak{B}_m,\hspace{0pt}\mathfrak{B}^{[r+1]};\hspace{0pt}0\circ i)\\
             \\
             \hspace{0pt}\hat{W}(\mathfrak{B}_m,\hspace{0pt}\mathfrak{B}^{[r+1]};\hspace{0pt}1\circ i)&&\hspace{0pt}{W}(\mathfrak{B}_n,\hspace{0pt}\mathfrak{B}^{[r+1]};\hspace{0pt}1\circ j)
             \end{array}
      \end{array}
\end{array}\]
 \end{center}
 \fcaption{The quotient algebra of rank $r$ $\{{\cal Q}(\mathfrak{B}^{[r]};2^{p+r}-1)\}$ in (a)
 and the co-quotient algebra of rank $r$ $\{{\cal Q}({W}(\mathfrak{B}_1,\mathfrak{B}^{[r]};i_1);2^{p+r}-1)\}$
 in (b) respectively given by an $r$-th maximal bi-subalgebra $\mathfrak{B}^{[r]}$
 of a Cartan subalgebra $\mathfrak{C}$ and a regular conditioned
 subspace ${W}(\mathfrak{B}_1,\mathfrak{B}^{[r]};s)\in\{{\cal Q}(\mathfrak{B}^{[r]})\}$,
 followed by two detached co-quotient algebra of rank $r+1$ acquired respectively by
 parallel detaching in (c) and by crossing detaching in (d);
 here
 $\mathfrak{B}_1=\mathfrak{B}_m\sqcap\mathfrak{B}_n\nsupseteq\mathfrak{B}^{[r]}$,
 $\mathfrak{B}^{[r+1]}=\mathfrak{B}_1\cap\mathfrak{B}^{[r]}$
 is a proper maximal bi-subalgebra of $\mathfrak{B}^{[r]}$,
 $1<m,n<2^p$, $s=i+j$ for $i,j,s\in{Z^r_2}$ and $\epsilon\circ i$ is the
 concatenation of two strings $\epsilon\in{Z_2}$ and $i\in{Z^r_2}$.
 \label{figdetQA}}
 \end{figure}

\newpage
\appendix{~~Tables of Quotient and Co-Quotient Algebras\label{appQAFigs}}

 \begin{figure}[htbp]
 \begin{center}
 \[
\begin{array}{c}
 \mathfrak{C}_{[\mathbf{0}]}\\
 \\
 {\cal S}^{000}_{000},\hspace{2pt}{\cal S}^{001}_{000},\hspace{2pt}{\cal S}^{010}_{000},\hspace{2pt}{\cal S}^{011}_{000},
 \hspace{2pt}{\cal S}^{100}_{000},\hspace{2pt}{\cal S}^{101}_{000},\hspace{2pt}{\cal S}^{110}_{000},\hspace{2pt}{\cal S}^{111}_{000}
 \end{array}\]
 \[\begin{array}{ccccc}
 {W}(\mathfrak{B}_{001})
 &{\cal S}^{000}_{001},\hspace{2pt}{\cal S}^{010}_{001},\hspace{2pt}{\cal S}^{100}_{001},\hspace{2pt}{\cal S}^{110}_{001}
 &\hspace{20pt}
 &{\cal S}^{001}_{001},\hspace{2pt}{\cal S}^{011}_{001},\hspace{2pt}{\cal S}^{101}_{001},\hspace{2pt}{\cal S}^{111}_{001}
 &\hat{W}(\mathfrak{B}_{001})\\
 &&&&\\
 W(\mathfrak{B}_{010})
 &{\cal S}^{000}_{010},\hspace{2pt}{\cal S}^{001}_{010},\hspace{2pt}{\cal S}^{100}_{010},\hspace{2pt}{\cal S}^{101}_{010}
 &
 &{\cal S}^{010}_{010},\hspace{2pt}{\cal S}^{011}_{010},\hspace{2pt}{\cal S}^{110}_{010},\hspace{2pt}{\cal S}^{111}_{010}
 &\hat{W}(\mathfrak{B}_{010})\\
 &&&&\\
 W(\mathfrak{B}_{011})
 &{\cal S}^{000}_{011},\hspace{2pt}{\cal S}^{011}_{011},\hspace{2pt}{\cal S}^{100}_{011},\hspace{2pt}{\cal S}^{111}_{011}
 &
 &{\cal S}^{001}_{011},\hspace{2pt}{\cal S}^{010}_{011},\hspace{2pt}{\cal S}^{101}_{011},\hspace{2pt}{\cal S}^{110}_{011}
 &\hat{W}(\mathfrak{B}_{011})\\
 &&&&\\
 W(\mathfrak{B}_{100})
 &{\cal S}^{000}_{100},\hspace{2pt}{\cal S}^{001}_{100},\hspace{2pt}{\cal S}^{010}_{100},\hspace{2pt}{\cal S}^{011}_{100}
 &
 &{\cal S}^{100}_{100},\hspace{2pt}{\cal S}^{101}_{100},\hspace{2pt}{\cal S}^{110}_{100},\hspace{2pt}{\cal S}^{111}_{100}
 &\hat{W}(\mathfrak{B}_{100})\\
 &&&&\\
\end{array} \]
 \vspace{15pt}
 \[ \hspace{10pt}\begin{array}{ccc}
 \mathfrak{B}_{001}=
 \{\hspace{2pt}{\cal S}^{000}_{000},{\cal S}^{010}_{000},{\cal S}^{100}_{000},{\cal S}^{110}_{000}\hspace{2pt}\},
 &\hspace{10pt}
 &\mathfrak{B}_{100}=
 \{\hspace{2pt}{\cal S}^{000}_{000},{\cal S}^{001}_{000},{\cal S}^{010}_{000},{\cal S}^{011}_{000}\hspace{2pt}\},
 \\
 \\
 \mathfrak{B}_{010}=
 \{\hspace{2pt}{\cal S}^{000}_{000},{\cal S}^{001}_{000},{\cal S}^{100}_{000},{\cal S}^{101}_{000}\hspace{2pt}\},
 &
 &\mathfrak{B}_{101}=
 \{\hspace{2pt}{\cal S}^{000}_{000},{\cal S}^{010}_{000},{\cal S}^{101}_{000},{\cal S}^{111}_{000}\hspace{2pt}\},
 \\
 \\
 \mathfrak{B}_{011}=
 \{\hspace{2pt}{\cal S}^{000}_{000},{\cal S}^{011}_{000},{\cal S}^{100}_{000},{\cal S}^{111}_{000}\hspace{2pt}\},
 &
 &\mathfrak{B}_{110}=
 \{\hspace{2pt}{\cal S}^{000}_{000},{\cal S}^{001}_{000},{\cal S}^{110}_{000},{\cal S}^{111}_{000}\hspace{2pt}\},\\
 &&\\
 &&\mathfrak{B}_{111}=
 \{\hspace{2pt}{\cal S}^{000}_{000},{\cal S}^{011}_{000},{\cal S}^{101}_{000},{\cal S}^{110}_{000}\hspace{2pt}\}.
 \end{array} \]
 \end{center}
 \fcaption{The intrinsic quotient algebra of rank zero given by the intrinsic Cartan subalgebra
 $\mathfrak{C}_{[{\bf 0}]}\subset su(8)$.\label{figsu8QArank0intr}}
 \end{figure}

 \begin{figure}[htbp]
 \begin{center}
  \[\begin{array}{c}
   \mathfrak{B}^{[1]}_{intr}\\
   \\
   {\cal S}^{000}_{000},{\cal S}^{001}_{000},{\cal S}^{010}_{000},{\cal S}^{011}_{000}
   \end{array}\]
  \[\begin{array}{ccccc}
  \hspace{0pt}\mathfrak{B}^{[1,1]}_{intr}=\mathfrak{C}_{[\mathbf{0}]}-\mathfrak{B}^{[1]}_{intr}
  &{\cal S}^{100}_{000},{\cal S}^{101}_{000},{\cal S}^{110}_{000},{\cal S}^{111}_{000}&&\{0\}&\\
  \\
  {W}(\mathfrak{B}_{100},\mathfrak{B}^{[1]}_{intr};0)
  &{\cal S}^{000}_{100},{\cal S}^{001}_{100},{\cal S}^{010}_{100},{\cal S}^{011}_{100}&&\{0\}&\hat{W}(\mathfrak{B}_{100},\mathfrak{B}^{[1]}_{intr};0)\\
  &&&&\\
  {W}(\mathfrak{B}_{100},\mathfrak{B}^{[1]}_{intr};1)&\{0\}&&{\cal S}^{100}_{100},{\cal S}^{101}_{100},{\cal S}^{110}_{100},{\cal S}^{111}_{100}&\hspace{0pt}\hat{W}(\mathfrak{B}_{100},\mathfrak{B}^{[1]}_{intr};1)\\
  \\
  W(\mathfrak{B}_{001},\mathfrak{B}^{[1]}_{intr};0)&{\cal S}^{000}_{001},{\cal S}^{010}_{001}
  &&{\cal S}^{001}_{001},{\cal S}^{011}_{001}&\hat{W}(\mathfrak{B}_{001},\mathfrak{B}^{[1]}_{intr};0)\\
  \\
\end{array}\]
 \end{center}
 \fcaption{The intrinsic quotient algebra of rank one given by the intrinsic
 center subalgebra $\mathfrak{B}^{[1]}_{intr}=\mathfrak{B}^{[1,0]}_{intr}=\mathfrak{B}_{100}\subset\mathfrak{C}_{[\mathbf{0}]}\subset su(8)$.\label{figsu8QArank1intr}}
 \end{figure}

 \begin{figure}[htbp]
 \begin{center}
  \[ \begin{array}{c}
   \mathfrak{B}^{[1,1]}_{intr}\\
   \\
   {\cal S}^{100}_{000},{\cal S}^{101}_{000},{\cal S}^{110}_{000},{\cal S}^{111}_{000}
   \end{array}\]
  \[\begin{array}{ccccc}
  \mathfrak{B}^{[1]}_{intr}
  &{\cal S}^{000}_{000},{\cal S}^{001}_{000},{\cal S}^{010}_{000},{\cal S}^{011}_{000}&&\{0\}&\\
  \\
  W(\mathfrak{B}_{100},\mathfrak{B}^{[1]}_{intr};0)&{\cal S}^{000}_{100},{\cal S}^{001}_{100},{\cal S}^{010}_{100},{\cal S}^{011}_{100}
  &&{\cal S}^{100}_{100},{\cal S}^{101}_{100},{\cal S}^{110}_{100},{\cal S}^{111}_{100}&\hspace{0pt}\hat{W}(\mathfrak{B}_{100},\mathfrak{B}^{[1]}_{intr};1)\\
  \\
  {W}(\mathfrak{B}_{100},\mathfrak{B}^{[1]}_{intr};1)
  &\{0\}&&\{0\}&\hat{W}(\mathfrak{B}_{100},\mathfrak{B}^{[1]}_{intr};0)\\
  \\
  W(\mathfrak{B}_{001},\mathfrak{B}^{[1]}_{intr};0)&{\cal S}^{000}_{001},{\cal S}^{010}_{001}
  &&{\cal S}^{101}_{001},{\cal S}^{111}_{001}&\hat{W}(\mathfrak{B}_{001},\mathfrak{B}^{[1]}_{intr};1)\\
  \\
\end{array} \]
 \end{center}
 \fcaption{The co-quotient algebra of rank one given by the coset
 $\mathfrak{B}^{[1,1]}_{intr}=\mathfrak{C}_{[{\bf 0}]}-\mathfrak{B}^{[1]}_{intr}\subset su(8)$.\label{figsu8coQArank1inC0}}
 \end{figure}

 \begin{figure}[htbp]
 \begin{center}
  \[\begin{array}{c}
   \mathfrak{B}^{[1]}_{can}\\
   \\
   {\cal S}^{000}_{000},{\cal S}^{010}_{000},{\cal S}^{100}_{000},{\cal S}^{110}_{000}
   \end{array}\]
  \[\begin{array}{ccccc}
  \hspace{0pt}\mathfrak{B}^{[1,1]}_{can}
  &{\cal S}^{001}_{000},{\cal S}^{011}_{000},{\cal S}^{101}_{000},{\cal S}^{111}_{000}&&\{0\}&\\
  \\
  W(\mathfrak{B}_{001},\mathfrak{B}^{[1]}_{can};0)&{\cal S}^{000}_{001},{\cal S}^{010}_{001},{\cal S}^{100}_{001},{\cal S}^{110}_{001}
  &&\{0\} &\hat{W}(\mathfrak{B}_{001},\mathfrak{B}^{[1]}_{can};0)\\
  \\
  W(\mathfrak{B}_{001},\mathfrak{B}^{[1]}_{can};1)&\{0\}
  &&{\cal S}^{101}_{001},{\cal S}^{111}_{001},{\cal S}^{001}_{001},{\cal S}^{011}_{001}&\hat{W}(\mathfrak{B}_{001},\mathfrak{B}^{[1]}_{can};1)\\
  \\
  W(\mathfrak{B}_{010},\mathfrak{B}^{[1]}_{can};0)&{\cal S}^{000}_{010},{\cal S}^{100}_{010}
  &&{\cal S}^{010}_{010},{\cal S}^{110}_{010}&\hat{W}(\mathfrak{B}_{010},\mathfrak{B}^{[1]}_{can};0)\\
  \\
\end{array}\]
 \end{center}
 \vspace{6pt}
 \fcaption{The canonical quotient algebra of rank one given by the canonical center subalgebra
 $\mathfrak{B}^{[1]}_{can}=\mathfrak{B}^{[1,0]}_{can}=\mathfrak{B}_{001}\subset\mathfrak{C}_{[\mathbf{0}]}\subset su(8)$.\label{figsu8canQArank1}}
 \end{figure}

 \begin{figure}[htbp]
 \begin{center}
  \[\begin{array}{c}
   \mathfrak{B}^{[1,1]}_{can}\\
   \\
   {\cal S}^{001}_{000},{\cal S}^{011}_{000},{\cal S}^{101}_{000},{\cal S}^{111}_{000}
   \end{array}\]
  \[\begin{array}{ccccc}
  \hspace{0pt}\mathfrak{B}^{[1]}_{can}
  &{\cal S}^{000}_{000},{\cal S}^{010}_{000},{\cal S}^{100}_{000},{\cal S}^{110}_{000}&&\{0\}&\\
  \\
  W(\mathfrak{B}_{001},\mathfrak{B}^{[1]}_{can};0)&{\cal S}^{000}_{001},{\cal S}^{010}_{001},{\cal S}^{100}_{001},{\cal S}^{110}_{001}
  &&{\cal S}^{101}_{001},{\cal S}^{111}_{001},{\cal S}^{001}_{001},{\cal S}^{011}_{001}&\hat{W}(\mathfrak{B}_{001},\mathfrak{B}^{[1]}_{can};1)\\
  \\
  W(\mathfrak{B}_{001},\mathfrak{B}^{[1]}_{can};1)&\{0\}
  && \{0\}  &\hat{W}(\mathfrak{B}_{001},\mathfrak{B}^{[1]}_{can};0)\\
  \\
  W(\mathfrak{B}_{010},\mathfrak{B}^{[1]}_{can};0)&{\cal S}^{000}_{010},{\cal S}^{100}_{010}
  &&{\cal S}^{011}_{010},{\cal S}^{111}_{010}&\hat{W}(\mathfrak{B}_{010},\mathfrak{B}^{[1]}_{can};1)\\
  \\
\end{array}\]
 \end{center}
 \vspace{6pt}
 \fcaption{The co-quotient algebra of rank one given by the coset
 $\mathfrak{B}^{[1,1]}_{can}=\mathfrak{C}_{[\mathbf{0}]}-\mathfrak{B}^{[1]}_{can}$.\label{figsu8cancoQArank1}}
 \end{figure}

 \begin{figure}[htbp]
 \begin{center}
 \[\begin{array}{c}
 \mathfrak{C}_{\rm III}\\
 \\
 {\cal S}^{000}_{000},\hspace{2pt}{\cal S}^{001}_{000},\hspace{2pt}{\cal S}^{010}_{000},\hspace{2pt}{\cal S}^{011}_{000},
 \hspace{2pt}{\cal S}^{100}_{100},\hspace{2pt}{\cal S}^{101}_{100},\hspace{2pt}{\cal S}^{110}_{100},\hspace{2pt}{\cal S}^{111}_{100}
 \end{array}\]
 \[\begin{array}{ccccc}
 W(\hat{\mathfrak{B}}_{100,100})
 &{\cal S}^{000}_{100},\hspace{2pt}{\cal S}^{001}_{100},\hspace{2pt} {\cal S}^{010}_{100},\hspace{2pt}{\cal S}^{011}_{100}
 &
 &{\cal S}^{100}_{000},\hspace{2pt}{\cal S}^{101}_{000},\hspace{2pt} {\cal S}^{110}_{000},\hspace{2pt}{\cal S}^{111}_{000}
 &\hat{W}(\hat{\mathfrak{B}}_{100,100})\\
 \\
 W(\hat{\mathfrak{B}}_{001,100})
 &{\cal S}^{000}_{001},\hspace{2pt}{\cal S}^{010}_{001},\hspace{2pt}{\cal S}^{100}_{101},\hspace{2pt}{\cal S}^{110}_{101}
 &\hspace{20pt}
 &{\cal S}^{001}_{001},\hspace{2pt}{\cal S}^{011}_{001},\hspace{2pt} {\cal S}^{101}_{101},\hspace{2pt}{\cal S}^{111}_{101}
 &\hat{W}(\hat{\mathfrak{B}}_{001,100})\\
 \\
 W(\hat{\mathfrak{B}}_{101,100})
 &{\cal S}^{101}_{001},\hspace{2pt}{\cal S}^{111}_{001},\hspace{2pt} {\cal S}^{000}_{101},\hspace{2pt}{\cal S}^{010}_{101}
 &
 &{\cal S}^{100}_{001},\hspace{2pt}{\cal S}^{110}_{001},\hspace{2pt}{\cal S}^{001}_{101},\hspace{2pt}{\cal S}^{011}_{101}
 &\hat{W}(\hat{\mathfrak{B}}_{101,100})\\
 \\
 W(\hat{\mathfrak{B}}_{010,100})
 &{\cal S}^{000}_{010},\hspace{2pt}{\cal S}^{001}_{010},\hspace{2pt} {\cal S}^{100}_{110},\hspace{2pt}{\cal S}^{101}_{110}
 &
 &{\cal S}^{010}_{010},\hspace{2pt}{\cal S}^{011}_{010},\hspace{2pt} {\cal S}^{110}_{110},\hspace{2pt}{\cal S}^{111}_{110}
 &\hat{W}(\hat{\mathfrak{B}}_{010,100})\\
 \\
\end{array} \]
 \vspace{15pt}
 \[\hspace{9pt}\begin{array}{ccc}
 \hspace{0pt}\hat{\mathfrak{B}}_{001,100}=
 \{\hspace{2pt}{\cal S}^{000}_{000},{\cal S}^{010}_{000},{\cal S}^{100}_{100},{\cal S}^{110}_{100}\hspace{2pt}\},
 &\hspace{10pt}
 &\hat{\mathfrak{B}}_{100,100}=
 \{\hspace{2pt}{\cal S}^{000}_{000},{\cal S}^{001}_{000},{\cal S}^{010}_{000},{\cal S}^{011}_{000}\hspace{2pt}\},
 \\
 \\
 \hat{\mathfrak{B}}_{010,100}=
 \{\hspace{2pt}{\cal S}^{000}_{000},{\cal S}^{001}_{000},{\cal S}^{100}_{100},{\cal S}^{101}_{100}\hspace{2pt}\},
 &
 &\hat{\mathfrak{B}}_{101,100}=
 \{\hspace{2pt}{\cal S}^{000}_{000},{\cal S}^{010}_{000},{\cal S}^{101}_{100},{\cal S}^{111}_{100}\hspace{2pt}\},
 \\
 \\
 \hat{\mathfrak{B}}_{011,100}=
 \{\hspace{2pt}{\cal S}^{000}_{000},{\cal S}^{011}_{000},{\cal S}^{100}_{100},{\cal S}^{111}_{100}\hspace{2pt}\},
 &
 &\hat{\mathfrak{B}}_{110,100}=
 \{\hspace{2pt}{\cal S}^{000}_{000},{\cal S}^{001}_{000},{\cal S}^{110}_{100},{\cal S}^{111}_{100}\hspace{2pt}\},
 \\
 \\
 &&\hat{\mathfrak{B}}_{111,100}=
 \{\hspace{2pt}{\cal S}^{000}_{000},{\cal S}^{011}_{000},{\cal S}^{101}_{100},{\cal S}^{110}_{100}\hspace{2pt}\}.
 \end{array} \]
 \end{center}
 \fcaption{The quotient algebra of rank zero given by a Cartan subalgebra $\mathfrak{C}_{\rm III}\subset su(8)$.\label{figsu8QArank0inCIII}}
 \end{figure}

 \begin{figure}[htbp]
  \begin{center}
   \[\begin{array}{c}
   \hat{\mathfrak{B}}^{[1]}_{intr}\\
   \\
   {\cal S}^{000}_{000},{\cal S}^{001}_{000},{\cal S}^{010}_{000},{\cal S}^{011}_{000}
   \end{array}\]
  \[\begin{array}{cccc}
  \hat{\mathfrak{B}}^{[1,1]}_{intr}=\mathfrak{C}_{\rm III}-\hat{\mathfrak{B}}^{[1]}_{intr}
  &{\cal S}^{100}_{100},{\cal S}^{101}_{100},{\cal S}^{110}_{100},{\cal S}^{111}_{100}
  &\{0\}&\\
  \\
  {W}(\hat{\mathfrak{B}}_{100,100},\hat{\mathfrak{B}}^{[1]}_{intr};0)
  &{\cal S}^{000}_{100},{\cal S}^{001}_{100},{\cal S}^{010}_{100},{\cal S}^{011}_{100}
  &\{0\}&\hat{W}(\hat{\mathfrak{B}}_{100,100},\hat{\mathfrak{B}}^{[1]}_{intr};0)\\
  \\
  {W}(\hat{\mathfrak{B}}_{100,100},\hat{\mathfrak{B}}^{[1]}_{intr};1)
  &\{0\}
  &{\cal S}^{100}_{000},{\cal S}^{101}_{000},{\cal S}^{110}_{000},{\cal S}^{111}_{000}&\hspace{0pt}\hat{W}(\hat{\mathfrak{B}}_{100,100},\hat{\mathfrak{B}}^{[1]}_{intr};1)\\
  \\
  {W}(\hat{\mathfrak{B}}_{001,100},\hat{\mathfrak{B}}^{[1]}_{intr};0)&{\cal S}^{000}_{001},{\cal S}^{010}_{001}
  &{\cal S}^{001}_{001},{\cal S}^{011}_{001}&\hat{W}(\hat{\mathfrak{B}}_{001,100},\hat{\mathfrak{B}}^{[1]}_{intr};0)\\
  \\
\end{array}\]
  \end{center}
 \fcaption{The quotient algebra of rank one given by a maximal bi-subalgebra
 $\hat{\mathfrak{B}}^{[1]}_{intr}=\hat{\mathfrak{B}}^{[1,0]}_{intr}=\hat{\mathfrak{B}}_{100,100}\subset\mathfrak{C}_{\rm III}\subset su(8)$.\label{figsu8QArank1inCIII}}
 \end{figure}

 \begin{figure}[htbp]
 \begin{center}
 \[\begin{array}{c}
   \hat{\mathfrak{B}}^{[1,1]}_{intr}\\
   \\
   {\cal S}^{100}_{100},{\cal S}^{101}_{100},{\cal S}^{110}_{100},{\cal S}^{111}_{100}
   \end{array}\]
  \[\begin{array}{cccc}
  \hat{\mathfrak{B}}^{[1]}_{intr}
  &{\cal S}^{000}_{000},{\cal S}^{001}_{000},{\cal S}^{010}_{000},{\cal S}^{011}_{000}
  &\{0\}&\\
  \\
  {W}(\hat{\mathfrak{B}}_{100,100},\hat{\mathfrak{B}}^{[1]}_{intr};0)
  &{\cal S}^{000}_{100},{\cal S}^{001}_{100},{\cal S}^{010}_{100},{\cal S}^{011}_{100}
  &{\cal S}^{100}_{000},{\cal S}^{101}_{000},{\cal S}^{110}_{000},{\cal S}^{111}_{000}&\hat{W}(\hat{\mathfrak{B}}_{100,100},\hat{\mathfrak{B}}^{[1]}_{intr};1)\\
  \\
  {W}(\hat{\mathfrak{B}}_{100,100},\hat{\mathfrak{B}}^{[1]}_{intr};1)
  &\{0\}
  &\{0\}&\hspace{0pt}\hat{W}(\hat{\mathfrak{B}}_{100,100},\hat{\mathfrak{B}}^{[1]}_{intr};0)\\
  \\
  {W}(\hat{\mathfrak{B}}_{001,100},\hat{\mathfrak{B}}^{[1]}_{intr};0)&{\cal S}^{000}_{001},{\cal S}^{010}_{001}
  &{\cal S}^{101}_{101},{\cal S}^{111}_{101}&\hat{W}(\hat{\mathfrak{B}}_{001,100},\hat{\mathfrak{B}}^{[1]}_{intr};1)\\
  \\
\end{array}\]
 \end{center}
 \fcaption{The intrinsic co-quotient algebra of rank one given by the
 intrinsic center subalgebra
 $\hat{\mathfrak{B}}^{[1,1]}_{intr}=\mathfrak{C}_{\rm III}-\hat{\mathfrak{B}}^{[1]}_{intr}\subset su(8)$.\label{figsu8coQArank1intr}}
 \end{figure}

 \begin{figure}[htbp]
 \begin{center}
 \[\begin{array}{c}
 \mathfrak{C}\\
 \\
 \hspace{20pt}{\cal S}^{000}_{000},\hspace{2pt}{\cal S}^{001}_{000},\hspace{2pt}{\cal S}^{010}_{010},\hspace{2pt}{\cal S}^{011}_{010},
 \hspace{2pt}{\cal S}^{100}_{100},\hspace{2pt}{\cal S}^{101}_{100},\hspace{2pt}{\cal S}^{110}_{110},\hspace{2pt}{\cal S}^{111}_{110}
 \end{array}\]
 \[\begin{array}{ccccc}
 W(\mathfrak{B}_{1})
 &{\cal S}^{100}_{000},\hspace{2pt}{\cal S}^{101}_{000},\hspace{2pt} {\cal S}^{110}_{010},\hspace{2pt}{\cal S}^{111}_{010}
 &
 &{\cal S}^{000}_{100},\hspace{2pt}{\cal S}^{001}_{100},\hspace{2pt} {\cal S}^{010}_{110},\hspace{2pt}{\cal S}^{011}_{110}
 &\hat{W}(\mathfrak{B}_{1})\\
 \\
 W(\mathfrak{B}_{2})
 &{\cal S}^{010}_{000},\hspace{2pt}{\cal S}^{011}_{000},\hspace{2pt}{\cal S}^{110}_{100},\hspace{2pt}{\cal S}^{111}_{100}
 &\hspace{20pt}
 &{\cal S}^{000}_{010},\hspace{2pt}{\cal S}^{001}_{010},\hspace{2pt} {\cal S}^{100}_{110},\hspace{2pt}{\cal S}^{101}_{110}
 &\hat{W}(\mathfrak{B}_{2})\\
 \\
 W(\mathfrak{B}_{3})
 &{\cal S}^{110}_{000},\hspace{2pt}{\cal S}^{111}_{000},\hspace{2pt}{\cal S}^{000}_{110},\hspace{2pt}{\cal S}^{001}_{110}
 &
 &{\cal S}^{100}_{010},\hspace{2pt}{\cal S}^{101}_{010},\hspace{2pt} {\cal S}^{010}_{100},\hspace{2pt}{\cal S}^{011}_{100}
 &\hat{W}(\mathfrak{B}_{3})\\
 \\
 W(\mathfrak{B}_{4})
 &{\cal S}^{000}_{001},\hspace{2pt}{\cal S}^{010}_{011},\hspace{2pt} {\cal S}^{100}_{101},\hspace{2pt}{\cal S}^{110}_{111}
 &
 &{\cal S}^{001}_{001},\hspace{2pt}{\cal S}^{011}_{011},\hspace{2pt} {\cal S}^{101}_{101},\hspace{2pt}{\cal S}^{111}_{111}
 &\hat{W}(\mathfrak{B}_{4})\\
 \\
\end{array} \]
 \vspace{15pt}
 \[ \hspace{10pt}\begin{array}{ccc}
 \mathfrak{B}_{1}=
 \{\hspace{2pt}{\cal S}^{000}_{000},{\cal S}^{001}_{000},{\cal S}^{010}_{010},{\cal S}^{011}_{010}\hspace{2pt}\},
 &\hspace{10pt}
 &\mathfrak{B}_{4}=
 \{\hspace{2pt}{\cal S}^{000}_{000},{\cal S}^{010}_{010},{\cal S}^{100}_{100},{\cal S}^{110}_{110}\hspace{2pt}\},
 \\
 \\
 \mathfrak{B}_{2}=
 \{\hspace{2pt}{\cal S}^{000}_{000},{\cal S}^{001}_{000},{\cal S}^{100}_{100},{\cal S}^{101}_{100}\hspace{2pt}\},
 &
 &
 \mathfrak{B}_{5}=
 \{\hspace{2pt}{\cal S}^{000}_{000},{\cal S}^{010}_{010},{\cal S}^{101}_{100},{\cal S}^{111}_{110}\hspace{2pt}\},
 \\
 \\
 \mathfrak{B}_{3}=
 \{\hspace{2pt}{\cal S}^{000}_{000},{\cal S}^{001}_{000},{\cal S}^{110}_{110},{\cal S}^{101}_{110}\hspace{2pt}\},
 &
 &
 \mathfrak{B}_{6}=
 \{\hspace{2pt}{\cal S}^{000}_{000},{\cal S}^{011}_{010},{\cal S}^{100}_{100},{\cal S}^{111}_{110}\hspace{2pt}\},
 \\
 \\
 &&
 \mathfrak{B}_{7}=
 \{\hspace{2pt}{\cal S}^{000}_{000},{\cal S}^{011}_{010},{\cal S}^{101}_{100},{\cal S}^{110}_{110}\hspace{2pt}\}.
 \end{array}\]
 \end{center}
 \fcaption{A quotient algebra of rank one given by a Cartan subalgebra $\mathfrak{C}\subset su(8)$.\label{figsu8QArank0inC}}
 \end{figure}

 \begin{figure}[htbp]
 \begin{center}
  \[\begin{array}{c}
   \mathfrak{B}^{[1]}\\
   \\
   {\cal S}^{000}_{000},{\cal S}^{001}_{000},{\cal S}^{010}_{010},{\cal S}^{011}_{010}
   \end{array}\]
  \[\begin{array}{ccccc}
  \mathfrak{B}^{[1,1]}=\mathfrak{C}-\mathfrak{B}^{[1]}
  &{\cal S}^{100}_{100},{\cal S}^{101}_{100},{\cal S}^{110}_{110},{\cal S}^{111}_{110}&&\{0\}&\\
  \\
  {W}(\mathfrak{B}_1,\mathfrak{B}^{[1]};0)
  &{\cal S}^{100}_{000},{\cal S}^{101}_{000},{\cal S}^{110}_{010}\,{\cal S}^{111}_{010}&&\{0\}&\hat{W}(\mathfrak{B}_1,\mathfrak{B}^{[1]};0)\\
  \\
  {W}(\mathfrak{B}_1,\mathfrak{B}^{[1]};1)&\{0\}&&{\cal S}^{000}_{100},{\cal S}^{001}_{100},{\cal S}^{010}_{110},{\cal S}^{011}_{110}&\hspace{0pt}\hat{W}(\mathfrak{B}_1,\mathfrak{B}^{[1]};1)\\
  \\
  {W}(\mathfrak{B}_2,\mathfrak{B}^{[1]};0)&{\cal S}^{010}_{000},{\cal S}^{011}_{000}
  &&{\cal S}^{000}_{010},{\cal S}^{001}_{010}&\hat{W}(\mathfrak{B}_2,\mathfrak{B}^{[1]};0)\\
  \\
\end{array} \]
 \end{center}
 \fcaption{A quotient algebra of rank one given by a maximal bi-subalgebra
 $\mathfrak{B}^{[1]}=\mathfrak{B}^{[1,0]}=\mathfrak{B}_1\subset\mathfrak{C}\subset su(8)$.\label{figsu8QArank1inC}}
 \end{figure}

 \begin{figure}[htbp]
 \begin{center}
 \[\begin{array}{c}
   \mathfrak{B}^{[1,1]}\\
   \\
   {\cal S}^{100}_{100},{\cal S}^{101}_{100},{\cal S}^{110}_{110},{\cal S}^{111}_{110}
   \end{array}\]
  \[\begin{array}{ccccc}
  \mathfrak{B}^{[1]}
  &{\cal S}^{000}_{000},{\cal S}^{001}_{000},{\cal S}^{010}_{010},{\cal S}^{011}_{010}
  &&\{0\}&\\
  \\
  {W}(\mathfrak{B}_1,\mathfrak{B}^{[1]};0)
  &{\cal S}^{100}_{000},{\cal S}^{101}_{000},{\cal S}^{110}_{010}\,{\cal S}^{111}_{010}
  &&
  {\cal S}^{000}_{100},{\cal S}^{001}_{100},{\cal S}^{010}_{110},{\cal S}^{011}_{110}&\hat{W}(\mathfrak{B}_1,\mathfrak{B}^{[1]};1)\\
  \\
  {W}(\mathfrak{B}_1,\mathfrak{B}^{[1]};1)&\{0\}&&\{0\}&\hspace{0pt}\hat{W}(\mathfrak{B}_1,\mathfrak{B}^{[1]};0)\\
  \\
  {W}(\mathfrak{B}_2,\mathfrak{B}^{[1]};0)&{\cal S}^{010}_{000},{\cal S}^{011}_{000}
  &&{\cal S}^{100}_{110},{\cal S}^{101}_{110}&\hat{W}(\mathfrak{B}_2,\mathfrak{B}^{[1]};1)\\
  \\
\end{array}\]
 \end{center}
 \fcaption{A co-quotient algebra of rank one given by the coset
 $\mathfrak{B}^{[1,1]}=\mathfrak{C}-\mathfrak{B}^{[1]}\subset su(8)$.\label{figsu8coQArank1inC}}
 \end{figure}

 \newpage
 \appendix{~~Tables of Co-Quotient Algebras with $\lambda$-generators\label{appAIIIQAFigs}}

 \begin{figure}[h]
 \begin{center}
 \[%
\]\\
 \end{center}
 \fcaption{The intrinsic merged co-quotient algebra of $su(8)$ with generators in the $\lambda$-representation;
 here $\mathfrak{C}_{[\mathbf{0}]}$ is the intrinsic Cartan subalgebra consisting of all diagonal generators in $su(8)$.  \label{figsu8QA4plus4}}
 \end{figure}

 \begin{figure}[p]
 \begin{center}
 \[ %
\]\\
 \end{center}
 \fcaption{A co-quotient algebra of $su(8)$, in the $\lambda$-representation,
 given by a center subalgebra consisting of three generators.\label{figsu8QA5plus3}}
 \end{figure}

 \begin{figure}[p]
 \begin{center}
 \[%
\]\\
 \end{center}
 \fcaption{A co-quotient algebra of $su(8)$ 
 given by a center subalgebra of two generators.\label{figsu8QA6plus2}}
 \end{figure}

 \begin{figure}[p]
 \begin{center}
 \[%
\]\\
 \end{center}
 \fcaption{A co-quotient algebra of $su(8)$ 
 given by a center subalgebra
 of only one generator.\label{figsu8QA7plus1}}
 \end{figure}

 \newpage
 \appendix{~~Tables of Cartan decompositions\label{apptpFigs}}

 \begin{figure}[!ht]
 \[%
 \end{array}
 \end{array}\]
 \fcaption{Three examples of Cartan decompositions determined from the intrinsic quotient algebra of $su(8)$ in Fig.~\ref{figsu8QArank1intr}:
 (a) the intrinsic decomposition of type {\bf AII}
 $\hat{\mathfrak{t}}_{\hspace{1pt}{\rm II}}\oplus\hat{\mathfrak{p}}_{\hspace{1pt}{\rm II}}$;
 (b) a decomposition of type {\bf AI}
 and (c) a type {\bf AII}. \label{figsu8tpintrrank1}}
 \end{figure}

 \begin{figure}[htbp]
 \[\begin{array}{ccc}
 \hspace{-142pt}
 \begin{array}{c}
 {\mathfrak{t}}_{\hspace{1pt}{\rm I}}=\\
 \hspace{120pt}\{\hspace{2pt}{W}(\mathfrak{B}_1,\mathfrak{B}^{[1]};0),\hspace{10pt}\hat{W}(\mathfrak{B}_1,\mathfrak{B}^{[1]};1),\\
 \hspace{127pt}{W}(\mathfrak{B}_2,\mathfrak{B}^{[1]};0),\hspace{10pt}\hat{W}(\mathfrak{B}_2,\mathfrak{B}^{[1]};1),\\
 \hspace{127pt}{W}(\mathfrak{B}_3,\mathfrak{B}^{[1]};1),\hspace{10pt}\hat{W}(\mathfrak{B}_3,\mathfrak{B}^{[1]};0),\\
 \hspace{127pt}{W}(\mathfrak{B}_4,\mathfrak{B}^{[1]};0),\hspace{10pt}\hat{W}(\mathfrak{B}_4,\mathfrak{B}^{[1]};1),\\
 \hspace{127pt}{W}(\mathfrak{B}_5,\mathfrak{B}^{[1]};1),\hspace{10pt}\hat{W}(\mathfrak{B}_5,\mathfrak{B}^{[1]};0),\\
 \hspace{127pt}{W}(\mathfrak{B}_6,\mathfrak{B}^{[1]};1),\hspace{10pt}\hat{W}(\mathfrak{B}_6,\mathfrak{B}^{[1]};0),\\
 \hspace{132pt}{W}(\mathfrak{B}_7,\mathfrak{B}^{[1]};0),\hspace{10pt}\hat{W}(\mathfrak{B}_7,\mathfrak{B}^{[1]};1)\hspace{2pt}\}\\
 \\
 \end{array}
 &\hspace{35pt}&\hspace{-153pt}
 \begin{array}{c}
 {\mathfrak{p}}_{\hspace{1pt}{\rm I}}=\\
 \hspace{63pt}\{\hspace{2pt}\mathfrak{C}=\mathfrak{B}^{[1]}\cup\mathfrak{B}^{[1,1]},\\
 \hspace{138pt}{W}(\mathfrak{B}_1,\mathfrak{B}^{[1]};1),\hspace{10pt}\hat{W}(\mathfrak{B}_1,\mathfrak{B}^{[1]};0),\\
 \hspace{138pt}{W}(\mathfrak{B}_2,\mathfrak{B}^{[1]};1),\hspace{10pt}\hat{W}(\mathfrak{B}_2,\mathfrak{B}^{[1]};0),\\
 \hspace{138pt}{W}(\mathfrak{B}_3,\mathfrak{B}^{[1]};0),\hspace{10pt}\hat{W}(\mathfrak{B}_3,\mathfrak{B}^{[1]};1),\\
 \hspace{138pt}{W}(\mathfrak{B}_4,\mathfrak{B}^{[1]};1),\hspace{10pt}\hat{W}(\mathfrak{B}_4,\mathfrak{B}^{[1]};0),\\
 \hspace{138pt}{W}(\mathfrak{B}_5,\mathfrak{B}^{[1]};0),\hspace{10pt}\hat{W}(\mathfrak{B}_5,\mathfrak{B}^{[1]};1),\\
 \hspace{138pt}{W}(\mathfrak{B}_6,\mathfrak{B}^{[1]};0),\hspace{10pt}\hat{W}(\mathfrak{B}_6,\mathfrak{B}^{[1]};1),\\
 \hspace{142pt}{W}(\mathfrak{B}_7,\mathfrak{B}^{[1]};1),\hspace{10pt}\hat{W}(\mathfrak{B}_7,\mathfrak{B}^{[1]};0)\hspace{2pt}\}
 \end{array}
 \end{array}\]
 \[\begin{array}{ccc}
 \hspace{-142pt}
 \begin{array}{c}
 {\mathfrak{t}}_{\hspace{1pt}{\rm II}}=\\
 \{\hspace{2pt}\mathfrak{B}^{[1,1]},\\
 \hspace{125pt}{W}(\mathfrak{B}_1,\mathfrak{B}^{[1]};0),\hspace{10pt}{W}(\mathfrak{B}_1,\mathfrak{B}^{[1]};1),\\
 \hspace{125pt}{W}(\mathfrak{B}_2,\mathfrak{B}^{[1]};0),\hspace{10pt}{W}(\mathfrak{B}_2,\mathfrak{B}^{[1]};1),\\
 \hspace{125pt}\hat{W}(\mathfrak{B}_3,\mathfrak{B}^{[1]};0),\hspace{10pt}\hat{W}(\mathfrak{B}_3,\mathfrak{B}^{[1]};1),\\
 \hspace{125pt}{W}(\mathfrak{B}_4,\mathfrak{B}^{[1]};0),\hspace{10pt}{W}(\mathfrak{B}_4,\mathfrak{B}^{[1]};1),\\
 \hspace{125pt}\hat{W}(\mathfrak{B}_5,\mathfrak{B}^{[1]};0),\hspace{10pt}\hat{W}(\mathfrak{B}_5,\mathfrak{B}^{[1]};1),\\
 \hspace{125pt}\hat{W}(\mathfrak{B}_6,\mathfrak{B}^{[1]};0),\hspace{10pt}\hat{W}(\mathfrak{B}_6,\mathfrak{B}^{[1]};1),\\
 \hspace{129pt}{W}(\mathfrak{B}_7,\mathfrak{B}^{[1]};0),\hspace{10pt}{W}(\mathfrak{B}_7,\mathfrak{B}^{[1]};1)\hspace{2pt}\}
 \end{array}
 &\hspace{35pt}&\hspace{-153pt}
 \begin{array}{c}
 \hspace{0pt}{\mathfrak{p}}_{\hspace{1pt}{\rm II}}=\\
 \{\hspace{2pt}\mathfrak{B}^{[1]},\\
 \hspace{132pt}\hat{W}(\mathfrak{B}_1,\mathfrak{B}^{[1]};0),\hspace{10pt}\hat{W}(\mathfrak{B}_1,\mathfrak{B}^{[1]};1),\\
 \hspace{132pt}\hat{W}(\mathfrak{B}_2,\mathfrak{B}^{[1]};0),\hspace{10pt}\hat{W}(\mathfrak{B}_2,\mathfrak{B}^{[1]};1),\\
 \hspace{132pt}{W}(\mathfrak{B}_3,\mathfrak{B}^{[1]};0),\hspace{10pt}{W}(\mathfrak{B}_3,\mathfrak{B}^{[1]};1),\\
 \hspace{132pt}\hat{W}(\mathfrak{B}_4,\mathfrak{B}^{[1]};0),\hspace{10pt}\hat{W}(\mathfrak{B}_4,\mathfrak{B}^{[1]};1),\\
 \hspace{132pt}{W}(\mathfrak{B}_5,\mathfrak{B}^{[1]};0),\hspace{10pt}{W}(\mathfrak{B}_5,\mathfrak{B}^{[1]};1),\\
 \hspace{132pt}{W}(\mathfrak{B}_6,\mathfrak{B}^{[1]};0),\hspace{10pt}{W}(\mathfrak{B}_6,\mathfrak{B}^{[1]};1),\\
 \hspace{136pt}\hat{W}(\mathfrak{B}_7,\mathfrak{B}^{[1]};0),\hspace{10pt}\hat{W}(\mathfrak{B}_7,\mathfrak{B}^{[1]};1)\hspace{2pt}\}
 \end{array}
 \end{array}\]
 \fcaption{Examples of a type-{\bf AI} and a type-{\bf AII} Cartan decompositions
 determined from the quotient algebra of $su(8)$ in Fig.~\ref{figsu8QArank1inC}. \label{figsu8tprank1}}
 \end{figure}


 \begin{figure}[htbp]
 \hspace{0pt}
 \[\hspace{-13pt}\begin{array}{ccc}
 \hspace{-246pt}(a)&&\\
 \hspace{-190pt}
 \begin{array}{c}
 \hat{\mathfrak{t}}_{\hspace{1pt}{\rm III}}=\\
 \{\hspace{2pt}\hat{\mathfrak{B}}^{[1]}_{intr},\\
 \hspace{182pt}{W}(\hat{\mathfrak{B}}_{001,100},\hat{\mathfrak{B}}^{[1]}_{intr};0),\hspace{10pt}\hat{W}(\mathfrak{B}_{001,100},\hat{\mathfrak{B}}^{[1]}_{intr};0),\\
 \hspace{182pt}{W}(\hat{\mathfrak{B}}_{010,100},\hat{\mathfrak{B}}^{[1]}_{intr};0),\hspace{10pt}\hat{W}(\mathfrak{B}_{010,100},\hat{\mathfrak{B}}^{[1]}_{intr};0),\\
 \hspace{182pt}{W}(\hat{\mathfrak{B}}_{011,100},\hat{\mathfrak{B}}^{[1]}_{intr};0),\hspace{10pt}\hat{W}(\mathfrak{B}_{011,100},\hat{\mathfrak{B}}^{[1]}_{intr};0),\\
 \hspace{182pt}{W}(\hat{\mathfrak{B}}_{100,100},\hat{\mathfrak{B}}^{[1]}_{intr};1),\hspace{10pt}\hat{W}(\mathfrak{B}_{100,100},\hat{\mathfrak{B}}^{[1]}_{intr};1),\\
 \hspace{182pt}{W}(\hat{\mathfrak{B}}_{101,100},\hat{\mathfrak{B}}^{[1]}_{intr};1),\hspace{10pt}\hat{W}(\mathfrak{B}_{101,100},\hat{\mathfrak{B}}^{[1]}_{intr};1),\\
 \hspace{182pt}{W}(\hat{\mathfrak{B}}_{110,100},\hat{\mathfrak{B}}^{[1]}_{intr};1),\hspace{10pt}\hat{W}(\mathfrak{B}_{110,100},\hat{\mathfrak{B}}^{[1]}_{intr};1),\\
 \hspace{186pt}{W}(\hat{\mathfrak{B}}_{111,100},\hat{\mathfrak{B}}^{[1]}_{intr};1),\hspace{10pt}\hat{W}(\mathfrak{B}_{111,100},\hat{\mathfrak{B}}^{[1]}_{intr};1)\hspace{2pt}\}
 \end{array}
 &&\hspace{-210pt}
 \begin{array}{c}
 \hat{\mathfrak{p}}_{\hspace{1pt}{\rm III}}=\\
 \{\hspace{2pt}\hat{\mathfrak{B}}^{[1,1]}_{intr},\\
 \hspace{182pt}{W}(\hat{\mathfrak{B}}_{001,100},\hat{\mathfrak{B}}^{[1]}_{intr};1),\hspace{10pt}\hat{W}(\mathfrak{B}_{001,100},\hat{\mathfrak{B}}^{[1]}_{intr};1),\\
 \hspace{182pt}{W}(\hat{\mathfrak{B}}_{010,100},\hat{\mathfrak{B}}^{[1]}_{intr};1),\hspace{10pt}\hat{W}(\mathfrak{B}_{010,100},\hat{\mathfrak{B}}^{[1]}_{intr};1),\\
 \hspace{182pt}{W}(\hat{\mathfrak{B}}_{011,100},\hat{\mathfrak{B}}^{[1]}_{intr};1),\hspace{10pt}\hat{W}(\mathfrak{B}_{011,100},\hat{\mathfrak{B}}^{[1]}_{intr};1),\\
 \hspace{182pt}{W}(\hat{\mathfrak{B}}_{100,100},\hat{\mathfrak{B}}^{[1]}_{intr};0),\hspace{10pt}\hat{W}(\mathfrak{B}_{100,100},\hat{\mathfrak{B}}^{[1]}_{intr};0),\\
 \hspace{182pt}{W}(\hat{\mathfrak{B}}_{101,100},\hat{\mathfrak{B}}^{[1]}_{intr};0),\hspace{10pt}\hat{W}(\mathfrak{B}_{101,100},\hat{\mathfrak{B}}^{[1]}_{intr};0),\\
 \hspace{182pt}{W}(\hat{\mathfrak{B}}_{110,100},\hat{\mathfrak{B}}^{[1]}_{intr};0),\hspace{10pt}\hat{W}(\mathfrak{B}_{110,100},\hat{\mathfrak{B}}^{[1]}_{intr};0),\\
 \hspace{186pt}{W}(\hat{\mathfrak{B}}_{111,100},\hat{\mathfrak{B}}^{[1]}_{intr};0),\hspace{10pt}\hat{W}(\mathfrak{B}_{111,100},\hat{\mathfrak{B}}^{[1]}_{intr};0)\hspace{2pt}\}
 \end{array}
 \end{array}\]
 \[\hspace{-13pt}\begin{array}{ccc}
 \hspace{-245pt}(b)&&\\
 \hspace{-180pt}
 \begin{array}{c}
 \hspace{-7pt}{\mathfrak{t}}_{\hspace{1pt}{\rm I}}=\\
 \hspace{167pt}\{\hspace{2pt}{W}(\hat{\mathfrak{B}}_{001,100},\hat{\mathfrak{B}}^{[1]}_{intr};1),\hspace{10pt}\hat{W}(\mathfrak{B}_{001,100},\hat{\mathfrak{B}}^{[1]}_{intr};1),\\
 \hspace{174pt}{W}(\hat{\mathfrak{B}}_{010,100},\hat{\mathfrak{B}}^{[1]}_{intr};0),\hspace{10pt}\hat{W}(\mathfrak{B}_{010,100},\hat{\mathfrak{B}}^{[1]}_{intr};0),\\
 \hspace{174pt}{W}(\hat{\mathfrak{B}}_{011,100},\hat{\mathfrak{B}}^{[1]}_{intr};1),\hspace{10pt}\hat{W}(\mathfrak{B}_{011,100},\hat{\mathfrak{B}}^{[1]}_{intr};1),\\
 \hspace{174pt}{W}(\hat{\mathfrak{B}}_{100,100},\hat{\mathfrak{B}}^{[1]}_{intr};1),\hspace{10pt}\hat{W}(\mathfrak{B}_{100,100},\hat{\mathfrak{B}}^{[1]}_{intr};1),\\
 \hspace{174pt}{W}(\hat{\mathfrak{B}}_{101,100},\hat{\mathfrak{B}}^{[1]}_{intr};0),\hspace{10pt}\hat{W}(\mathfrak{B}_{101,100},\hat{\mathfrak{B}}^{[1]}_{intr};0),\\
 \hspace{174pt}{W}(\hat{\mathfrak{B}}_{110,100},\hat{\mathfrak{B}}^{[1]}_{intr};0),\hspace{10pt}\hat{W}(\mathfrak{B}_{110,100},\hat{\mathfrak{B}}^{[1]}_{intr};0),\\
 \hspace{178pt}{W}(\hat{\mathfrak{B}}_{111,100},\hat{\mathfrak{B}}^{[1]}_{intr};0),\hspace{10pt}\hat{W}(\mathfrak{B}_{111,100},\hat{\mathfrak{B}}^{[1]}_{intr};0)\hspace{2pt}\}\\
 \\
 \end{array}
 &&\hspace{-213pt}
 \begin{array}{c}
 {\mathfrak{p}}_{\hspace{1pt}{\rm I}}=\\
 \hspace{68pt}\{\hspace{2pt}\mathfrak{C}_{\rm III}=\hat{\mathfrak{B}}^{[1]}_{intr}\cup\hat{\mathfrak{B}}^{[1,1]}_{intr},\\
 \hspace{183pt}{W}(\hat{\mathfrak{B}}_{001,100},\hat{\mathfrak{B}}^{[1]}_{intr};0),\hspace{10pt}\hat{W}(\mathfrak{B}_{001,100},\hat{\mathfrak{B}}^{[1]}_{intr};0),\\
 \hspace{183pt}{W}(\hat{\mathfrak{B}}_{010,100},\hat{\mathfrak{B}}^{[1]}_{intr};1),\hspace{10pt}\hat{W}(\mathfrak{B}_{010,100},\hat{\mathfrak{B}}^{[1]}_{intr};1),\\
 \hspace{183pt}{W}(\hat{\mathfrak{B}}_{011,100},\hat{\mathfrak{B}}^{[1]}_{intr};0),\hspace{10pt}\hat{W}(\mathfrak{B}_{011,100},\hat{\mathfrak{B}}^{[1]}_{intr};0),\\
 \hspace{183pt}{W}(\hat{\mathfrak{B}}_{100,100},\hat{\mathfrak{B}}^{[1]}_{intr};0),\hspace{10pt}\hat{W}(\mathfrak{B}_{100,100},\hat{\mathfrak{B}}^{[1]}_{intr};0),\\
 \hspace{183pt}{W}(\hat{\mathfrak{B}}_{101,100},\hat{\mathfrak{B}}^{[1]}_{intr};1),\hspace{10pt}\hat{W}(\mathfrak{B}_{101,100},\hat{\mathfrak{B}}^{[1]}_{intr};1),\\
 \hspace{183pt}{W}(\hat{\mathfrak{B}}_{110,100},\hat{\mathfrak{B}}^{[1]}_{intr};1),\hspace{10pt}\hat{W}(\mathfrak{B}_{110,100},\hat{\mathfrak{B}}^{[1]}_{intr};1),\\
 \hspace{187pt}{W}(\hat{\mathfrak{B}}_{111,100},\hat{\mathfrak{B}}^{[1]}_{intr};1),\hspace{10pt}\hat{W}(\mathfrak{B}_{111,100},\hat{\mathfrak{B}}^{[1]}_{intr};1)\hspace{2pt}\}
 \end{array}
 \end{array}\]
 \[\hspace{-10pt}\begin{array}{ccc}
 \hspace{-190pt}
 \begin{array}{c}
 {\mathfrak{t}}_{\hspace{1pt}{\rm III}}=\\
 \{\hspace{2pt}\hat{\mathfrak{B}}^{[1]}_{intr},\\
 \hspace{182pt}{W}(\hat{\mathfrak{B}}_{001,100},\hat{\mathfrak{B}}^{[1]}_{intr};0),\hspace{10pt}\hat{W}(\mathfrak{B}_{001,100},\hat{\mathfrak{B}}^{[1]}_{intr};0),\\
 \hspace{182pt}{W}(\hat{\mathfrak{B}}_{010,100},\hat{\mathfrak{B}}^{[1]}_{intr};1),\hspace{10pt}\hat{W}(\mathfrak{B}_{010,100},\hat{\mathfrak{B}}^{[1]}_{intr};1),\\
 \hspace{182pt}{W}(\hat{\mathfrak{B}}_{011,100},\hat{\mathfrak{B}}^{[1]}_{intr};1),\hspace{10pt}\hat{W}(\mathfrak{B}_{011,100},\hat{\mathfrak{B}}^{[1]}_{intr};1),\\
 \hspace{182pt}{W}(\hat{\mathfrak{B}}_{100,100},\hat{\mathfrak{B}}^{[1]}_{intr};0),\hspace{10pt}\hat{W}(\mathfrak{B}_{100,100},\hat{\mathfrak{B}}^{[1]}_{intr};0),\\
 \hspace{182pt}{W}(\hat{\mathfrak{B}}_{101,100},\hat{\mathfrak{B}}^{[1]}_{intr};0),\hspace{10pt}\hat{W}(\mathfrak{B}_{101,100},\hat{\mathfrak{B}}^{[1]}_{intr};0),\\
 \hspace{182pt}{W}(\hat{\mathfrak{B}}_{110,100},\hat{\mathfrak{B}}^{[1]}_{intr};1),\hspace{10pt}\hat{W}(\mathfrak{B}_{110,100},\hat{\mathfrak{B}}^{[1]}_{intr};1),\\
 \hspace{186pt}{W}(\hat{\mathfrak{B}}_{111,100},\hat{\mathfrak{B}}^{[1]}_{intr};1),\hspace{10pt}\hat{W}(\mathfrak{B}_{111,100},\hat{\mathfrak{B}}^{[1]}_{intr};1)\hspace{2pt}\}
 \end{array}
 &&\hspace{-210pt}
 \begin{array}{c}
 {\mathfrak{p}}_{\hspace{1pt}{\rm III}}=\\
 \{\hspace{2pt}\hat{\mathfrak{B}}^{[1,1]}_{intr},\\
 \hspace{182pt}{W}(\hat{\mathfrak{B}}_{001,100},\hat{\mathfrak{B}}^{[1]}_{intr};1),\hspace{10pt}\hat{W}(\mathfrak{B}_{001,100},\hat{\mathfrak{B}}^{[1]}_{intr};1),\\
 \hspace{182pt}{W}(\hat{\mathfrak{B}}_{010,100},\hat{\mathfrak{B}}^{[1]}_{intr};0),\hspace{10pt}\hat{W}(\mathfrak{B}_{010,100},\hat{\mathfrak{B}}^{[1]}_{intr};0),\\
 \hspace{182pt}{W}(\hat{\mathfrak{B}}_{011,100},\hat{\mathfrak{B}}^{[1]}_{intr};0),\hspace{10pt}\hat{W}(\mathfrak{B}_{011,100},\hat{\mathfrak{B}}^{[1]}_{intr};0),\\
 \hspace{182pt}{W}(\hat{\mathfrak{B}}_{100,100},\hat{\mathfrak{B}}^{[1]}_{intr};1),\hspace{10pt}\hat{W}(\mathfrak{B}_{100,100},\hat{\mathfrak{B}}^{[1]}_{intr};1),\\
 \hspace{182pt}{W}(\hat{\mathfrak{B}}_{101,100},\hat{\mathfrak{B}}^{[1]}_{intr};1),\hspace{10pt}\hat{W}(\mathfrak{B}_{101,100},\hat{\mathfrak{B}}^{[1]}_{intr};1),\\
 \hspace{182pt}{W}(\hat{\mathfrak{B}}_{110,100},\hat{\mathfrak{B}}^{[1]}_{intr};0),\hspace{10pt}\hat{W}(\mathfrak{B}_{110,100},\hat{\mathfrak{B}}^{[1]}_{intr};0),\\
 \hspace{186pt}{W}(\hat{\mathfrak{B}}_{111,100},\hat{\mathfrak{B}}^{[1]}_{intr};0),\hspace{10pt}\hat{W}(\mathfrak{B}_{111,100},\hat{\mathfrak{B}}^{[1]}_{intr};0)\hspace{2pt}\}
 \end{array}
 \end{array}\]
 \fcaption{Three examples of Cartan decompositions determined from the intrinsic co-quotient algebra of $su(8)$ in
 Fig.~\ref{figsu8coQArank1intr}:
 (a) the intrinsic decomposition of type {\bf AIII}
 $\hat{\mathfrak{t}}_{\hspace{1pt}{\rm III}}\oplus\hat{\mathfrak{p}}_{\hspace{1pt}{\rm III}}$;
 (b) a type {\bf AI} and a type {\bf AIII}. \label{figsu8tpintrcorank1}}
 \end{figure}

 \begin{figure}[htbp]
 \[\begin{array}{ccc}
 \hspace{-132pt}
 \begin{array}{c}
 \hspace{-8pt}{\mathfrak{t}}_{\hspace{1pt}{\rm I}}=\\
 \hspace{110pt}\{\hspace{2pt}\hat{W}(\mathfrak{B}_1,\mathfrak{B}^{[1]};0),\hspace{10pt}\hat{W}(\mathfrak{B}_1,\mathfrak{B}^{[1]};0),\\
 \hspace{117pt}\hat{W}(\mathfrak{B}_2,\mathfrak{B}^{[1]};1),\hspace{10pt}\hat{W}(\mathfrak{B}_2,\mathfrak{B}^{[1]};1),\\
 \hspace{117pt}\hat{W}(\mathfrak{B}_3,\mathfrak{B}^{[1]};1),\hspace{10pt}\hat{W}(\mathfrak{B}_3,\mathfrak{B}^{[1]};1),\\
 \hspace{117pt}{W}(\mathfrak{B}_4,\mathfrak{B}^{[1]};1),\hspace{10pt}{W}(\mathfrak{B}_4,\mathfrak{B}^{[1]};1),\\
 \hspace{117pt}{W}(\mathfrak{B}_5,\mathfrak{B}^{[1]};1),\hspace{10pt}{W}(\mathfrak{B}_5,\mathfrak{B}^{[1]};1),\\
 \hspace{117pt}{W}(\mathfrak{B}_6,\mathfrak{B}^{[1]};0),\hspace{10pt}{W}(\mathfrak{B}_6,\mathfrak{B}^{[1]};0),\\
 \hspace{122pt}{W}(\mathfrak{B}_7,\mathfrak{B}^{[1]};0),\hspace{10pt}{W}(\mathfrak{B}_7,\mathfrak{B}^{[1]};0)\hspace{2pt}\}\\
 \\
 \end{array}
 &\hspace{5pt}&\hspace{-153pt}
 \begin{array}{c}
 \hspace{15pt}{\mathfrak{p}}_{\hspace{1pt}{\rm I}}=\\
 \hspace{63pt}\{\hspace{2pt}\mathfrak{C}=\mathfrak{B}^{[1]}\cup\mathfrak{B}^{[1,1]},\\
 \hspace{137pt}{W}(\mathfrak{B}_1,\mathfrak{B}^{[1]};1),\hspace{10pt}{W}(\mathfrak{B}_1,\mathfrak{B}^{[1]};1),\\
 \hspace{137pt}{W}(\mathfrak{B}_2,\mathfrak{B}^{[1]};0),\hspace{10pt}{W}(\mathfrak{B}_2,\mathfrak{B}^{[1]};0),\\
 \hspace{137pt}{W}(\mathfrak{B}_3,\mathfrak{B}^{[1]};0),\hspace{10pt}{W}(\mathfrak{B}_3,\mathfrak{B}^{[1]};0),\\
 \hspace{137pt}\hat{W}(\mathfrak{B}_4,\mathfrak{B}^{[1]};0),\hspace{10pt}\hat{W}(\mathfrak{B}_4,\mathfrak{B}^{[1]};0),\\
 \hspace{137pt}\hat{W}(\mathfrak{B}_5,\mathfrak{B}^{[1]};0),\hspace{10pt}\hat{W}(\mathfrak{B}_5,\mathfrak{B}^{[1]};0),\\
 \hspace{137pt}\hat{W}(\mathfrak{B}_6,\mathfrak{B}^{[1]};1),\hspace{10pt}\hat{W}(\mathfrak{B}_6,\mathfrak{B}^{[1]};1),\\
 \hspace{141pt}\hat{W}(\mathfrak{B}_7,\mathfrak{B}^{[1]};1),\hspace{10pt}\hat{W}(\mathfrak{B}_7,\mathfrak{B}^{[1]};1)\hspace{2pt}\}
 \end{array}
 \end{array}\]
 \[\begin{array}{ccc}
 \hspace{-147pt}
 \begin{array}{c}
 \hspace{10pt}{\mathfrak{t}}_{\hspace{1pt}{\rm III}}=\\
 \{\hspace{2pt}\mathfrak{B}^{[1]},\\
 \hspace{131pt}{W}(\mathfrak{B}_1,\mathfrak{B}^{[1]};1),\hspace{10pt}\hat{W}(\mathfrak{B}_1,\mathfrak{B}^{[1]};1),\\
 \hspace{131pt}{W}(\mathfrak{B}_2,\mathfrak{B}^{[1]};1),\hspace{10pt}\hat{W}(\mathfrak{B}_2,\mathfrak{B}^{[1]};1),\\
 \hspace{131pt}{W}(\mathfrak{B}_3,\mathfrak{B}^{[1]};0),\hspace{10pt}\hat{W}(\mathfrak{B}_3,\mathfrak{B}^{[1]};0),\\
 \hspace{131pt}{W}(\mathfrak{B}_4,\mathfrak{B}^{[1]};1),\hspace{10pt}\hat{W}(\mathfrak{B}_4,\mathfrak{B}^{[1]};1),\\
 \hspace{131pt}{W}(\mathfrak{B}_5,\mathfrak{B}^{[1]};0),\hspace{10pt}\hat{W}(\mathfrak{B}_5,\mathfrak{B}^{[1]};0),\\
 \hspace{131pt}{W}(\mathfrak{B}_6,\mathfrak{B}^{[1]};0),\hspace{10pt}\hat{W}(\mathfrak{B}_6,\mathfrak{B}^{[1]};0),\\
 \hspace{135pt}{W}(\mathfrak{B}_7,\mathfrak{B}^{[1]};1),\hspace{10pt}\hat{W}(\mathfrak{B}_7,\mathfrak{B}^{[1]};1)\hspace{2pt}\}
 \end{array}
 &\hspace{35pt}&\hspace{-173pt}
 \begin{array}{c}
 \hspace{5pt}{\mathfrak{p}}_{\hspace{1pt}{\rm III}}=\\
 \{\hspace{2pt}\mathfrak{B}^{[1,1]},\\
 \hspace{124pt}{W}(\mathfrak{B}_1,\mathfrak{B}^{[1]};0),\hspace{10pt}\hat{W}(\mathfrak{B}_1,\mathfrak{B}^{[1]};0),\\
 \hspace{124pt}{W}(\mathfrak{B}_2,\mathfrak{B}^{[1]};0),\hspace{10pt}\hat{W}(\mathfrak{B}_2,\mathfrak{B}^{[1]};0),\\
 \hspace{124pt}{W}(\mathfrak{B}_3,\mathfrak{B}^{[1]};1),\hspace{10pt}\hat{W}(\mathfrak{B}_3,\mathfrak{B}^{[1]};1),\\
 \hspace{124pt}{W}(\mathfrak{B}_4,\mathfrak{B}^{[1]};0),\hspace{10pt}\hat{W}(\mathfrak{B}_4,\mathfrak{B}^{[1]};0),\\
 \hspace{124pt}{W}(\mathfrak{B}_5,\mathfrak{B}^{[1]};1),\hspace{10pt}\hat{W}(\mathfrak{B}_5,\mathfrak{B}^{[1]};1),\\
 \hspace{124pt}{W}(\mathfrak{B}_6,\mathfrak{B}^{[1]};1),\hspace{10pt}\hat{W}(\mathfrak{B}_6,\mathfrak{B}^{[1]};1),\\
 \hspace{128pt}{W}(\mathfrak{B}_7,\mathfrak{B}^{[1]};0),\hspace{10pt}\hat{W}(\mathfrak{B}_7,\mathfrak{B}^{[1]};0)\hspace{2pt}\}
 \end{array}
 \end{array}\]
 \fcaption{Examples of a type-{\bf AI} and a type-{\bf AIII} Cartan decompositions determined from the co-quotient
 algebra of $su(8)$ in Fig.~\ref{figsu8coQArank1inC}. \label{figsu8tpcorank1}}
 \end{figure}

 \newpage
 \begin{figure}[p]
 \begin{center}
 \[\begin{array}{c}
 \begin{array}{cc}
 \begin{array}{c}
 \hat{\mathfrak{t}}_{\hspace{1pt}{\rm III}}=
 \{\hspace{2pt}\mathfrak{C}_{[\mathbf{0}]},
 \hspace{2pt}{W}(\mathfrak{B}_{001};0),\hspace{5pt}\hat{W}(\mathfrak{B}_{001};0),\\
 \hspace{55pt}{W}(\mathfrak{B}_{010};0),\hspace{5pt}\hat{W}(\mathfrak{B}_{010};0),\\
 \hspace{55pt}{W}(\mathfrak{B}_{011};0),\hspace{5pt}\hat{W}(\mathfrak{B}_{011};0),\\
 \hspace{55pt}{W}(\mathfrak{B}_{100};1),\hspace{5pt}\hat{W}(\mathfrak{B}_{100};1),\\
 \hspace{55pt}{W}(\mathfrak{B}_{101};1),\hspace{5pt}\hat{W}(\mathfrak{B}_{101};1),\\
 \hspace{55pt}{W}(\mathfrak{B}_{110};1),\hspace{5pt}\hat{W}(\mathfrak{B}_{110};1),\\
 \hspace{57pt}{W}(\mathfrak{B}_{111};1),\hspace{5pt}\hat{W}(\mathfrak{B}_{111};1)\hspace{2pt}\}
 \end{array}
 &\hspace{-18pt}
 \begin{array}{c}
 =\hspace{3pt}\{\mathfrak{C}_{[\mathbf{0}]},
 \hspace{2pt}\lambda_{12},\lambda_{34},\lambda_{56},\hat{\lambda}_{12},\hat{\lambda}_{34},\hat{\lambda}_{56},\\
 \hspace{41pt}\lambda_{13},\lambda_{24},\lambda_{68},\hat{\lambda}_{13},\hat{\lambda}_{24},\hat{\lambda}_{68},\\
 \hspace{41pt}\lambda_{14},\lambda_{23},\lambda_{58},\hat{\lambda}_{14},\hat{\lambda}_{23},\hat{\lambda}_{58},\\
 \hspace{-34pt}\lambda_{47}, \hat{\lambda}_{47},\\
 \hspace{-34pt} \lambda_{37},\hat{\lambda}_{37},\\
 \hspace{-34pt}\lambda_{27},\hat{\lambda}_{27},\\
 \hspace{-32pt} \lambda_{17},\hat{\lambda}_{17}\}
 \end{array}
 \end{array}
 \\
 \\
 \begin{array}{cc}
 \hspace{0pt}
 \begin{array}{c}
 \hat{\mathfrak{p}}_{\hspace{1pt}{\rm III}}=
 \{\hspace{21pt}{W}(\mathfrak{B}_{001};1),\hspace{5pt}\hat{W}(\mathfrak{B}_{001};1),\\
 \hspace{55pt}{W}(\mathfrak{B}_{010};1),\hspace{5pt}\hat{W}(\mathfrak{B}_{010};1),\\
 \hspace{55pt}{W}(\mathfrak{B}_{011};1),\hspace{5pt}\hat{W}(\mathfrak{B}_{011};1),\\
 \hspace{55pt}{W}(\mathfrak{B}_{100};0),\hspace{5pt}\hat{W}(\mathfrak{B}_{100};0),\\
 \hspace{55pt}{W}(\mathfrak{B}_{101};0),\hspace{5pt}\hat{W}(\mathfrak{B}_{101};0),\\
 \hspace{55pt}{W}(\mathfrak{B}_{110};0),\hspace{5pt}\hat{W}(\mathfrak{B}_{110};0),\\
 \hspace{57pt}{W}(\mathfrak{B}_{111};0),\hspace{5pt}\hat{W}(\mathfrak{B}_{111};0)\}
 \end{array}
 &\hspace{-93pt}
 \begin{array}{c}
 =\hspace{3pt}\{ \hspace{21pt}\lambda_{78},\hat{\lambda}_{78},\\
 \hspace{41pt}\lambda_{57},\hat{\lambda}_{57},\\
 \hspace{41pt}\lambda_{67},\hat{\lambda}_{67},\\
 \hspace{116pt}\lambda_{16},\lambda_{25},\lambda_{38},\hat{\lambda}_{16},\hat{\lambda}_{25},\hat{\lambda}_{38},\\
 \hspace{116pt}\lambda_{26},\lambda_{15},\lambda_{48},\hat{\lambda}_{26},\hat{\lambda}_{15},\hat{\lambda}_{48},\\
 \hspace{116pt}\lambda_{36},\lambda_{45},\lambda_{18},\hat{\lambda}_{36},\hat{\lambda}_{45},\hat{\lambda}_{18},\\
 \hspace{118pt}\lambda_{46},\lambda_{35},\lambda_{28},\hat{\lambda}_{46},\hat{\lambda}_{35},\hat{\lambda}_{28}\}
 \end{array}
 \end{array}
 \end{array}\]
 \fcaption{The intrinsic Cartan decomposition of type {\bf AIII}
 $\hat{\mathfrak{t}}_{\hspace{1pt}{\rm III}}\oplus\hat{\mathfrak{p}}_{\hspace{1pt}{\rm III}}$
 for $su(5+3)$ determined from the co-quotient algebra of $su(8)$ in
 Fig.~\ref{figsu8QA5plus3}, and the version written in the permuted $\lambda$-generators.\label{figsu8tp5+3}}
 \end{center}
 \end{figure}

 \begin{figure}[htbp]
 \begin{center}
 \[\begin{array}{c}
 \hspace{-380pt}(a)\\
 \hspace{-20pt}
 \begin{array}{cc}
 \begin{array}{c}
 \hat{\mathfrak{t}}_{\hspace{1pt}{\rm III}}=
 \{\hspace{2pt}\mathfrak{C}_{[\mathbf{0}]},
 \hspace{2pt}{W}(\mathfrak{B}_{001};0),\hspace{5pt}\hat{W}(\mathfrak{B}_{001};0),\\
 \hspace{55pt}{W}(\mathfrak{B}_{010};0),\hspace{5pt}\hat{W}(\mathfrak{B}_{010};0),\\
 \hspace{55pt}{W}(\mathfrak{B}_{011};0),\hspace{5pt}\hat{W}(\mathfrak{B}_{011};0),\\
 \hspace{55pt}{W}(\mathfrak{B}_{100};1),\hspace{5pt}\hat{W}(\mathfrak{B}_{100};1),\\
 \hspace{55pt}{W}(\mathfrak{B}_{101};1),\hspace{5pt}\hat{W}(\mathfrak{B}_{101};1),\\
 \hspace{55pt}{W}(\mathfrak{B}_{110};1),\hspace{5pt}\hat{W}(\mathfrak{B}_{110};1),\\
 \hspace{57pt}{W}(\mathfrak{B}_{111};1),\hspace{5pt}\hat{W}(\mathfrak{B}_{111};1)\hspace{2pt}\}
 \\
 \\
 \end{array}
 &\hspace{-25pt}
 \begin{array}{c}
 =\hspace{3pt}\{\hspace{2pt}\mathfrak{C}_{[\mathbf{0}]},
 \hspace{2pt}\lambda_{12},\lambda_{34},\lambda_{56},\lambda_{78},\\
 \hspace{43pt}\hat{\lambda}_{12},\hat{\lambda}_{34},\hat{\lambda}_{56},\hat{\lambda}_{78},\\
 \hspace{43pt}\lambda_{13},\lambda_{24},\hat{\lambda}_{13},\hat{\lambda}_{24},\\
 \hspace{43pt}\lambda_{14},\lambda_{23},\hat{\lambda}_{14},\hat{\lambda}_{23},\\
 \hspace{43pt}\lambda_{35},\lambda_{46}, \hat{\lambda}_{35},\hat{\lambda}_{46},\\
 \hspace{43pt}\lambda_{36},\lambda_{45}, \hat{\lambda}_{36},\hat{\lambda}_{45},\\
 \hspace{43pt}\lambda_{15},\lambda_{26}, \hat{\lambda}_{15},\hat{\lambda}_{26},\\
 \hspace{52pt}\lambda_{16},\lambda_{25}, \hat{\lambda}_{16},\hat{\lambda}_{25},\hspace{2pt}\}
 \end{array}
 \end{array}
 \\
 \\
 \hspace{38pt}
 \begin{array}{cc}
 \hspace{-70pt}
 \begin{array}{c}
 \hat{\mathfrak{p}}_{\hspace{1pt}{\rm III}}=
 \{\hspace{19pt}{W}(\mathfrak{B}_{001};1),\hspace{5pt}\hat{W}(\mathfrak{B}_{001};1),\\
 \hspace{53pt}{W}(\mathfrak{B}_{010};1),\hspace{5pt}\hat{W}(\mathfrak{B}_{010};1),\\
 \hspace{53pt}{W}(\mathfrak{B}_{011};1),\hspace{5pt}\hat{W}(\mathfrak{B}_{011};1),\\
 \hspace{53pt}{W}(\mathfrak{B}_{100};0),\hspace{5pt}\hat{W}(\mathfrak{B}_{100};0),\\
 \hspace{53pt}{W}(\mathfrak{B}_{101};0),\hspace{5pt}\hat{W}(\mathfrak{B}_{101};0),\\
 \hspace{53pt}{W}(\mathfrak{B}_{110};0),\hspace{5pt}\hat{W}(\mathfrak{B}_{110};0),\\
 \hspace{55pt}{W}(\mathfrak{B}_{111};0),\hspace{5pt}\hat{W}(\mathfrak{B}_{111};0)\}
 \end{array}
 &\hspace{-60pt}
 \begin{array}{c}
 =\hspace{3pt}\{ \hspace{52pt}\\
 \hspace{83pt}\lambda_{57},\lambda_{68},\hat{\lambda}_{57},\hat{\lambda}_{68},\\
 \hspace{83pt}\lambda_{58},\lambda_{67},\hat{\lambda}_{58},\hat{\lambda}_{67},\\
 \hspace{83pt}\lambda_{17},\lambda_{28},\hat{\lambda}_{17},\hat{\lambda}_{28},\\
 \hspace{83pt}\lambda_{18},\lambda_{27},\hat{\lambda}_{18},\hat{\lambda}_{27},\\
 \hspace{83pt}\lambda_{37},\lambda_{48},\hat{\lambda}_{37},\hat{\lambda}_{48},\\
 \hspace{87pt}\lambda_{38},\lambda_{47},\hat{\lambda}_{38},\hat{\lambda}_{47}\hspace{2pt}\}
 \end{array}
 \end{array}
 \end{array}\]
 \[\begin{array}{ccc}
 \hspace{-205pt}(b)&&\\
 \hspace{-33pt}
 \begin{array}{c}
 \hspace{38pt}\mathfrak{t}_{\hspace{1pt}{\rm III}}=
 \{\mathfrak{C}_{[\mathbf{0}]},
 \hspace{2pt}{W}(\mathfrak{B}_{001};1),\hspace{5pt}\hat{W}(\mathfrak{B}_{001};1),\\
 \hspace{92pt}{W}(\mathfrak{B}_{010};0),\hspace{5pt}\hat{W}(\mathfrak{B}_{010};0),\\
 \hspace{92pt}{W}(\mathfrak{B}_{011};1),\hspace{5pt}\hat{W}(\mathfrak{B}_{011};1),\\
 \hspace{92pt}{W}(\mathfrak{B}_{100};1),\hspace{5pt}\hat{W}(\mathfrak{B}_{100};1),\\
 \hspace{92pt}{W}(\mathfrak{B}_{101};0),\hspace{5pt}\hat{W}(\mathfrak{B}_{101};0),\\
 \hspace{92pt}{W}(\mathfrak{B}_{110};1),\hspace{5pt}\hat{W}(\mathfrak{B}_{110};1),\\
 \hspace{94pt}{W}(\mathfrak{B}_{111};1),\hspace{5pt}\hat{W}(\mathfrak{B}_{111};1)\}
 \end{array}
 &\hspace{10pt}&
 \hspace{-0pt}\begin{array}{c}
 \hspace{-28pt}\mathfrak{p}_{\hspace{1pt}{\rm III}}=
 \{ \hspace{0pt}{W}(\mathfrak{B}_{001};0),\hspace{5pt}\hat{W}(\mathfrak{B}_{001};0),\\
 \hspace{6pt}{W}(\mathfrak{B}_{010};1),\hspace{5pt}\hat{W}(\mathfrak{B}_{010};1),\\
 \hspace{6pt}{W}(\mathfrak{B}_{011};0),\hspace{5pt}\hat{W}(\mathfrak{B}_{011};0),\\
 \hspace{6pt}{W}(\mathfrak{B}_{100};0),\hspace{5pt}\hat{W}(\mathfrak{B}_{100};0),\\
 \hspace{6pt}{W}(\mathfrak{B}_{101};1),\hspace{5pt}\hat{W}(\mathfrak{B}_{101};1),\\
 \hspace{6pt}{W}(\mathfrak{B}_{110};0),\hspace{5pt}\hat{W}(\mathfrak{B}_{110};0),\\
 \hspace{8pt}{W}(\mathfrak{B}_{111};1),\hspace{5pt}\hat{W}(\mathfrak{B}_{111};1)\}
 \\
 \end{array}
 \end{array}\]
 \fcaption{Two examples of Cartan decompositions of type {\bf AIII}
 determined from the co-quotient algebra of $su(8)$ in
 Fig.~\ref{figsu8QA6plus2}: (a) the intrinsic decomposition
 $\hat{\mathfrak{t}}_{\hspace{1pt}{\rm III}}\oplus\hat{\mathfrak{p}}_{\hspace{1pt}{\rm III}}$ for $su(6+2)$
 and that in the permuted $\lambda$-generators; (b) a type-{\bf AIII} decomposition for $su(4+4)$.\label{figsu8tp6+2}}
 \end{center}
 \end{figure}

 \newpage
 \begin{figure}[p]
 \begin{center}
 \[\begin{array}{c}
 \hspace{-360pt}(a)\\
 \begin{array}{cc}
 \begin{array}{c}
 \hat{\mathfrak{t}}_{\hspace{1pt}{\rm III}}=
 \{\hspace{2pt}\mathfrak{C}_{[\mathbf{0}]},
 \hspace{2pt}{W}(\mathfrak{B}_{001};0),\hspace{5pt}\hat{W}(\mathfrak{B}_{001};0),\\
 \hspace{55pt}{W}(\mathfrak{B}_{010};0),\hspace{5pt}\hat{W}(\mathfrak{B}_{010};0),\\
 \hspace{55pt}{W}(\mathfrak{B}_{011};0),\hspace{5pt}\hat{W}(\mathfrak{B}_{011};0),\\
 \hspace{55pt}{W}(\mathfrak{B}_{100};1),\hspace{5pt}\hat{W}(\mathfrak{B}_{100};1),\\
 \hspace{55pt}{W}(\mathfrak{B}_{101};1),\hspace{5pt}\hat{W}(\mathfrak{B}_{101};1),\\
 \hspace{55pt}{W}(\mathfrak{B}_{110};1),\hspace{5pt}\hat{W}(\mathfrak{B}_{110};1),\\
 \hspace{59pt}{W}(\mathfrak{B}_{111};1),\hspace{5pt}\hat{W}(\mathfrak{B}_{111};1)\hspace{2pt}\}
 \end{array}
 &\hspace{-23pt}
 \begin{array}{c}
 =\hspace{3pt}\{\hspace{2pt}\mathfrak{C}_{[\mathbf{0}]},
 \hspace{2pt}\lambda_{12},\lambda_{34},\lambda_{57},\hat{\lambda}_{12},\hat{\lambda}_{34},\hat{\lambda}_{57},\\
 \hspace{43pt}\lambda_{13},\lambda_{24},\lambda_{56},\hat{\lambda}_{13},\hat{\lambda}_{24},\hat{\lambda}_{56},\\
 \hspace{43pt}\lambda_{14},\lambda_{23},\lambda_{67},\hat{\lambda}_{14},\hat{\lambda}_{23},\hat{\lambda}_{67},\\
 \hspace{43pt}\lambda_{26},\lambda_{37},\lambda_{45}, \hat{\lambda}_{26},\hat{\lambda}_{37},\hat{\lambda}_{45},\\
 \hspace{43pt}\lambda_{16},\lambda_{35},\lambda_{47},\hat{\lambda}_{16},\hat{\lambda}_{35},\hat{\lambda}_{47},\\
 \hspace{43pt}\lambda_{17},\lambda_{25},\lambda_{46},\hat{\lambda}_{17},\hat{\lambda}_{25},\hat{\lambda}_{46},\\
 \hspace{52pt}\lambda_{15},\lambda_{27},\lambda_{36},\hat{\lambda}_{15},\hat{\lambda}_{27},\hat{\lambda}_{36},\hspace{2pt}\}
 \end{array}
 \end{array}
 \\
 \\
 \begin{array}{cc}
 \hspace{-77pt}
 \begin{array}{c}
 \hat{\mathfrak{p}}_{\hspace{1pt}{\rm III}}=
 \{\hspace{21pt}{W}(\mathfrak{B}_{001};1),\hspace{5pt}\hat{W}(\mathfrak{B}_{001};1),\\
 \hspace{55pt}{W}(\mathfrak{B}_{010};1),\hspace{5pt}\hat{W}(\mathfrak{B}_{010};1),\\
 \hspace{55pt}{W}(\mathfrak{B}_{011};1),\hspace{5pt}\hat{W}(\mathfrak{B}_{011};1),\\
 \hspace{55pt}{W}(\mathfrak{B}_{100};0),\hspace{5pt}\hat{W}(\mathfrak{B}_{100};0),\\
 \hspace{55pt}{W}(\mathfrak{B}_{101};0),\hspace{5pt}\hat{W}(\mathfrak{B}_{101};0),\\
 \hspace{55pt}{W}(\mathfrak{B}_{110};0),\hspace{5pt}\hat{W}(\mathfrak{B}_{110};0),\\
 \hspace{59pt}{W}(\mathfrak{B}_{111};0),\hspace{5pt}\hat{W}(\mathfrak{B}_{111};0)\hspace{2pt}\}
 \end{array}
 &\hspace{-20pt}
 \begin{array}{c}
 =\hspace{3pt}\{\hspace{21pt}\lambda_{68},\hat{\lambda}_{68}\\
 \hspace{43pt}\lambda_{78},\hat{\lambda}_{78},\\
 \hspace{43pt}\lambda_{58},\hat{\lambda}_{58},\\
 \hspace{43pt}\lambda_{18},\hat{\lambda}_{18},\\
 \hspace{43pt}\lambda_{28},\hat{\lambda}_{28},\\
 \hspace{43pt}\lambda_{38},\hat{\lambda}_{38},\\
 \hspace{47pt}\lambda_{48},\hat{\lambda}_{48}\hspace{2pt}\}
 \end{array}
 \end{array}
 \end{array}\]
 \[\begin{array}{ccc}
 \hspace{-190pt}(b)&&\\
 \hspace{-30pt}
 \begin{array}{c}
 \hspace{38pt}\mathfrak{t}_{\hspace{1pt}{\rm III}}=
 \{\hspace{2pt}\mathfrak{C}_{[\mathbf{0}]},
 \hspace{2pt}{W}(\mathfrak{B}_{001};1),\hspace{5pt}\hat{W}(\mathfrak{B}_{001};1),\\
 \hspace{95pt}{W}(\mathfrak{B}_{010};0),\hspace{5pt}\hat{W}(\mathfrak{B}_{010};0),\\
 \hspace{95pt}{W}(\mathfrak{B}_{011};1),\hspace{5pt}\hat{W}(\mathfrak{B}_{011};1),\\
 \hspace{95pt}{W}(\mathfrak{B}_{100};1),\hspace{5pt}\hat{W}(\mathfrak{B}_{100};1),\\
 \hspace{95pt}{W}(\mathfrak{B}_{101};0),\hspace{5pt}\hat{W}(\mathfrak{B}_{101};0),\\
 \hspace{95pt}{W}(\mathfrak{B}_{110};1),\hspace{5pt}\hat{W}(\mathfrak{B}_{110};1),\\
 \hspace{99pt}{W}(\mathfrak{B}_{111};1),\hspace{5pt}\hat{W}(\mathfrak{B}_{111};1)\hspace{2pt}\}
 \end{array}
 &\hspace{10pt}&
 \hspace{0pt}\begin{array}{c}
 \hspace{-28pt}\mathfrak{p}_{\hspace{1pt}{\rm III}}=
 \{ \hspace{2pt}{W}(\mathfrak{B}_{001};0),\hspace{5pt}\hat{W}(\mathfrak{B}_{001};0),\\
 \hspace{8pt}{W}(\mathfrak{B}_{010};1),\hspace{5pt}\hat{W}(\mathfrak{B}_{010};1),\\
 \hspace{8pt}{W}(\mathfrak{B}_{011};0),\hspace{5pt}\hat{W}(\mathfrak{B}_{011};0),\\
 \hspace{8pt}{W}(\mathfrak{B}_{100};0),\hspace{5pt}\hat{W}(\mathfrak{B}_{100};0),\\
 \hspace{8pt}{W}(\mathfrak{B}_{101};1),\hspace{5pt}\hat{W}(\mathfrak{B}_{101};1),\\
 \hspace{8pt}{W}(\mathfrak{B}_{110};0),\hspace{5pt}\hat{W}(\mathfrak{B}_{110};0),\\
 \hspace{12pt}{W}(\mathfrak{B}_{111};1),\hspace{5pt}\hat{W}(\mathfrak{B}_{111};1)\hspace{2pt}\}
 \end{array}
 \end{array}\]
 \fcaption{Two examples of Cartan decompositions of type {\bf AIII} determined from
 the co-quotient algebra of $su(8)$ in
 Fig.~\ref{figsu8QA7plus1}: (a) the intrinsic decomposition
 $\hat{\mathfrak{t}}_{\hspace{1pt}{\rm III}}\oplus\hat{\mathfrak{p}}_{\hspace{1pt}{\rm III}}$
 for $su(7+1)$ and that in the permuted $\lambda$-generators;
 (b) a type-{\bf AIII} decomposition for $su(5+3)$.    \label{figsu8tp7+1}}
 \end{center}
 \end{figure}

\end{document}